\documentclass{jpp}

\usepackage[toc, xindy, acronym]{glossaries}
\usepackage{natbib}

\usepackage[title]{appendix}

\usepackage[setpagesize=false]{hyperref} 
\hypersetup{
    colorlinks,
    linkcolor={blue!80!black},
    citecolor={blue!80!black},
    urlcolor={blue!80!black}
}

\usepackage{zref-savepos}
\usepackage{cleveref}
\crefformat{section}{\S#2#1#3} 
\crefformat{subsection}{\S#2#1#3}
\crefformat{subsubsection}{\S#2#1#3}
\crefformat{figure}{figure~#2#1#3}
\crefformat{appendix}{appendix~#2#1#3}
\crefformat{equation}{(#2#1#3)}

\usepackage[makeroom]{cancel}
\usepackage{scrextend} 
\usepackage{bold-extra} 
\usepackage{color} 
\usepackage{fancyhdr}
\usepackage{ulem}
\usepackage{xcolor}
\usepackage{relsize}
\usepackage{textcomp}
\usepackage{lipsum}
\usepackage{setspace}
\usepackage{empheq}

\usepackage{float} 

\usepackage{epstopdf, epsfig}

\usepackage{pdfpages} 
\usepackage{graphicx} 
\usepackage{graphics} 

\usepackage{caption}
\usepackage{tabu} 
\usepackage{tabulary} 

\usepackage{framed} 
\usepackage{listings} 

\usepackage{amsmath} 
\usepackage{amssymb} 
\usepackage{mathrsfs} 

\usepackage{algorithm} 
\usepackage{algpseudocode} 

\usepackage{tensor}
\usepackage{upgreek}
\usepackage{physics}

\usepackage{tikz}
\usepackage{circuitikz}
\usetikzlibrary{calc}
\usetikzlibrary{patterns}
\usetikzlibrary{arrows}
\usetikzlibrary{shapes}
\usetikzlibrary{plotmarks}
\usetikzlibrary{decorations.markings}

\usepackage{pgfplots}
\usepgfplotslibrary{polar}
\usepgflibrary{shapes.geometric}
\usepackage[]{pstricks}

\tikzset{
	set arrow inside/.code={\pgfqkeys{/tikz/arrow inside}{#1}},
	set arrow inside={end/.initial=>, opt/.initial=},
	/pgf/decoration/Mark/.style={
		mark/.expanded=at position #1 with
		{
			\noexpand\arrow[\pgfkeysvalueof{/tikz/arrow inside/opt}]{\pgfkeysvalueof{/tikz/arrow inside/end}}
		}
	},
	arrow inside/.style 2 args={
		set arrow inside={#1},
		postaction={
			decorate,decoration={
				markings,Mark/.list={#2}
			}
		}
	},
}

\newcommand{\gradd}{\vec{\nabla}}

\renewcommand{\vec}[1]{{\boldsymbol{#1}}}

\newcommand{\spc}{\:\:\:\:}

\newcommand{\taubar}{\bar{\tau}}
\newcommand{\lambdae}{\lambda_{ei}}

\newcommand{\df}{\delta \! f}

\newcommand{\dBpar}{\delta \! B_\parallel}
\newcommand{\dBperp}{\delta \! \vec{B}_\perp}
\newcommand{\dTe}{\delta T_{e}}
\newcommand{\dne}{\delta n_{e}}
\newcommand{\rmd}{\mathrm{d}}
\newcommand{\sgn}{\mathrm{sgn}}
\renewcommand{\Re}{\mathrm{Re}}
\renewcommand{\Im}{\mathrm{Im}}
\newcommand{\s}{s}

\newcommand{\bea}{\begin{eqnarray}}
\newcommand{\eea}{\end{eqnarray}}
\newcommand{\beq}{\begin{equation}}
\newcommand{\eeq}{\end{equation}}

\newcommand{\kperp}{k_\perp}

\newcommand{\kpar}{k_\parallel}

\newcommand{\vperp}{v_\perp}
\newcommand{\vpar}{v_\parallel}

\newcommand{\vths}{v_{{\rm th}\s}}

\newcommand{\vthe}{v_{{\rm th}e}}

\newcommand{\dn}{\delta n}

\renewcommand{\tt}{\tilde t}

\newcommand{\gpar}{\gamma_\parallel}

\newcommand{\amp}{\bar}

\newcounter{NoTableEntry}
\renewcommand*{\theNoTableEntry}{NTE-\the\value{NoTableEntry}}

\newcommand{\customlabel}[2]{%
   \protected@write \@auxout {}{\string \newlabel {#1}{{#2}{\thepage}{#2}{#1}{}} }%
   \hypertarget{#1}{#2}
}

\renewcommand{\gpar}{\gradd_\parallel}

\newcommand{\wpar}{{\omega_\parallel}}
\newcommand{\rhoperp}{\rho_\perp}
\newcommand{\arbnorm}{\sigma}
\newcommand{\hypervisscoeff}{{N_\nu}}
\newcommand{\dTebar}{\delta \bar{T}_e}

\title[]{Scale invariance and critical balance in electrostatic drift-kinetic turbulence}
\author[T. Adkins et al.] 
{
T.~Adkins$^{1,2,3}$\thanks{Email: toby.adkins@physics.ox.ac.uk}
,
P.~G.~Ivanov$^{1}$
,
and
A.~A.~Schekochihin$^{1,2}$
}

\affiliation{
$^1$Rudolf Peierls Centre for Theoretical Physics, University of Oxford,\\ 
Oxford, OX1 3PU, UK
\\[\affilskip]
$^2$Merton College, Oxford, OX1 4JD, UK
\\[\affilskip]
$^3$Culham Centre for Fusion Energy, United Kingdom Atomic Energy Authority,\\ Abingdon, OX14 3DB, UK
}

\begin{document}

\maketitle

\begin{abstract}
The equations of electrostatic drift kinetics are observed to possess a symmetry associated with their intrinsic scale invariance. Under the assumptions of spatial periodicity, stationarity, and locality, this symmetry implies a particular scaling of the turbulent heat flux with the system's parallel size, from which its scaling with the equilibrium temperature gradient can be deduced under some additional assumptions. This macroscopic transport prediction is then confirmed numerically for a reduced model of electron-temperature-gradient-driven turbulence in slab geometry. The system realises this scaling through a turbulent cascade from large to small perpendicular spatial scales. The route of this cascade through wavenumber space (i.e., the relationship between parallel and perpendicular scales in the inertial range) is shown to be determined by a balance between nonlinear-decorrelation and parallel-dissipation timescales. This type of ``critically balanced" cascade, which maintains a constant energy flux despite the presence of parallel dissipation throughout the inertial range (as well as order-unity dissipative losses at the outer scale) is expected to be a generic feature of plasma turbulence. The outer scale of the turbulence, on which the turbulent heat flux depends, is determined by the breaking of drift-kinetic scale invariance due to the existence of large-scale parallel inhomogeneity (the parallel system size).
\end{abstract}

\section{Introduction}
\label{sec:introduction}
In many plasmas, energy is injected into the system on some large, system-specific macroscale (the ``outer scale"). In order for such a system to reach a steady state, this energy must be dissipated. The usual route to this dissipation in kinetic plasmas is via a turbulent cascade of this energy to fine scales in both position and velocity space, where it is eventually thermalised by collisions (the ``inner scale"). Given that there is often a large separation between these inner and outer scales [such as in, e.g., astrophysical systems, where energy is often injected by magnetohydrodynamic (MHD) instabilities], many studies of plasma turbulence are able to consider the dynamics of this turbulent cascade separately from the specific mechanisms of injection, simply assuming that there is some energy arriving from large scales that needs to be processed (see, e.g., \citealt{sch09} and references therein).

There are, however, a variety of plasma systems for which such a scale separation is not \textit{a priori} obvious. This is often due to the existence of gradients associated with an equilibrium (whether gravitational or magnetic) that, thermodynamically speaking, provide sources of free energy for unstable, microscale perturbations that can engender a turbulent cascade well below the usual macroscopic outer scale. In fact, the most (linearly) unstable perturbations in such systems often occur at the smallest scales. This is the case in tokamaks, in which the turbulent heat and particle transport is dominated by the (microscale) instabilities driven by the gradients of the plasma pressure between the inner core of the tokamak and its edge. The most important of these instabilities are the ion-temperature gradient (ITG) \citep[see, e.g.,][]{waltz88, cowley91, kotschenreuther95} and electron-temperature gradient (ETG) ones \citep[see, e.g.,][]{liu71,lee87,dorland00,jenko00}. The relationship between the macroscopic scales associated with the plasma equilibrium and the microscopic scales on which turbulent fluctuations grow --- and how the interaction between these two scales determines the heat and particle transport properties of the confined plasma --- remains a topic of active research and great consequence. 

In this paper, we consider electrostatic, drift-kinetic plasma turbulence --- applicable to many regimes of tokamak operation --- with a particular focus on the connection between its macroscopic transport properties and microscale dynamics. In the presence of constant perpendicular equilibrium gradients, it is observed that the equations of electrostatic drift kinetics possess a symmetry associated with their intrinsic scale invariance, in both the collisionless and collisional limits. We then show that this symmetry implies a particular scaling of the turbulent heat flux with equilibrium-scale quantities, in particular the parallel system size, provided one can assume spatial periodicity, stationarity (that the system has reached a statistical steady state), and locality (that the heat flux is independent of the system's perpendicular size, as it should be for any valid local model of plasma turbulence, provided its perpendicular size is large enough). This macroscopic transport prediction is then confirmed numerically in the context of an electron-scale, collisional model of electrostatic turbulence driven by the ETG instability {in slab geometry}. The choice to focus on ETG-driven turbulence was motivated, in part, by the fact that, despite significant recent progress \citep[see, e.g.,][]{ren17,hatch19,parisi20,guttenfelder21,guttenfelder22,parisi22,chapman22,field23}, the saturation of such turbulence remains significantly less well-understood than its ITG cousin.   

Further consideration of the microscale dynamics of our system of equations reveals that this heat flux scaling is enabled by a critically-balanced, \cite{K41} style cascade of energy from large to small spatial scales. The (approximately) constant flux of energy is that which survives the parallel dissipation present at the largest scales due, in our model, to thermal conduction. The existence of this parallel dissipation is also shown to play a key role in determining the saturated state of the system, limiting the cascade of free energy in wavenumber space. The outer scale of the turbulence is found to be determined by the breaking of the drift-kinetic scale invariance due to the existence of some large-scale parallel inhomogeneity, viz., the parallel system size, rather than by the smallest scales on which the ETG instability's growth rate peaks. It is thus the largest scales that are the most important in determining the saturated amplitudes to which the fluctuations grow, and the resultant turbulent transport. This is the first detailed demonstration of a critically balanced cascade in a temperature-gradient-driven plasma system since \cite{barnes11} proposed such a cascade for ITG turbulence.

The rest of this paper is organised as follows. In \cref{sec:drift_kinetic_scale_invariance}, the scaling of the turbulent heat flux with parallel system size is derived from considerations of the scale invariance of the electrostatic drift-kinetic system of equations. Our model system of fluid equations is introduced in \cref{sec:collisional_fluid_model}, and the aforementioned heat flux scaling is verified in \cref{sec:numerical_results}. The dynamics of the inertial range are considered extensively in \cref{sec:inertial_range_dynamics}, including the free-energy budget (\cref{sec:free_energy_budget}), the existence of a constant-flux cascade and dynamical critical balance (\cref{sec:constant_flux}), the nature of the outer scale (\cref{sec:outer_scale}), the two-dimensional $(\kperp, \kpar)$ spectra (\cref{sec:two_dimensional_spectra}), and the perpendicular isotropy in wavenumber space (\cref{sec:perpendicular_isotropy}). Lastly, we summarise our results and generic conclusions in \cref{sec:summary_and_discussion}, and discuss the limits of their applicability to plasma systems in which finite-Larmor-radius (FLR) or electromagnetic effects are thought to be important.

\section{Electrostatic drift-kinetic scale invariance}
\label{sec:drift_kinetic_scale_invariance}
For systems adequately described by electrostatic drift kinetics, the heat flux through some volume $V$ is given by
\begin{align}
	Q = \sum_\s Q_\s, \quad Q_s = n_{0\s} T_{0\s}\int \frac{\rmd^3 \vec{r}}{V} \: \left( \vec{v}_E \cdot \gradd x \right)  \frac{\delta T_\s}{T_{0\s}},
	\label{eq:dk_heat_flux}
\end{align}
where $\gradd x$ is the (radial) direction of the equilibrium gradients, $n_{0\s}$ and $T_{0\s}$ are the equilibrium density and temperature, respectively, of species $\s$, $\delta T_\s$ is the corresponding temperature perturbation [see \cref{eq:invariance_temperature_perturbation}], and
\begin{align}
	\vec{v}_E = \frac{c}{B_0} \vec{b}_0 \times \gradd \phi,
	\label{eq:dk_exb_flow}
\end{align}
is the $\vec{E}\times \vec{B}$ drift velocity due to the perturbed electrostatic potential $\phi$, $B_0$ and $\vec{b}_0$ being the magnitude and direction of the equilibrium magnetic field, respectively. In what follows, $L_{n_\s}$ and $L_{T_\s}$ denote the characteristic scale lengths associated with the gradients of the equilibrium density and temperature, respectively, while the equilibrium energy scale of species $\s$ is set by its thermal speed $\vths = \sqrt{2T_{0\s}/m_\s}$, with $m_\s$ being the particle mass.

In \cref{app:derivation_of_drift_kinetic_scale_invariance}, we show that, for constant perpendicular equilibrium gradients, the electrostatic, drift-kinetic system of equations is invariant under a particular one-parameter transformation. Under this transformation, the perturbed temperature and electrostatic potential transform as, for any $\lambda$, 
\begin{align}
	\delta \tilde{T}_\s = \lambda^2 \delta T_\s(x/\lambda^2, y/\lambda^2, z/\lambda^{2/\alpha}, t/\lambda^2), \quad \tilde{\phi} = \lambda^2 \phi(x/\lambda^2, y/\lambda^2, z/\lambda^{2/\alpha}, t/\lambda^2).
	\label{eq:dk_invariance}
\end{align}
Here, $x$, $y$ and $z$ are the radial, binormal and parallel coordinates, respectively, the tildes indicate the transformed fields, and $\alpha = 1,2$ in the collisionless and collisional limits, respectively. We have assumed that the collisional limit corresponds to the case where the frequency of the perturbations $\omega$ is comparable to rate of thermal conduction, but much smaller than $\nu_{\s \s'}$, the characteristic collision frequency between species $\s$ and $\s'$, viz., $\omega \sim (\kpar \vths)^2/\nu_{\s \s'} \ll \nu_{ss'}$, as in \cite{braginskii65} (where $\kpar$ is the characteristic wavenumber of the perturbations along the direction of the equilibrium magnetic field). Mathematically, the existence of the symmetry \cref{eq:dk_invariance} is a consequence of the scale invariance of electrostatic drift kinetics: in the absence of finite-Larmor-radius effects associated with $\rho_\s$ --- the Larmor radius of species $\s$, manifest in the gyroaverages and the resultant Bessel functions appearing in gyrokinetics \citep[see, e.g.,][]{abel13} --- there is no intrinsic perpendicular scale in the system, with nothing to distinguish any perpendicular scale from any other. 

Under \cref{eq:dk_invariance}, {and noting the presence of the perpendicular derivative in \cref{eq:dk_exb_flow},} the heat flux \cref{eq:dk_heat_flux} transforms as 
\begin{align}
	\tilde{Q}_\s = \lambda^2 Q_\s.
	\label{eq:dk_heat_flux_transformation}
\end{align}
Now suppose that our original solutions for $\delta T_\s$ and $\phi$ were periodic in $x$, $y$ and $z$ with domain sizes $L_x$, $L_y$, and $L_\parallel$, respectively. Then, the transformed solutions $\delta \tilde{T}_\s$ and $\tilde{\phi}$ are still periodic in $x$, $y$ and $z$, except with domain sizes $\lambda^2 L_x$, $\lambda^2 L_y$, and $\lambda^{2/\alpha} L_\parallel$, implying that 
\begin{align}
	\tilde{Q}_\s(\lambda^2 L_x, \lambda^2 L_y, \lambda^{2/\alpha} L_\parallel, t/\lambda^2) = \lambda^{2} Q_\s( L_x,  L_y, L_\parallel, t).
	\label{eq:dk_heat_flux_parameters}
\end{align}
The heat flux will, of course, depend on other parameters of the system, e.g., equilibrium gradients and collisionality. These, however, remain unchanged under the transformation (by construction), and so we did not write them explicitly in \cref{eq:dk_heat_flux_parameters}. In a strongly magnetised (gyrokinetic) plasma, structures generated by the turbulent fluctuations are ordered comparable to the equilibrium scales in the parallel direction ($\kpar^{-1} \sim L_\parallel \sim L_{T_\s}$), but remain microscopic in the perpendicular direction ($\kperp^{-1} \sim \rho_\s$). This means that, as the perpendicular domain size $L_\perp$ (ordered as $L_\perp \sim L_x \sim L_y \sim \rho_\s$) is increased, there must come a point at which the turbulence, and the resultant heat flux, become independent of the perpendicular domain size; if this were not the case, then the heat flux would diverge as $L_\perp/\rho_\s \rightarrow \infty$, implying that drift kinetics is not a valid local model of the plasma. We thus assume that the heat flux is independent of the perpendicular domain size, viz., independent of $L_x$ and $L_y$. We also assume that the heat flux is independent of time, in the sense that it has been able to reach a statistical steady state. Then, given that $\lambda$ can be chosen arbitrarily, \cref{eq:dk_heat_flux_parameters} directly implies that
\begin{align}
	Q_\s \propto L_\parallel^{\alpha},
	\label{eq:dk_heat_flux_final}
\end{align}
where once again $\alpha = 1,2$ in the collisionless and collisional limits, respectively. Physically, $L_\parallel$ can be thought of either as a measure of a quantity analogous to the connection length $2\pi q R$ in tokamak geometry (where $q$ is the safety factor and $R$ the major radius) or, in the absence of any other gradients, as a proxy for the temperature-gradient scale length. The latter follows from dimensional analysis: without loss of generality, 
	\begin{align}
		\frac{Q_\s}{Q_{\text{gB}\s}} =  \left( \frac{L_\parallel}{L_{T_\s}} \right)^\alpha G \left( \nu_{*\s}, \frac{L_{n_\s}}{L_{T_\s}}, \frac{R}{L_{T_\s}}, \dots \right), 
		\label{eq:dk_dimensional_analysis}
	\end{align}
where $Q_{\text{gB}\s} = n_{0\s} T_{0\s} \vths (\rho_\s/L_{T_\s})^2$ is the ``gyro-Bohm" heat flux, $G$ is an unknown function, $\nu_{*\s} = (L_T/\vths) \sum_{\s '} \nu_{\s \s'}$ is the normalised collisionality, and ``$\dots$" stands for other equilibrium parameters on which the heat flux can depend, normalised, wherever a scale is required, using the temperature-gradient scale length $L_{T_\s}$.  If the dependence of $G$ on these other parameters can be ignored in \cref{eq:dk_dimensional_analysis} {--- due either to the absence of other gradients in, e.g., slab geometry, or the system being driven far above marginality where such dependences are typically weak ---} then the scaling of the heat flux with $L_{T_\s}$ follows directly from its dependence on $L_\parallel$.

This is perhaps a surprising result. Under the assumptions that the system is spatially periodic, that it is able to reach a statistical steady state (stationarity), and that the heat flux is independent of the system's perpendicular size, as it should be for any valid local model of a plasma (spatial locality), the scale invariance of electrostatic drift kinetics enforces the scaling \cref{eq:dk_heat_flux_final}, which is a non-trivial prediction about the scaling of the heat flux with equilibrium parameters. The key physics question, then, is how the system organises itself in order to obey this scaling. Namely, it has to find a way to process the free energy injected by the equilibrium gradients at a steady rate (stationarity) and to choose a spatial scale independent of $L_\perp$ (locality). 
In the remainder of this paper, we investigate a particular example of a system that should exhibit this scaling, being derived in an asymptotic limit of electrostatic drift kinetics, and find that a critically balanced, \cite{K41}-style cascade of energy from large to small spatial scales is the dynamical means by which the formal constraint imposed by~\cref{eq:dk_invariance} is realised.

\section{Collisional fluid model}
\label{sec:collisional_fluid_model}
Fluid models are capable of providing remarkable insight about the dynamics of more general physical systems, while retaining the advantage of being (comparatively) simple to handle both numerically and analytically \citep[e.g.,][]{cowley91,newton10,ivanov20,ivanov22}. The ITG and ETG instabilities in tokamaks rely on destabilisation mechanisms that are fundamentally fluid (i.e., they are not resonant instabilities) and, even in kinetic regimes, they tend to be described adequately by fluid closures \citep{hammett90,hammett92,hammett93,dorland93,beer96,snyder97}.
Here we consider an electron-scale, collisional ($ \nu_{*e} \gg 1$) fluid model of electrostatic turbulence driven by the electron-temperature gradient. Despite its simplicity, we expect that many of the results reported below are qualitatively applicable to more general turbulent plasma systems.

We take the local plasma equilibrium to be that of conventional slab gyrokinetics \cite[see, e.g.,][]{howes06}. The (homogeneous) equilibrium magnetic field is in the $\vec{b}_0 = \hat{\vec{z}}$ direction; perturbations to both its direction and magnitude are assumed negligible, consistent with the electrostatic limit. The electric field is then related to the electrostatic potential $\phi$ by $\vec{E} = - \gradd \phi $, and has no mean part. The equilibrium profile of the electron temperature $T_{0e}$ varies radially, with the scale length
\begin{align}
	L_T^{-1} \equiv L_{T_e}^{-1} = - \frac{1}{T_{0e}} \frac{\rmd T_{0e}}{\rmd x},
	\label{eq:etg}
\end{align}
which is assumed to be constant over the domain of the system. The equilibrium gradients of density and ion temperature are assumed to be negligibly small. {The omission of an equilibrium density gradient means that our system will be unstable for any finite $L_T^{-1}$; the resultant dynamics can thus be considered to apply to a plasma driven strongly above marginality (unlike the cases considered in, e.g., \citealt{guttenfelder21,hatch21etg,chapman22,field23}, where a strong dependence of the heat flux on the equilibrium density gradient was identified).}

\subsection{Moment equations}
\label{sec:moment_equations}
With this local equilibrium, we derive, in \cref{app:derivation_of_collisional_fluid_model}, evolution equations for the density ($\dne$), parallel velocity ($u_{\parallel e}$), and temperature ($\dTe$) perturbations of the electrons:
\begin{align}
	&\frac{\rmd }{\rmd t} \frac{\dne}{n_{0e}} + \frac{\partial u_{\parallel e}}{\partial z}  = 0, \label{eq:density_moment} \\
	& \frac{\nu_{ei}}{c_1} \frac{u_{\parallel e}}{\vthe} = - \frac{\vthe}{2} \frac{\partial}{\partial z} \left[\frac{\dne}{n_{0e}} - \varphi + \left( 1 + \frac{c_2}{c_1} \right) \frac{\dTe}{T_{0e}} \right], \label{eq:velocity_moment} \\
	&\frac{\rmd}{\rmd t} \frac{\dTe}{T_{0e}} + \frac{2}{3} \frac{\partial}{\partial z} \frac{\delta q_e}{n_{0e} T_{0e}} + \frac{2}{3} \left(1 + \frac{c_2}{c_1} \right) \frac{\partial u_{\parallel e}}{\partial z}  = - \frac{\rho_e \vthe}{2 L_T} \frac{\partial \varphi}{\partial y}. \label{eq:temperature_moment}
\end{align}
Let us discuss what these equations represent. 

Equation \cref{eq:density_moment} is the familiar continuity equation. It describes the advection of the density perturbation by the $\vec{E}\times\vec{B}$ motion \cref{eq:dk_exb_flow} of the electrons:
\begin{align}
	\frac{\rmd}{\rmd t}  = \frac{\partial}{\partial t} + \vec{v}_E\cdot \gradd_\perp = \frac{\partial}{\partial t}  + \frac{\rho_e \vthe}{2} \left\{ \varphi, \dots\right\}, \quad \varphi = \frac{e \phi}{T_{0e}},
	\label{eq:convective_derivative}
\end{align}
and their compression or rarefaction due to the perturbed electron flow $u_{\parallel e} \vec{b}_0$ parallel to the equilibrium magnetic-field direction.

This flow velocity is determined (instantaneously) from a balance between the electron-ion frictional force --- proportional to the electron-ion collision frequency $\nu_{ei}$ [see \cref{eq:definition_collision_frequencies}] and appearing on the left-hand side of \cref{eq:velocity_moment} --- and the forces on the right-hand side of \cref{eq:velocity_moment}: the parallel pressure gradient, the electrostatic part of the parallel electric field, and the collisional `thermal forces' \citep{braginskii65,helander05}, proportional to $c_2/c_1$, that arise due to the velocity dependence of the collision frequency associated with the Landau collision operator [see \cref{eq:landau_operator}]. The order-unity constants $c_1$, $c_2$ and $c_3$ [the latter appearing in \cref{eq:collisional_heatflux}] arise from the inversion of said operator, and depend on the magnitude of the ion charge $Z$ [see \cref{eq:charge_coefficients}]: e.g., for $Z=1$, $c_1 \approx 1.94$, $c_2 \approx 1.39$ and $c_3 \approx 3.16$, in agreement with \cite{braginskii65}. 

The temperature perturbation in \cref{eq:temperature_moment} is advected by the local $\vec{E}\times \vec{B}$ flow \cref{eq:dk_exb_flow}, again according to \cref{eq:convective_derivative}, and is locally increased (or decreased) by compressional heating (or rarefaction cooling) due to $u_{\parallel e}$, as well as by the perturbed parallel collisional heat flux
\begin{align}
	\frac{\delta q_e}{n_{0e} T_{0e}} = - c_3\frac{ \vthe^2}{2 \nu_{ei}} \frac{\partial}{\partial z} \frac{\dTe}{T_{0e}}
	\label{eq:collisional_heatflux}
\end{align}
caused by the gradient of the temperature perturbation along the equilibrium magnetic field direction. The term on the right-hand side of \cref{eq:temperature_moment} is the familiar linear drive (advection of the equilibrium temperature profile by the perturbed $\vec{E} \times \vec{B}$ flow) responsible for extracting free energy from the equilibrium temperature gradient $L_T^{-1}$, defined in \cref{eq:etg}. 

Finally, the electron-density perturbation is related to the non-dimensionalised potential $\varphi$ via quasineutrality:
\begin{align}
	\frac{\delta n_e}{n_{0e}} = - \taubar^{-1} \varphi, \quad \taubar = \frac{\tau}{Z},
	\label{eq:quasineutrality_final}
\end{align}
where $\tau = T_{0i}/T_{0e}$ is the ratio of the ion to electron equilibrium temperatures. This describes an adiabatic ion response at electron scales: at scales much smaller than their Larmor radius $\rho_i$, ions can be viewed as motionless rings of charge, and their density response is Boltzmann.  

Given \cref{eq:velocity_moment}, \cref{eq:collisional_heatflux} and \cref{eq:quasineutrality_final}, we can contract our system to two evolution equations written entirely in terms of the electrostatic potential and temperature perturbations:
\begin{align}
	&\frac{\partial}{\partial t} \taubar^{-1} \varphi - \frac{c_1 \vthe^2}{2 \nu_{ei}}  \frac{\partial^2 }{\partial z^2}\left[\left(1 + \frac{1}{\taubar} \right)\varphi   - \left( 1 + \frac{c_2}{c_1} \right) \frac{\dTe}{T_{0e}} \right] = 0, \label{eq:density_moment_num}\\
	&\frac{\rmd}{\rmd t} \frac{\dTe}{T_{0e}} + \frac{2}{3 } \frac{c_1 \vthe^2}{2 \nu_{ei}} \frac{\partial^2}{\partial z^2} \left\{\left( 1 + \frac{1}{\taubar} \right)\left( 1 + \frac{c_2}{c_1} \right) \varphi - \left[\frac{c_3}{c_1} + \left( 1 + \frac{c_2}{c_1} \right)^2 \right]  \frac{\dTe}{T_{0e}} \right\} \label{eq:temperature_moment_num} \\
	& \quad \quad \quad = - \frac{\rho_e \vthe}{2 L_T}\frac{\partial \varphi}{\partial y}. \nonumber 
\end{align}
Note that, due to the Boltzmann density response \cref{eq:quasineutrality_final}, the advection term in \cref{eq:density_moment_num} has vanished, leaving a purely linear relationship between $\varphi$ and $\dTe$. The only nonlinearity left in the system is thus the advection of $\dTe$ by the $\vec{E}\times \vec{B}$ flow in \cref{eq:temperature_moment_num}. Note that we find finite-amplitude nonlinear saturation in our numerical simulations despite this absence of any nonlinearity in \cref{eq:density_moment_num}; see \cref{sec:perpendicular_isotropy} for further discussion. If finite magnetic drifts (associated with an inhomogeneous equilibrium magnetic field) are included in our model, however, we find that simulations fail to saturate; this is discussed in \cref{sec:summary_and_discussion} and \cref{app:finite_magnetic_field_gradients}.

\subsection{Scale invariance}
\label{sec:scale_invariance}
Given that \cref{eq:density_moment_num} and \cref{eq:temperature_moment_num} were derived in an asymptotic subsidiary limit of drift kinetics (see \cref{app:electron_fluid_equations}), they are necessarily invariant under the transformation~\cref{eq:dk_invariance} with $\alpha = 2$, and must therefore exhibit the scaling \cref{eq:dk_heat_flux_final}  of the heat flux with parallel system size --- this is confirmed numerically in \cref{sec:scan_in_parallel_system_size}.

Physically, this scale invariance is a consequence of the fact that \cref{eq:density_moment_num} and \cref{eq:temperature_moment_num} are valid within the wavenumber range (see \cref{app:collisional_electron_scale_ordering}), 
\begin{align}
	\sqrt{\beta_e} \ll \kpar L_T \ll 1, \quad \beta_e \frac{\lambdae}{L_T} \ll \kperp \rho_e  \ll \frac{\lambdae}{L_T},  
	\label{eq:col_wavenumber_range}
\end{align} 
i.e., at perpendicular scales much smaller that those at which electromagnetic effects become important \citep[the ``flux-freezing scale"; see][]{adkins22}, but much larger than those on which one encounters the effects of electron thermal diffusion due to the finite Larmor motion of the electrons \citep{hardman22,adkins23thesis} --- both of these bring in a special perpendicular scale that would break the drift-kinetic scale invariance\footnote{Implicit in \cref{eq:col_wavenumber_range} is the assumption that the electron inertial scale $d_e = \rho_e/\sqrt{\beta_e}$ is smaller than the ion Larmor radius $\rho_i$; this is only the case if $\beta_e \gtrsim m_e/m_i$, which is well-satisfied in most systems of interest. Should $d_e$ lie on the large-scale side of $\rho_i$ (i.e., $\beta_e \lesssim m_e/m_i$), the lower bound in \cref{eq:arbnorm_range} must be replaced with~$\sqrt{m_e/m_i}$.}. In other words, \cref{eq:density_moment_num} and \cref{eq:temperature_moment_num} describe physics on scales 
\begin{align}
	\kpar L_T \sim \sqrt{\arbnorm}, \quad \kperp \rhoperp \sim 1, \quad \rhoperp = \frac{\rho_e}{\arbnorm} \frac{L_T}{\lambdae} ,
	\label{eq:col_wavenumber_range_rhoperp}
\end{align}
where 
$\arbnorm$ is, formally, some arbitrary constant satisfying
\begin{align}
	\beta_e \ll \arbnorm \ll 1.
	\label{eq:arbnorm_range}
\end{align}
The fact that it should be arbitrary follows from the fact that there is no special scale within the wavenumber ranges \cref{eq:col_wavenumber_range_rhoperp}. Our normalisation of perpendicular and parallel wavenumbers in \cref{eq:density_moment_num}-\cref{eq:temperature_moment_num} will thus also be arbitrary, up to the definition of $\arbnorm$. 

\subsection{Collisional slab ETG instability}
\label{sec:collisional_slab_etg}
Let us now summarise briefly the linear stability properties of the system of equations \cref{eq:density_moment_num}-\cref{eq:temperature_moment_num}. Linearising and Fourier-transforming these equations, one obtains the dispersion relation: 
\begin{align}
	\omega^2  & + \left[ 1+ \taubar + \frac{2}{3} \left(1 + \frac{c_2}{c_1}\right)^2 + \frac{2}{3} \frac{c_3}{c_1} \right]  i \wpar \omega \nonumber \\
	& \quad \quad \quad\quad - \left[\frac{2}{3}(1+\taubar) \frac{c_3}{c_1} \wpar + \left(1 + \frac{c_2}{c_1} \right) i\omega_{*e}\taubar \right] \wpar = 0,
	\label{eq:lin_dispersion} 
\end{align}
where we have introduced the characteristic parallel and perpendicular frequencies
\begin{align}
	\wpar = c_1 \frac{\left(\kpar \vthe \right)^2}{2 \nu_{ei}}, \quad \omega_{*e} = \frac{k_y \rho_e \vthe}{2 L_T}.
	\label{eq:lin_omega_parallel}
\end{align}
These are, respectively, the rate of parallel thermal conduction and the drift frequency associated with the electron-temperature gradient.
Note that the dispersion relation~\cref{eq:lin_dispersion} is quadratic in the frequency $\omega$ because the parallel velocity $u_{\parallel e}$ is determined instantaneously in terms of the other fields, by \cref{eq:velocity_moment}, unlike in collisionless ETG theory \citep{adkins22}. 

\begin{figure}
	
	\begin{tabular}{cc}
		\multicolumn{2}{c}{\includegraphics[width=1\textwidth]{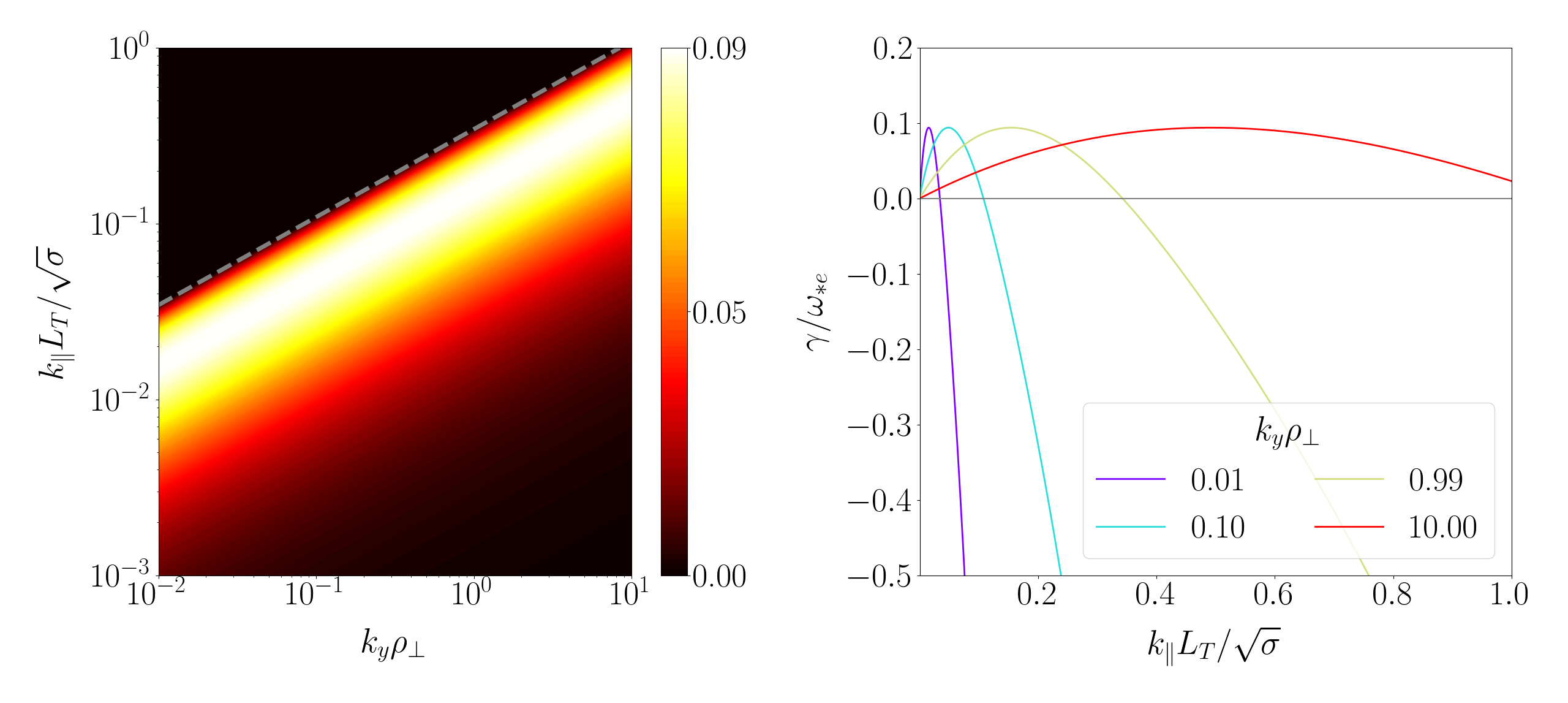}}   \\
		\hspace{2.3cm}	(a) $k_x \rho_\perp = 0$ &  \hspace{3.4cm} (b) 
	\end{tabular}
	
	\caption[]{The growth rate of the collisional sETG instability: these are solutions to \cref{eq:lin_dispersion} for $\tau = Z = 1$, normalised to $\omega_{*e}$. Panel (a) is a contour plot of the positive growth rates ($\Im \: \omega >0)$ in the $(k_y, \kpar)$ plane; panel (b) shows cuts of the growth rate at constant $k_y \rhoperp$, plotted as a function of $\kpar L_T/\sqrt{\sigma}$. The normalisations $\rhoperp$ and $\sigma$ are defined in \cref{eq:col_wavenumber_range_rhoperp}. The stability boundary \cref{eq:lin_stab_boundary} is indicated by grey dashed line in panel (a).}
	\label{fig:lin_setg}
\end{figure} 

If we consider the limit of long parallel wavelengths, viz., 
\begin{align}
	 \wpar \ll \omega \ll \omega_{*e},
	 \label{eq:lin_strong_drive}
\end{align}
then the balance of the first and last terms in \cref{eq:lin_dispersion} gives us
\begin{align}
	\omega^2 = \left( 1 + \frac{c_2 }{c_1}\right) i \wpar \omega_{*e}  \taubar  \quad \Rightarrow \quad  \omega = \pm \frac{1+ i \sgn(k_y)}{\sqrt{2}} \left( 1 + \frac{c_2 }{c_1}\right)^{1/2} \left( \wpar|\omega_{*e}| \taubar \right)^{1/2}.
	\label{eq:lin_setg}
\end{align}
We recognise this as the collisional slab ETG (sETG) instability \citep{adkins22}, a cousin of the collisionless sETG \citep{lee87}.

The minimal set of equations that elucidate this process physically can be obtained from \cref{eq:density_moment}-\cref{eq:temperature_moment} under the ordering \cref{eq:lin_strong_drive}: using \cref{eq:quasineutrality_final} to express the density perturbations in terms of the electrostatic potential, we have:
\begin{align}
	\frac{\partial}{\partial t} \taubar^{-1} \varphi = \frac{\partial u_{\parallel e}}{\partial z} , \quad \frac{\nu_{ei}}{c_1} u_{\parallel e} = - \left(1 + \frac{c_2}{c_1} \right) \frac{\vthe^2}{2} \frac{\partial}{\partial z} \frac{\delta T_e}{T_{0e}}, \quad \frac{\rmd}{\rmd t} \frac{\delta T_e}{T_{0e} } = - \frac{\rho_e \vthe}{2 L_T} \frac{\partial \varphi}{\partial y}.
	\label{eq:lin_setg_equations} 
\end{align}
In this limit, the instability works as follows. Suppose that a small perturbation of the electron temperature is created with $k_y \neq 0$ and $\kpar \neq  0$, bringing the plasma from regions with higher $T_{0e}$ to those with lower $T_{0e}$ ($\dTe >0$), and vice-versa ($\dTe < 0 $). This temperature perturbation produces alternating hot and cold regions along the equilibrium magnetic field. The resulting perturbed temperature (and, therefore, pressure) gradients drive electron flows --- determined instantaneously by the balance between the pressure gradient and collisional drag --- from the hot regions to the cold regions [the second equation in~\cref{eq:lin_setg_equations}], giving rise to increased electron density in the cold regions [the first equation in~\cref{eq:lin_setg_equations}]. By quasineutrality, the electron density perturbation gives rise to an exactly equal ion density perturbation, and that, via the Boltzmann response \cref{eq:quasineutrality_final}, creates an electric field that produces an $\vec{E} \times \vec{B}$ drift that in turn pushes hotter particles further into the colder region, and vice-versa [the third equation in~\cref{eq:lin_setg_equations}], reinforcing the initial temperature perturbation and thus completing the feedback loop required for the instability. 

At short enough parallel wavelengths, the collisional sETG instability is quenched by rapid thermal conduction that leads to the damping of the associated temperature perturbation. To see this, we relax the assumption \cref{eq:lin_strong_drive} and consider the exact stability boundary of \cref{eq:lin_dispersion}, determined by the requirement that, assuming $\omega$ to be purely real, the real and imaginary parts of \cref{eq:lin_dispersion} must vanish individually. The resultant equations can be straightforwardly combined to yield
\begin{align}
 \left( \frac{ \wpar}{\omega_{*e}} \right)^2 = \frac{\displaystyle \frac{3}{2} \left(1 + \frac{c_2}{c_1} \right)^2 \taubar^2}{\displaystyle (1+ \taubar) \frac{c_3 }{c_1}\left[ 1+ \taubar + \frac{2}{3} \left(1 + \frac{c_2}{c_1}\right)^2 + \frac{2}{3} \frac{c_3}{c_1} \right]^2}.
	\label{eq:lin_stab_boundary}
\end{align}
This is a curve $k_y \propto \kpar^2$ in wavenumber space, plotted as the grey dashed line in \cref{fig:lin_setg}(a). Above this line, corresponding to the limit $\wpar \gg \omega_{*e}$, all modes are purely damped due to rapid thermal conduction, as in \cref{fig:lin_setg}(b). 

At any given $k_y$, the maximum growth rate of the collisional sETG is, therefore, reached when
\begin{align}
	\wpar \sim \omega \sim \omega_{*e},
	\label{eq:lin_max}
\end{align}
which is a balance between dissipation (through conduction) and energy injection due to the background temperature gradient. Indeed, maximising the growth rate from \cref{eq:lin_dispersion} with respect to $\wpar$, one finds $\gamma_\text{max} = C(\tau,Z) \omega_{*e}$, where $C(\tau,Z)$ is a constant formally of order unity, e.g., $C(1,1) \approx 0.094$ [cf. the maximum values in \cref{fig:lin_setg}(b)]. The increase of the maximum growth rate of the sETG instability with the perpendicular wavenumber, $\omega_{*e} \propto k_y$, can only be checked by the effects of the electron perpendicular thermal diffusion due to finite electron Larmor motion, which, as discussed in \cref{sec:scale_invariance}, occurs outside the range of wavenumbers in which \cref{eq:density_moment_num}-\cref{eq:temperature_moment_num} are valid [see \cref{eq:col_wavenumber_range_rhoperp}], meaning that the instability grows fastest at the smallest perpendicular scales. The consequences of the intrinsic reliance of the collisional sETG on dissipative physics will be discussed in a nonlinear setting in \cref{sec:free_energy_budget}.

\section{Numerical verification of scale invariance}
\label{sec:numerical_results}

\subsection{Numerical setup}
\label{sec:numerical_setup}
In what follows, the system \cref{eq:density_moment_num}-\cref{eq:temperature_moment_num} is solved numerically in a triply periodic box of size $L_x \times L_y \times L_\parallel$ using a pseudo-spectral algorithm. Numerical integration is done in Fourier space ($N_x$, $N_y$ and $N_\parallel$ are the number of Fourier harmonics in the respective directions) with the nonlinear term calculated in real space using the 2/3 rule for de-aliasing \citep{orszag71}. We integrate the linear terms implicitly in time using the Crank-Nicolson method, while the nonlinear term is integrated explicitly using the Adams-Bashforth three-step method. This integration scheme is similar to the one implemented in the popular gyrokinetic code \texttt{GS2} \citep{kotschenreuther95GS2,dorland00}. 

Perpendicular hyperviscosity is introduced in order to provide an ultraviolet (large-wavenumber) cutoff for the instabilities, achieved by the replacement of the time derivative on the left-hand sides of \cref{eq:density_moment_num} and \cref{eq:temperature_moment_num} with
\begin{align}
	\frac{\partial}{\partial t} + (-1)^\hypervisscoeff \nu_\perp \left(\rhoperp \gradd_\perp \right)^{2 \hypervisscoeff},
	\label{eq:hypervisc}
\end{align}
where $\nu_\perp$ is the ``hypercollision" frequency and $\hypervisscoeff \geqslant 2$. With this change, our equations now depend only on the following dimensionless parameters: the perpendicular and parallel box sizes $L_x/\rhoperp$, $L_y/\rhoperp$ and $L_\parallel \sqrt{\sigma}/L_T$, the hyper-collision frequency $2\sigma(\rhoperp/\rho_e)^{2}\nu_\perp/\nu_{ei}$, and the power of the hyperviscous diffusion operator $\hypervisscoeff$. Convergence scans in $N_x$, $N_y$, and the perpendicular box size $L_x = L_y = L_\perp$ were carried out on a baseline simulation (see \cref{tab:simulation_parameters}) to ensure that the chosen resolution adequately captured the dynamics, and to verify that $L_\perp$ was large enough so that it did not significantly affect the simulation results, as was required for the arguments of \cref{sec:drift_kinetic_scale_invariance}.

\begin{table}
	
	\centering

	\begin{tabular}{l  c c c c c c c c}
		& $L_x/\rhoperp$ & $L_y/\rhoperp$ & $L_\parallel \sqrt{\arbnorm}/L_T$ & $N_x$ & $N_y$ & $N_\parallel$ & $2\arbnorm(\rhoperp/\rho_e)^2\nu_\perp/\nu_{ei}$ & $\hypervisscoeff$ \vspace*{1mm}\\
		Baseline & 40 & 40 & 20 & 191 & 191 & 31 & 0.00050 & 2 \\
		Higher-resolution & 40 & 40 & 20 & 383 & 383 & 63 & 0.00015 & 2 
	\end{tabular}

	\caption{The parameters used in the ``baseline" and ``higher-resolution" simulations. Both simulations had $\tau = Z = 1$.}
	\label{tab:simulation_parameters}
	
\end{table}

\begin{figure}
	
	\includegraphics[width=1\textwidth]{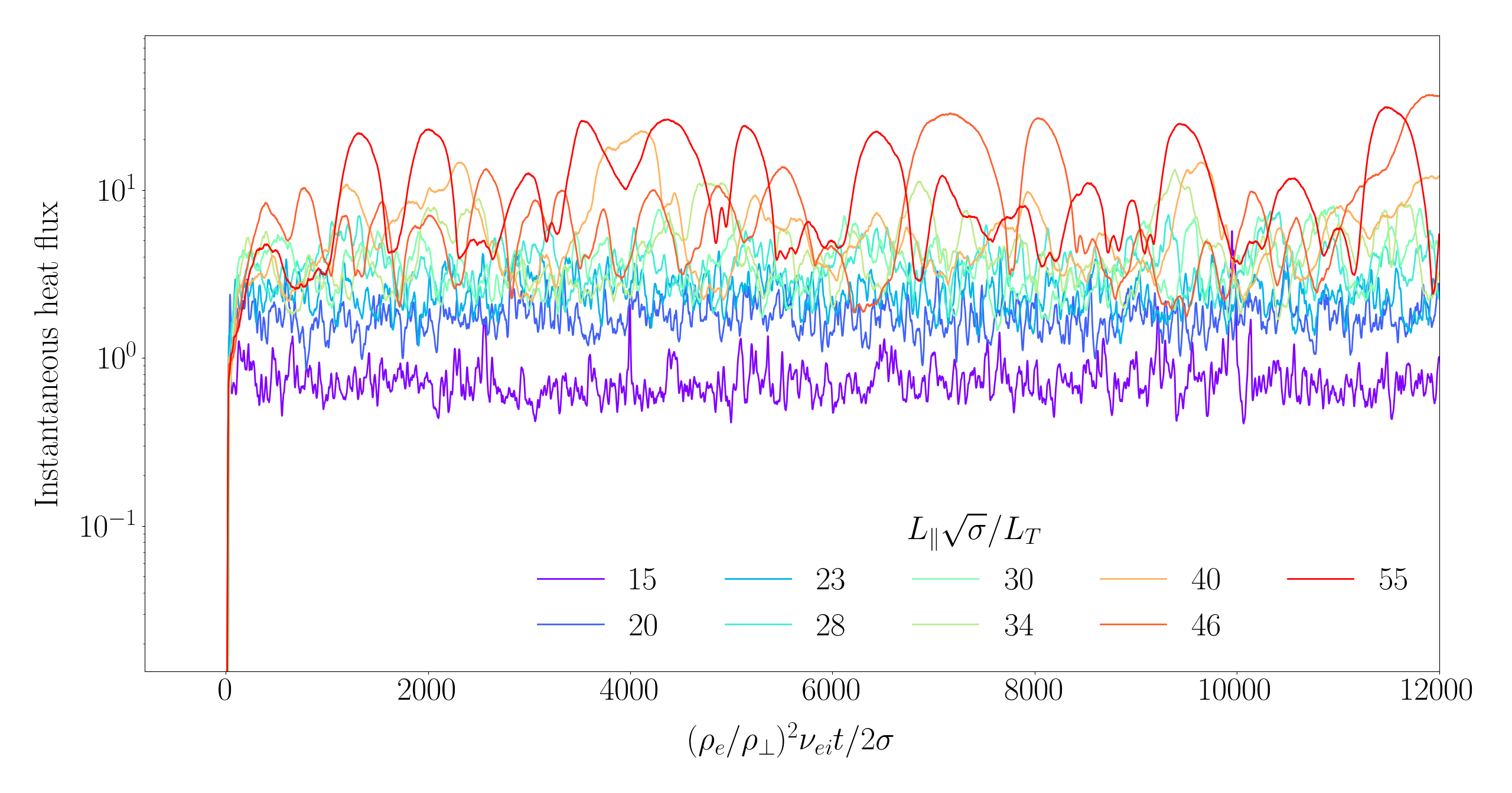}
	
	\centering
	
	\caption[]{Time traces of the instantaneous heat flux from simulations in which $L_\parallel\sqrt{\sigma}/L_T$ was varied from 15 to 55, normalised to $(\rhoperp/\rho_e) Q_{\text{gB}e}$.}
	\label{fig:heat_flux_scan_timetraces}
	
\end{figure} 

We have also found that our results do not depend on the specific details of the hyperviscosity, viz., on the values of $\nu_\perp$ and $\hypervisscoeff$. It can be viewed as a numerical tool that allows us to capture the dynamics of the system within a finite simulation domain and resolution, and is not intended to model a specific physical process. Ultimately, \cref{eq:hypervisc} is a stand-in for the physical sinks of energy that exist at higher perpendicular wavenumbers. The fact that our results end up being independent of hyperviscosity is, however, significant. The addition of \cref{eq:hypervisc} breaks the scale invariance associated with the transformation \cref{eq:dk_invariance}, similarly to the way in which FLR effects would break the drift-kinetic scale invariance in the context of gyrokinetics, a point that we shall revisit in \cref{sec:summary_and_discussion}. One could thus question the inevitability of obtaining the scaling of the heat flux \cref{eq:dk_heat_flux_final} in our system of equations with the modification \cref{eq:hypervisc}. Furthermore, the fact that the growth rate of the sETG instability peaks at a perpendicular scale determined by the hyperviscosity --- since \cref{eq:density_moment_num} and \cref{eq:temperature_moment_num} contain no intrinsic perpendicular wavenumber cutoff --- may also be a cause for concern, as the most unstable perpendicular scale is often thought to play a central role in determining turbulent transport. Both of these concerns can be dispelled by the realisation that the arguments of \cref{sec:drift_kinetic_scale_invariance} did not rely on the details of the state of the system at small perpendicular scales; indeed, the behaviour of the heat flux is determined by the parallel system size $L_\parallel$, which is manifestly an equilibrium-scale quantity. In \cref{sec:constant_flux}, we will show that this is a consequence of the fact that the \textit{outer scale} is central in (dynamically) determining the transport and that this outer scale turns out to be independent of hyperviscosity.

\subsection{Scan in $L_\parallel/L_T$}
\label{sec:scan_in_parallel_system_size}

\begin{figure}
	
	\includegraphics[width=1\textwidth]{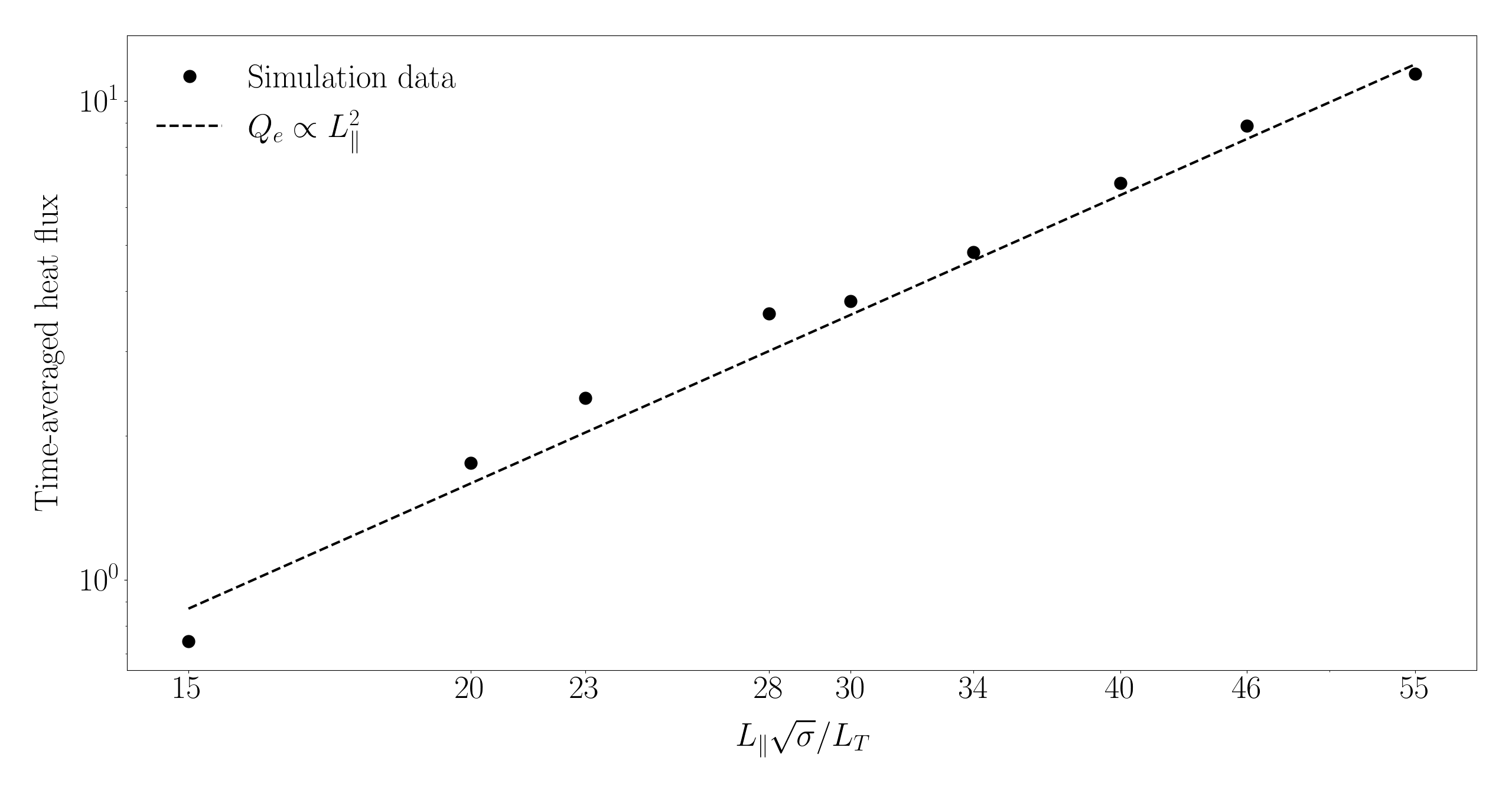}
	
	\centering
	
	\caption[]{The scaling of the turbulent heat flux with $L_\parallel /L_T$, normalised to $(\rhoperp/\rho_e) Q_{\text{gB}e}$ and plotted against logarithmic axes. The points are the simulation data, while the theoretical prediction [see \cref{eq:dk_heat_flux_final}] is shown by the dashed black line. A logarithmic fit to the data gives the slope of $2.02$.}
	\label{fig:heat_flux_scan}
	
\end{figure} 
In order to test the dependence of the turbulent heat flux on $L_\parallel$ predicted by \cref{eq:dk_heat_flux_final}, we performed a series of simulations in which $L_\parallel \sqrt{\arbnorm}/L_T$ was varied between 15 and 55 at fixed parallel resolution (viz., fixed ratio of $L_\parallel \sqrt{\arbnorm}/L_T$ to $N_\parallel$), while keeping all other parameters the same as in the baseline simulation (see \cref{tab:simulation_parameters}). Each simulation was run to long enough times for it to reach saturation and stay in a statistically stationary state for a while, as can be seen from the time traces of the instantaneous heat fluxes plotted in \cref{fig:heat_flux_scan_timetraces}. That such a stationary state exists confirms one of the assumptions necessary for \cref{eq:dk_heat_flux_final}\footnote{Refining our consideration beyond this assumption of stationarity, we observe that the characteristic timescale of the fluctuations of the instantaneous heat flux increases with the parallel system size --- this is manifest in \cref{fig:heat_flux_scan_timetraces}, where the simulations with larger $L_\parallel \sqrt{\arbnorm}/L_T$ exhibit higher-amplitude, longer-timescale fluctuations. The origin of this trend can be understood as follows. Relaxing the assumption of stationarity, instead of~\cref{eq:dk_heat_flux_final}, we have, from~\cref{eq:dk_heat_flux_parameters}, $\tilde{Q}_\s(\lambda^{2/\alpha} L_\parallel, t/\lambda^2) = \lambda^2 Q_\s (L_\parallel, t)$. If $Q_\s$ exhibits fluctuations on some characteristic timescale $\tau$, then, if we assume that that both solutions must be periodic with the same period, the corresponding timescale for the transformed heat flux will be $\tilde{\tau} = \lambda^2 \tau$. Given that the parallel system sizes for both solutions are related by $\tilde{L}_{\parallel} = \lambda^{2/\alpha}L_\parallel$, it follows that $\tau \propto L_\parallel^\alpha$. This dependence was confirmed numerically for the set of simulations shown in \cref{fig:heat_flux_scan_timetraces}.}, the other assumption, also confirmed numerically, being that this state is independent of $L_x$ and $L_y$.

In \cref{fig:heat_flux_scan}, we plot the time average of the turbulent heat flux --- as defined in \cref{eq:dk_heat_flux} for $s=e$, and normalised to $(\rhoperp/\rho_e) Q_{\text{gB}e}$, where $Q_{\text{gB}e} = n_{0e} T_{0e} \vthe (\rho_e/L_T)^2$ is the (electron) ``gyro-Bohm" flux. It is clear that the simulation data agrees extremely well with the theoretical scaling \cref{eq:dk_heat_flux_final}. This agreement, however, should not be a cause for complacency: though these results suggest that \cref{eq:dk_heat_flux_final} correctly predicts the transport, we would like to understand how the system manages this, i.e., how it contrives to satisfy the assumptions underpinning the prediction \cref{eq:dk_heat_flux_final}. To explain this, we shall consider the dynamics in the \textit{inertial range}. This is the subject of the following section. 

\section{Inertial-range dynamics}
\label{sec:inertial_range_dynamics}
\begin{figure}
	
	\includegraphics[width=1\textwidth]{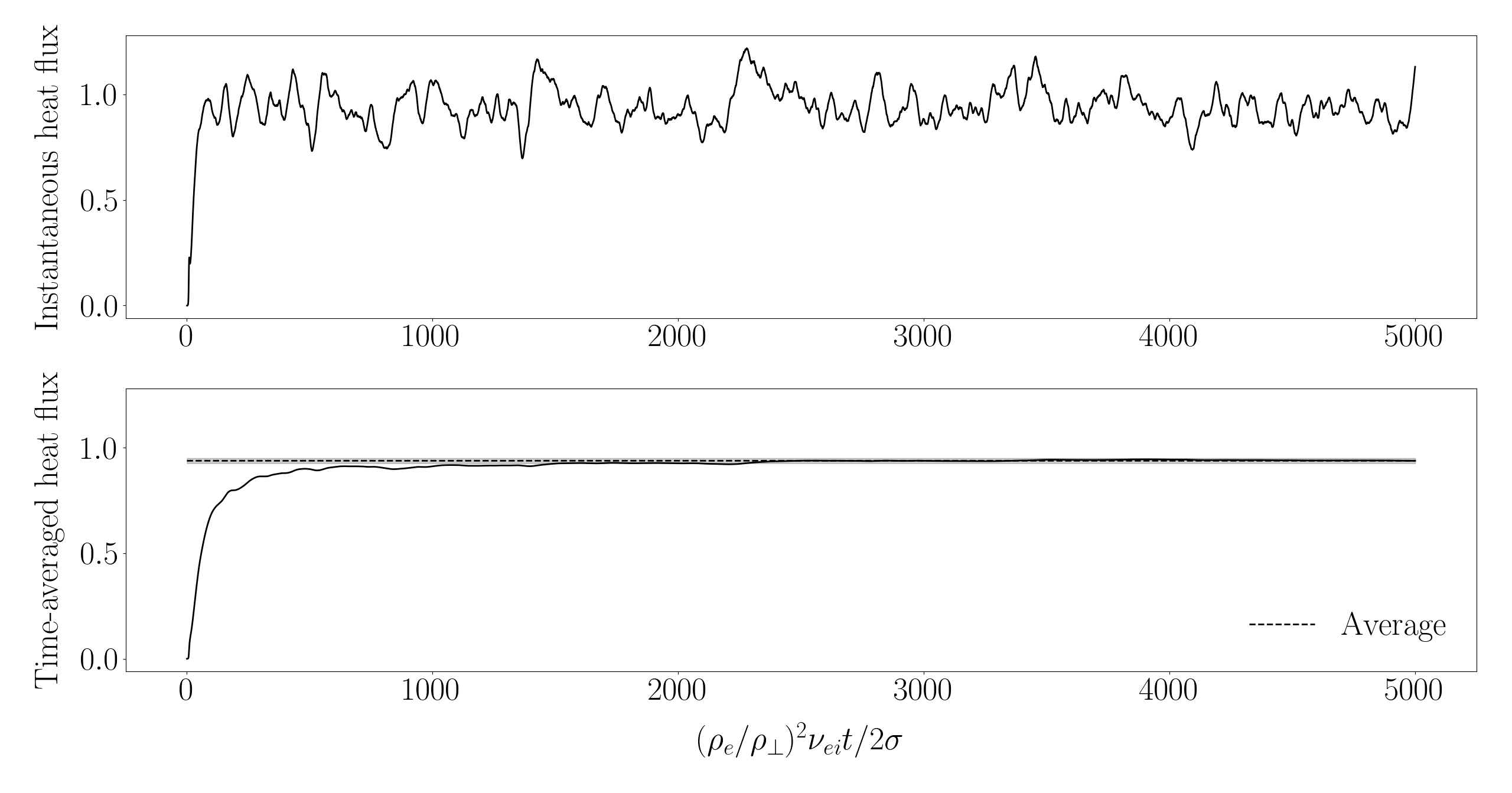}
	
	\centering
	
	\caption[]{Turbulent heat flux in the higher-resolution simulation (see \cref{tab:simulation_parameters}), normalised to $(\rhoperp/\rho_e) Q_{\text{gB}e}$. The upper and lower panels show, respectively, the instantaneous and (rolling) time-averaged heat fluxes in solid black. The dashed horizontal line in the lower panel is the average value --- as calculated over the entire time interval --- while the transparent grey region around this value shows the error bar associated with the mean, calculated by means of a moving window average. The time-averaged heat flux converges to within the final error bar by $(\rho_e/\rhoperp)^2\nu_{ei} t/2\arbnorm \sim 2000$.}
	\label{fig:high_res_heat_flux}
	
\end{figure}

To ensure that we had sufficient numerical resolution to resolve adequately the dynamics of the inertial range, we conducted a ``higher-resolution" simulation (see \cref{tab:simulation_parameters}), on which we shall now focus. Due to the computational demands introduced by the higher resolution, this simulation was run only up to 5000 $(\rho_e/\rhoperp)^2\nu_{ei} t/2\arbnorm$; this was sufficient to ensure that the heat flux had converged to a well-defined average value (see \cref{fig:high_res_heat_flux}). 

\subsection{Free-energy budget}
\label{sec:free_energy_budget}
Magnetised plasma systems containing small perturbations around a Maxwellian equilibrium nonlinearly conserve free energy, which is a quadratic norm of the magnetic perturbations and the perturbations of the distribution functions of both ions and electrons away from the Maxwellian \citep[see, e.g.,][]{abel13}. In the system of equations that we are considering, the (normalised) free energy reduces to the form 
\begin{align}
	\frac{W}{n_{0e} T_{0e}} = \int \frac{\rmd^3 \vec{r}}{V} \left[ \frac{1}{2\taubar}\left(1 + \frac{1}{\taubar} \right) \varphi^2 + \frac{3}{4} \frac{\dTe^2}{T_{0e}^2} \right].
	\label{eq:cb_free_energy}
\end{align}
The free energy is a nonlinear invariant, i.e., it is conserved by nonlinear interactions, but can be injected into the system by equilibrium gradients and is dissipated by collisions. It is straightforward to show from \cref{eq:density_moment_num} and \cref{eq:temperature_moment_num} [with the hyperviscosity \cref{eq:hypervisc} appended] that the free-energy budget is 
\begin{align}
	\frac{1}{n_{0e} T_{0e}} \frac{\rmd W}{\rmd t} = \varepsilon - D_\parallel - D_\perp,
	\label{eq:cb_free_energy_time_derivative}
\end{align}
where
\begin{align}
	\varepsilon = \frac{1}{L_T} \int \frac{\rmd^3 \vec{r}}{V} \: \frac{3}{2} \frac{\dTe}{T_{0e}} v_{Ex}, \quad  v_{Ex} = -\frac{\rho_e \vthe}{2} \frac{\partial \varphi }{\partial y},
	\label{eq:cb_injection}
\end{align}
is the energy-injection rate from the equilibrium temperature gradient, and 
\begin{align}
	D_\parallel & = \frac{c_1\vthe^2}{2 \nu_{ei}} \int \frac{\rmd^3 \vec{r}}{V} \left\{\left[\left(1 + \frac{1}{\taubar} \right) \frac{\partial \varphi}{\partial z} - \left(1 + \frac{c_2}{c_1} \right) \frac{\partial}{\partial z} \frac{\dTe}{T_{0e}} \right]^2  + \frac{c_3}{c_1} \left(\frac{\partial}{\partial z} \frac{\dTe}{T_{0e}} \right)^2\right\}, \label{eq:cb_dissipation_parallel} \\
	D_\perp & = \nu_\perp \int \frac{\rmd^3 \vec{r}}{V} \left[\left( \rhoperp^{\hypervisscoeff} \gradd_\perp^{\hypervisscoeff} \varphi \right)^2 + \frac{3}{2} \left( \rhoperp^{\hypervisscoeff} \gradd_\perp^{\hypervisscoeff}  \frac{\dTe}{T_{0e}}\right)^2 \right], \label{eq:cb_dissipation_perp}
\end{align}
are the dissipation rates due to (parallel) thermal conduction and (perpendicular) hyperviscosity, respectively. The corresponding 1D perpendicular wavenumber spectrum of the energy injection is 
\begin{align}
	\varepsilon_{\vec{k}} (\kperp) =  2 \pi \kperp  \int_{-\infty}^\infty \rmd k_\parallel \: \frac{3}{2} \Re \left< i \omega_{*e} \varphi_{\vec{k}}^* \frac{{\dTe}_{\vec{k}}}{T_{0e}} \right>,
	\label{eq:injection_rate}
\end{align}  
while those of the parallel and perpendicular dissipation are 
\begin{align}
	{D_\parallel}_{\vec{k}}(\kperp) & = 2 \pi \kperp  \int_{-\infty}^\infty \rmd k_\parallel \left< \omega_\parallel\left[ c_{1}\left| \left(1+ \frac{1}{\taubar}\right) \varphi_{\vec{k}} - \left(1 + \frac{c_2}{c_1} \right)  \frac{{\dTe}_{\vec{k}}}{T_{0e}}\right|^2 + c_3 \left|\frac{{\dTe}_{\vec{k}}}{T_{0e}}\right|^2 \right] \right>,  \label{eq:cb_dissipation_parallel_spectrum}  \\
	{D_\perp}_{\vec{k}}(\kperp) & = 2 \pi \kperp  \int_{-\infty}^\infty \rmd k_\parallel \left< (\kperp \rho_\perp)^{2\hypervisscoeff} \nu_\perp \left(\left|\varphi_{\vec{k}}\right|^2 + \frac{3}{2} \left|\frac{{\dTe}_{\vec{k}}}{T_{0e}}\right|^2 \right)\right> \label{eq:cb_dissipation_perpvisc_spectrum}.
\end{align}
In \cref{eq:injection_rate}, the asterisk denotes complex conjugation, and the angle brackets an ensemble average. Note that when analysing the output of simulations, we consider ensemble averages to be equal to time averages over a period following saturation and the establishment of a statistical steady state [e.g., after $(\rho_e/\rhoperp)^2\nu_{ei} t/2\arbnorm \sim 2000$ in \cref{fig:heat_flux_scan_timetraces}]. 

\begin{figure}
	
	\begin{tabular}{cc}
		\multicolumn{2}{c}{\includegraphics[width=1\textwidth]{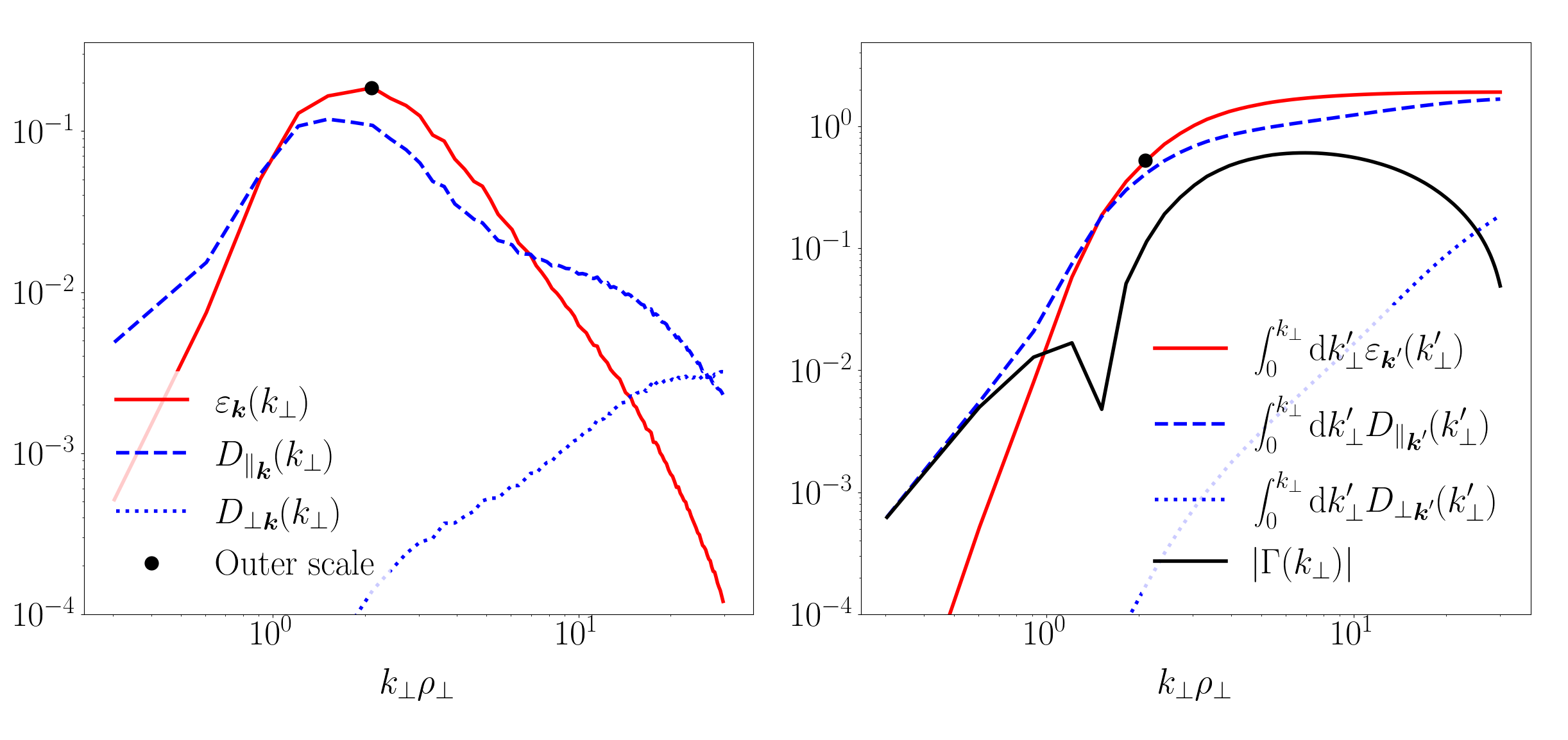}}   \\
		\hspace{3.3cm}	(a) &  \hspace{3cm} (b)
	\end{tabular}
	
	\caption[]{(a) 1D perpendicular spectra of the energy injection \cref{eq:injection_rate} (solid red), parallel dissipation \cref{eq:cb_dissipation_parallel_spectrum} (dashed blue) and perpendicular dissipation \cref{eq:cb_dissipation_perpvisc_spectrum} (dotted blue), normalised to $(\rho_e/L_T)^2 \nu_{ei}/2$. The location of the outer scale is shown by the black dot. The rate of parallel dissipation is significant at the largest scales, while perpendicular dissipation takes over at the smallest scales. (b) The cumulative perpendicular wavenumber integrals of the quantities plotted in (a), as well as the nonlinear energy flux \cref{eq:cb_nonlinear_flux} (solid black line). The latter is approximately constant in the inertial range, displaying only an order-unity variation, due to the finite simulation domain.}
	\label{fig:high_res_dissipation}
\end{figure}

Plotting the injection and dissipation spectra \cref{eq:injection_rate}-\cref{eq:cb_dissipation_perpvisc_spectrum} in \cref{fig:high_res_dissipation}(a) allows us to make a series of important observations. The first, and unsurprising, one is that the perpendicular dissipation due to hyperviscosity is dominant only at the very smallest scales, where ${D_\perp}_{\vec{k}}$ peaks. This confirms the assertion made in \cref{sec:numerical_setup} that it can be viewed as a sink of energy that exists at higher perpendicular wavenumbers and has no significant effect on the dynamics. The outer scale --- at which energy is primarily injected into the turbulence and which we define as corresponding to the perpendicular wavenumber where the maximum of~\cref{eq:injection_rate} is achieved\footnote{In standard turbulence literature, the outer scale is often defined to be the integral scale of the 1D perpendicular energy spectrum~\cref{eq:cb_varphi_definition}, viz., 
		$
		\kperp^o \equiv \int_0^\infty \rmd \kperp \:  E_\perp^\varphi(\kperp)/ \int_0^\infty \rmd \kperp \:  \kperp^{-1} E_\perp^\varphi(\kperp).
		$
		However, given that, physically, we are interested in the outer scale as the scale at which the free energy is predominantly injected, the choice to maximise \cref{eq:injection_rate} seems to be better motivated physically.} --- appears to be independent of hyperviscosity, being localised on much larger scales, where ${D_\perp}_{\vec{k}}$ is negligible. The arguments of \cref{sec:scale_invariance} leading to the heat-flux scaling \cref{eq:dk_heat_flux_final} relied on the scale invariance of the drift-kinetic system, which, as we have discussed previously, is broken by the introduction of hypervisocisty. The fact that the energy injection is both independent of hypervisocisty and localised at the largest scales supports the prediction of \cref{eq:dk_heat_flux_final} that the heat flux should be determined by the inviscid dynamics at scales where scale-invariant drift kinetics is valid. 

Considering scales that are larger than the injection scale, it is clear from \cref{fig:high_res_dissipation} that the parallel dissipation ${D_\parallel}_{\vec{k}}$ is dominant there, peaking on scales comparable to the outer scale. This is because the existence of the collisional sETG instability depends intrinsically on the presence of thermal conduction (see \cref{sec:collisional_slab_etg}), which is a dissipative effect. Indeed, the maximum growth rate \cref{eq:lin_max} occurs where the rates of thermal conduction and energy injection are comparable, $\wpar \sim \omega_{*e}$. Thus, in order to inject energy, the system has to dissipate a finite fraction of it. The energy that survives this dissipation then cascades to small scales through a constant-flux inertial range. This can be seen in \cref{fig:high_res_dissipation}(b), where we plot the cumulative perpendicular wavenumber integrals of \cref{eq:injection_rate}, \cref{eq:cb_dissipation_parallel_spectrum}, and \cref{eq:cb_dissipation_perpvisc_spectrum}, as well as the nonlinear energy flux, which can be inferred from the difference between injection and dissipation:
\begin{align}
	\Gamma(\kperp) = \int_0^{\kperp} \rmd \kperp' \: \left[\varepsilon_{\vec{k}'}(\kperp') -{D_\parallel}_{\vec{k}'}(\kperp') - {D_\perp}_{\vec{k}'}(\kperp')\right].
	\label{eq:cb_nonlinear_flux}
\end{align}
Both the injection and parallel-dissipation rates reach an approximate plateau at scales smaller than the outer scale and are much larger than the nonlinear energy flux, which is approximately constant in the inertial range, displaying an order-unity variation due to the finite width of the latter in our numerical simulations. The remainder of \cref{sec:inertial_range_dynamics} is devoted to characterising the dynamics in the inertial range in order to explain how the system organises itself to maintain a constant-flux cascade to small scales despite the presence of significant (parallel) dissipation.

\subsection{Constant flux and critical balance in the inertial range}
\label{sec:constant_flux}
The results of the previous section suggest that our fully developed electrostatic turbulence organises itself into a state wherein there is a local cascade of the free energy~\cref{eq:cb_free_energy} that carries the injected power from the outer scale, through an inertial range, to the (perpendicular) dissipation scale. This injected power is the (order-unity) fraction of $\varepsilon$ that survives the parallel dissipation at larger scales, viz., $\varepsilon - D_\parallel$, which, for brevity, we shall call $\varepsilon$ in the scaling arguments that follow.

The only nonlinearity in our equations is the advection of the temperature fluctuations by the fluctuating $\vec{E}\times \vec{B}$ flows in \cref{eq:temperature_moment_num}. Therefore, we take the nonlinear cascade time to be the nonlinear $\vec{E}\times \vec{B}$ advection time: 
\begin{align}
	t_\text{nl}^{-1} \sim k_\perp v_E \sim  \rho_e \vthe k_\perp^2 \amp{\varphi} \sim \Omega_e (k_\perp \rho_e)^2 \amp{\varphi}.
	\label{eq:cb_nonlinear_time}
\end{align}
Here and in what follows, $\amp{\varphi}$ refers to the characteristic amplitude of the electrostatic potential at the scale $k_\perp^{-1}$. Formally, $\amp{\varphi}$ can be defined by
\begin{align}
	\amp{\varphi}^2 = \int_{k_\perp}^\infty \rmd k_\perp' \: E_\perp^\varphi(k_\perp'), \quad E_\perp^\varphi(k_\perp) \equiv \int_{-\infty}^\infty \rmd k_\parallel \: 2\pi k_\perp  \left<|\varphi_{\vec{k}}|^2 \right>,
	\label{eq:cb_varphi_definition}
\end{align}
where $ E_\perp^\varphi(k_\perp) $ is the 1D perpendicular spectrum of $\varphi$, $\varphi_{\vec{k}}$ is the spatial Fourier transform of the potential, and the angle brackets denote an ensemble average. The corresponding quantities for the temperature perturbations, $\dTebar$, $E_\perp^T(\kperp)$, and ${\dTe}_{\vec{k}}$, are defined analogously.

Assuming that any possible anisotropy in the perpendicular plane can be neglected (an assumption that will be verified in \cref{sec:perpendicular_isotropy}), a Kolmogorov-style constant-flux argument leads to a scaling of the amplitudes in the inertial range:
\begin{align}
	t_\text{nl}^{-1} \frac{\dTebar^2}{T_{0e}^2 } \sim \varepsilon = \text{const} \quad \Rightarrow \quad \bar{\varphi} \frac{\dTebar^2}{T_{0e}^2} \sim \frac{\varepsilon}{\Omega_e} (\kperp \rho_e)^{-2}.
	\label{eq:cb_constant_flux}
\end{align}
We are using the $\dTe/T_{0e}$ part of the free energy \cref{eq:cb_free_energy} because, as we noted earlier, in \cref{eq:density_moment_num}-\cref{eq:temperature_moment_num}, $\dTe/T_{0e}$ is the only field that is advected nonlinearly whereas $\varphi$ is `sourced' by the temperature perturbations through the second term on the left-hand side of \cref{eq:density_moment_num}.

To estimate the size of the electrostatic potential, we therefore balance the two terms in \cref{eq:density_moment_num}, yielding
\begin{align}
	\bar{\varphi} \sim \frac{\wpar}{\omega } \frac{\bar{\dTe}}{T_{0e}},
	\label{eq:cb_phi_scaling}
\end{align}
which should hold at every scale. This implies that the potential and temperature perturbations will be comparable in magnitude and have the same wavenumber scaling throughout the inertial range if we posit, scale by scale, that
\begin{align}
	t_\text{nl}^{-1} \sim \omega \sim \wpar.
	\label{eq:cb_critical_balance}
\end{align}
This is the conjecture of \textit{critical balance}, whereby the characteristic time associated with parallel dynamics along the field lines is assumed comparable to the nonlinear advection rate $t_\text{nl}^{-1}$ at each perpendicular scale $k_\perp^{-1}$, as in \cite{barnes11} and \cite{adkins22}. The original rationale for this conjecture comes from the causality argument proposed in the context of MHD turbulence \citep{GS95,boldyrev05,sch22}: two points along a field line can only remain correlated with one another if information can propagate between them faster than they are decorrelated by the (perpendicular) nonlinearity; in MHD, this information is carried by Alfv\'en waves (similarly, it can be carried by other waves in different plasma and hydro-dynamical systems: see \citealt{cho04, nazarenko11, adkins22}). In our system, the parallel dynamics are dissipative, with the relevant timescale being set by the parallel conduction rate $\wpar$. Since there is no mechanism to preserve the parallel coherence of structures created by perpendicular mechanisms (via injection due to the sETG instability, or nonlinear cascade), one expects them to break up in the parallel direction to as fine scales as the system will allow, i.e., structures for which $\wpar \ll t_\text{nl}^{-1}$ should be immediately decorrelated by the nonlinearity and broken up into shorter pieces in the parallel direction. The limiting factor for this parallel refinement is that if structures reach parallel scales such that $\wpar \gg t_\text{nl}^{-1}$, they are wiped out by heat conduction. As a result, the ``dissipation ridge" (the line of critical balance) $\wpar \sim t_\text{nl}^{-1}$ will form a natural locus for turbulent structures. This is a version of critical balance that is appropriate for a system where parallel dissipation is present everywhere [which may also be true for collisionless plasmas, where $\wpar$ is instead the \cite{landau46} damping rate\footnote{Although it remains to be seen whether the dominant effect in enforcing \cref{eq:cb_critical_balance} is plain linear dissipation or its suppression via stochastic echos \citep{sch16,adkins18}.}]. In \cref{sec:free_energy_budget}, we saw that the actual amount of parallel dissipation that happens in the inertial range is small --- free energy chooses to stay just shy of the dissipation region ($\wpar > t_\text{nl}^{-1}$) and instead cascade, at an approximately constant rate, along the dissipation ridge ($\wpar \sim t_\text{nl}^{-1}$).

Combining \cref{eq:cb_constant_flux}, \cref{eq:cb_phi_scaling} and \cref{eq:cb_critical_balance}, we find the following scaling of the amplitudes in the inertial range
\begin{align}
	\bar{\varphi} \sim \frac{\bar{\dTe}}{T_{0e}} \sim \left( \frac{\varepsilon}{\Omega_e} \right)^{1/3} (\kperp \rho_e)^{-2/3}.
	\label{eq:cb_field_scaling}
\end{align}
Then, recalling \cref{eq:cb_varphi_definition}, the 1D perpendicular energy spectra in the inertial range are: 
\begin{align}
	E_\perp^\varphi(\kperp) \sim E_\perp^T(\kperp) \sim \frac{\bar{\varphi}^2}{\kperp} \propto \kperp^{-7/3}.
	\label{eq:cb_perp_spectrum_scaling}
\end{align}

Using \cref{eq:cb_nonlinear_time} and \cref{eq:cb_field_scaling}, the critical balance \cref{eq:cb_critical_balance} translates into the following relationship between parallel and perpendicular scales in the inertial range:
\begin{align}
	\kpar \lambdae \sim  \frac{\Omega_e^{1/3} \varepsilon^{1/6}}{\nu_{ei}^{1/2}} (\kperp \rho_e)^{2/3}.
	\label{eq:cb_critical_balance_wavenumbers}
\end{align}
If we define the 1D parallel spectrum
\begin{align}
	E_\parallel^\varphi(\kpar) \equiv  \int_0^\infty \rmd \kperp  \: 2\pi\kperp \left<|\varphi_{\vec{k}}|^2 \right>,
	\label{eq:cb_parallel_spectrum}
\end{align}
and the corresponding temperature spectrum $E_\parallel^T(\kpar)$ analogously, \cref{eq:cb_critical_balance_wavenumbers} and~\cref{eq:cb_field_scaling} imply the following inertial-range scaling of amplitudes with parallel wavenumbers:
\begin{align}
	\bar{\varphi} \propto \kpar^{-1} \quad \Rightarrow \quad E_\parallel^\varphi(\kpar) \sim E_\parallel^T(\kpar) \sim \frac{\bar{\varphi}^2}{\kpar} \propto \kpar^{-3}. 
	\label{eq:cb_parallel_spectrum_scaling}
\end{align}

\begin{figure}
	
	\centering
	
	\begin{tabular}{cc}
		\multicolumn{2}{c}{\includegraphics[width=1\textwidth]{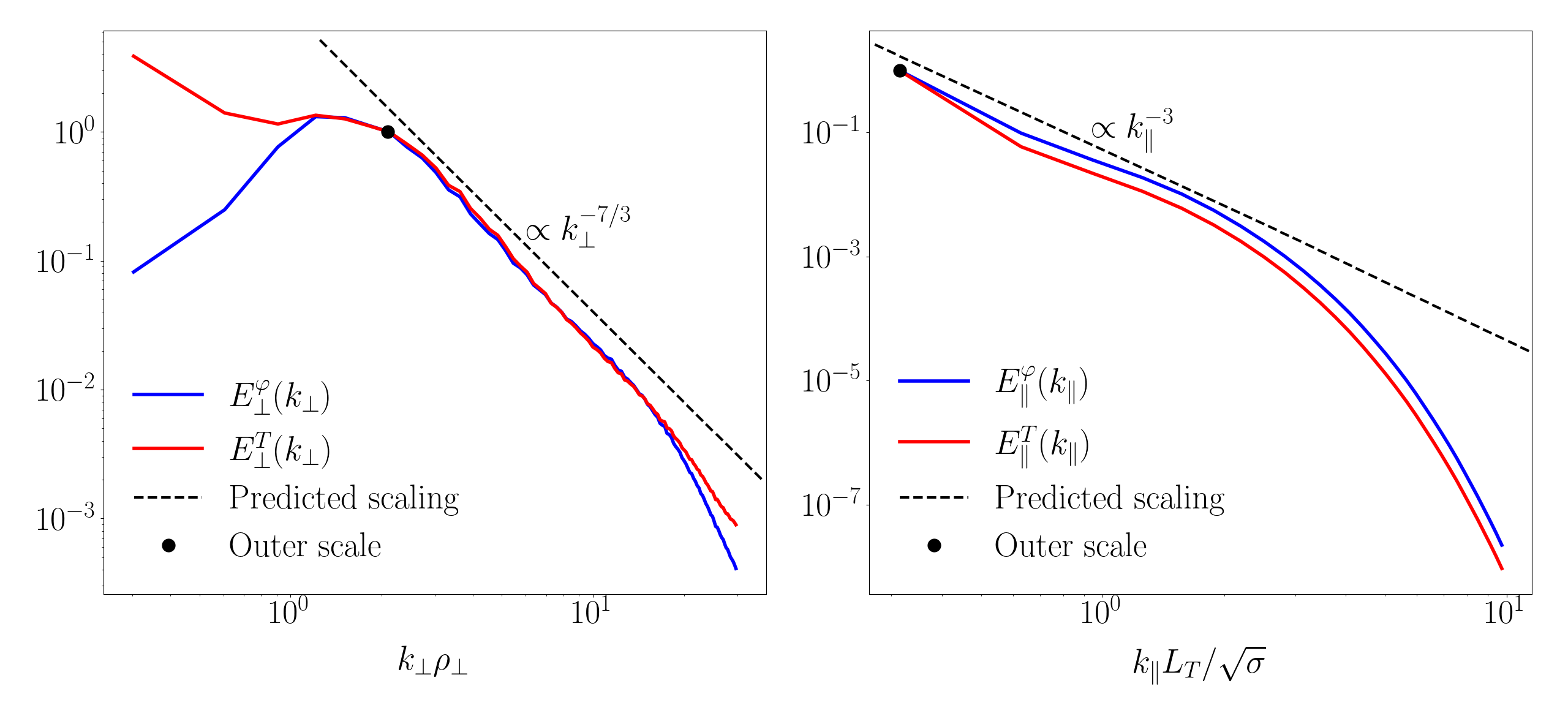}}   \\
		\hspace{3.4cm}	(a)  &  \hspace{2.9cm} (b) 
	\end{tabular}

	\caption[]{The 1D (a) perpendicular \cref{eq:cb_varphi_definition} and (b) parallel \cref{eq:cb_parallel_spectrum} spectra, normalised to their value at the outer scale. The spectra of the electrostatic potential are plotted in blue, those of the temperature perturbations are in red. The predicted inertial-range scalings \cref{eq:cb_perp_spectrum_scaling} and \cref{eq:cb_parallel_spectrum_scaling} are shown by the dashed black lines. The location of the outer scale (see \cref{sec:outer_scale}) is indicated by the black dot. In (a), this is calculated from the maximum of \cref{eq:injection_rate}, while in (b), it is calculated from the maximum of the 1D parallel spectrum of the energy injection, defined analogously to \cref{eq:injection_rate}.}
	\label{fig:high_res_spectra_1d}
	
\end{figure} 

These simple scaling arguments are vindicated by simulation data. The 1D spectra \cref{eq:cb_varphi_definition} and \cref{eq:cb_parallel_spectrum} for both $\varphi$ and $\dTe$ are plotted in \cref{fig:high_res_spectra_1d}. They follow quite well the predicted scalings \cref{eq:cb_perp_spectrum_scaling} and \cref{eq:cb_parallel_spectrum_scaling}, respectively, below the outer scale and up to the wavenumbers at which the spectra begin to steepen due to perpendicular dissipation. 


We shall return to these inertial-range scalings in \cref{sec:two_dimensional_spectra}, where we will study the full 2D spectra of the turbulence and provide further support for the argument that the cascade follows the dissipation ridge (the line of critical balance), but first let us demonstrate how the simple scaling theory developed above allows one to recover --- now on physically motivated dynamical grounds --- the scaling of the heat flux \cref{eq:dk_heat_flux_final} that was previously inferred from a formal scaling symmetry of our equations.

\subsection{Outer scale and scaling of heat flux}
\label{sec:outer_scale}
From \cref{eq:lin_max}, we know that, for a given $k_y$, the most unstable collisional sETG modes satisfy
\begin{align}
	\wpar \sim \omega_{*e} \sim k_y \rho_e \frac{ \vthe}{L_T},
	\label{eq:cb_setg_max}
\end{align}
and thus grow at a rate $\sim \omega_{*e} \propto k_y$. This means that the linear instability will be overwhelmed by nonlinear interactions in the inertial range, because their characteristic rate increases more quickly with perpendicular wavenumber: from \cref{eq:cb_nonlinear_time} and \cref{eq:cb_field_scaling},
\begin{align}
	t_\text{nl}^{-1} \sim \Omega_e \left( \frac{\varepsilon}{\Omega_e} \right)^{1/3} (k_\perp \rho_e)^{4/3}.
	\label{eq:cb_nonlinear_time_scaling}
\end{align}
The outer scale is then the scale at which these two rates are comparable: balancing \cref{eq:cb_nonlinear_time_scaling} and \cref{eq:cb_setg_max}, we get 
\begin{align}
	\Omega_e (k_\perp^o \rho_e )^2 \amp{\varphi}^o \sim \omega_\parallel^o \sim \omega_{*e} \quad \Rightarrow \quad \amp{\varphi}^o \sim (k_\perp^o L_T)^{-1}, \quad k_y^o \rho_e \sim (k_\parallel^o)^2 L_T \lambdae,
	\label{eq:cb_outerscale_balance}
\end{align}
where the superscript `$o$' refers to outer-scale quantities. Thus, we have two relationships between $k_\perp^o$, $\amp{\varphi}^o$ and $k_\parallel^o$, but in order to determine the outer-scale quantities uniquely, we need a third constraint. Given that our system \cref{eq:density_moment_num}-\cref{eq:temperature_moment_num} is scale invariant, there is no special (microscopic) perpendicular scale that can be used to fix $k_y^o$. Then, assuming that the heat flux is independent of the perpendicular system size, the only remaining physically meaningful length scale that can set the outer scale is the parallel system size $L_\parallel$. The same should be true for more general systems described by electrostatic drift kinetics, as the scaling \cref{eq:dk_heat_flux_final} would suggest and as we shall discuss shortly. 

\begin{figure}
	
	\begin{tabular}{cc}
		\multicolumn{2}{c}{\includegraphics[width=1\textwidth]{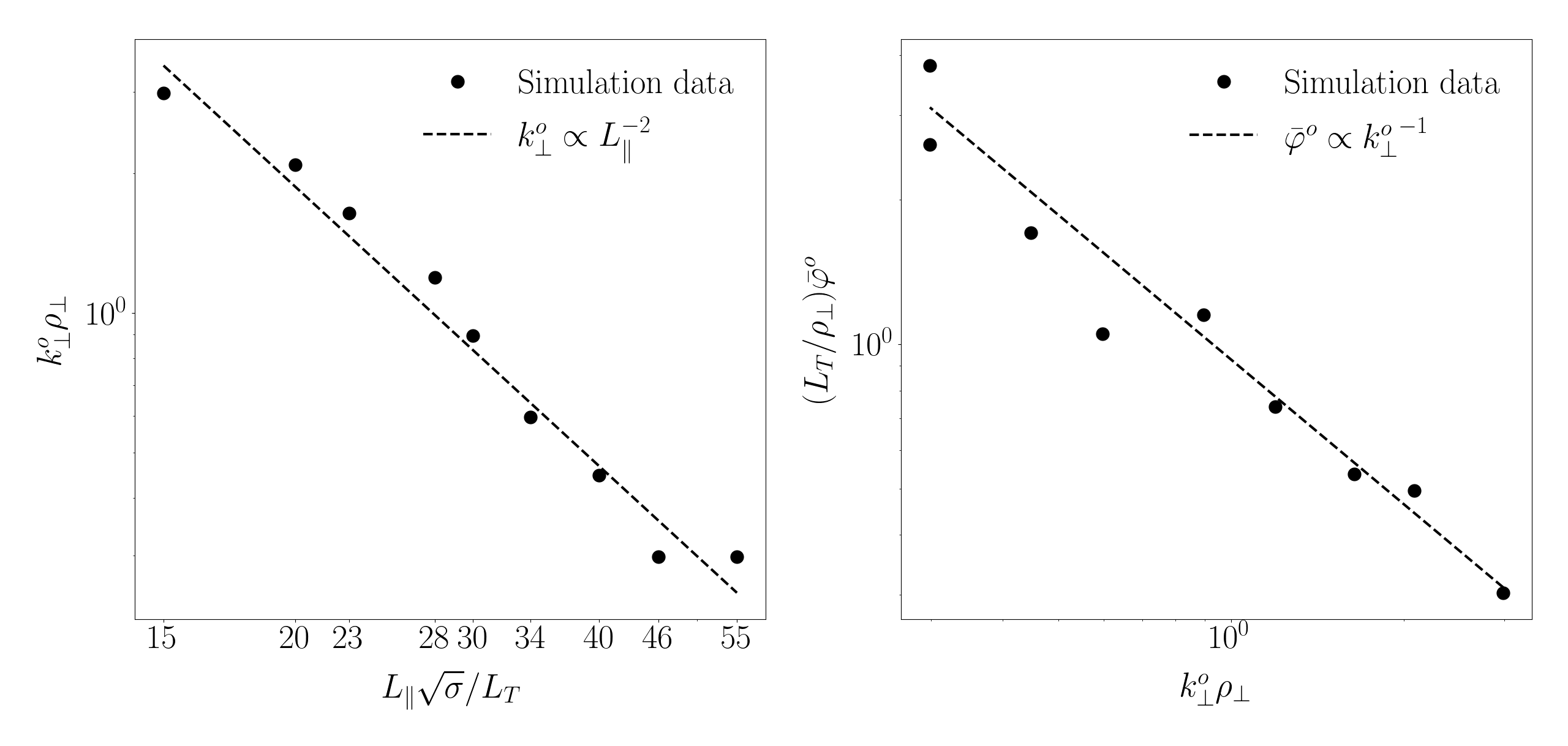}}   \\
		\hspace{3.6cm}	(a) &  \hspace{3cm} (b)
	\end{tabular}
	
	\caption[]{(a) The scaling of the perpendicular outer scale $\kperp^o$ [defined as the peak wavenumber of the energy injection \cref{eq:injection_rate}] with $L_\parallel/L_T$. (b) The scaling of the amplitude of the electrostatic potential $\bar{\varphi}^o$ [defined as the amplitude of $\varphi$ at $\kperp = \kperp^o$, via \cref{eq:cb_varphi_definition}] with the perpendicular outer scale. The black points are the simulation data, while the theoretical predictions \cref{eq:cb_outerscale_balance_explicit} are shown by the black dashed lines. A logarithmic fit to the data gives the slopes of -1.99 and -0.95 in (a) and (b), respectively.}
	\label{fig:outer_scale_scan}
\end{figure}

Assuming, then, that the outer scale is indeed set by the parallel system size, we find from \cref{eq:cb_outerscale_balance}:
\begin{align}
	\kpar^o L_\parallel \sim 1 \quad  \Rightarrow \quad \left( \frac{L_T}{\rhoperp} \right) \bar{\varphi}^o \sim \left( k_\perp^o \rhoperp \right)^{-1}, \quad  k_\perp^o \rhoperp \sim \left( \frac{L_T}{L_\parallel \sqrt{\sigma}} \right)^2,
	\label{eq:cb_outerscale_balance_explicit}
\end{align}
where $\rhoperp$ and $\sigma$ are defined in \cref{eq:col_wavenumber_range_rhoperp}, and the magnitude of $\arbnorm$ only matters for the purposes of normalising amplitudes and wavenumbers in plots. \Cref{fig:outer_scale_scan} shows that these theoretical predictions agree very well with the data from the scan in $L_\parallel/L_T$ that was presented in \cref{sec:scan_in_parallel_system_size}.

Let us now estimate the energy flux that is injected by the collisional sETG instability at the outer scale \cref{eq:cb_outerscale_balance_explicit}: using its definition \cref{eq:cb_injection}, and ignoring any possibility of a non-order-unity contribution from phase factors, we have, from \cref{eq:cb_phi_scaling}, \cref{eq:cb_outerscale_balance} and \cref{eq:cb_outerscale_balance_explicit},
\begin{align}
	\varepsilon \sim \omega_{*e}^o \amp{\varphi}^o \frac{\delta \amp{T}_{e}^o}{T_{0e}} \sim \frac{\vthe \rho_e^2}{L_T^3 }(k_\perp^o \rho_e)^{-1} \sim  \frac{\vthe \rho_e^2 L_\parallel^2}{\lambdae L_T^4}.
	\label{eq:cb_energy_flux}
\end{align}
Recalling \cref{eq:dk_heat_flux}, the combination of \cref{eq:cb_energy_flux} and \cref{eq:cb_outerscale_balance} yields the following expression for the turbulent heat flux:
\begin{align}
	Q_e \sim n_{0e}T_{0e} \varepsilon L_T \sim Q_{\text{gB}e}\frac{L_T}{\lambdae} \left( \frac{L_\parallel}{L_T} \right)^2 \propto \frac{L_\parallel^2}{L_T^3},
	\label{eq:cb_heat_flux}
\end{align}
where once again $Q_{\text{gB}e} = n_{0e} T_{0e} \vthe (\rho_e/L_T)^2$ is the ``gyro-Bohm" flux.
Unsurprisingly, this reproduces the scaling with $L_\parallel$ given by \cref{eq:dk_heat_flux_final} for $\alpha = 2$ [cf., also, \cref{eq:dk_dimensional_analysis} for $G = L_T/\lambdae$]. Note that, apart from the inevitable dimensional factors, the $L_\parallel$ scaling determines (the nontrivial part of) the dependence of the turbulent heat flux on the temperature gradient, as we anticipated following \cref{eq:dk_heat_flux_final}. The fact that our equations \cref{eq:density_moment_num} and \cref{eq:temperature_moment_num} are invariant under the same transformation as drift kinetics \cref{eq:dk_invariance} means that obtaining this scaling was, in a sense, a foregone conclusion. That being said, in arriving at \cref{eq:cb_heat_flux} via this alternative route, we have been able to elucidate the \textit{dynamical} origin of this scaling, viz., that it is consistent with a critically balanced, constant-flux nonlinear cascade of free energy to small perpendicular scales. This conclusion is not exclusive to the collisional model considered in this paper. Starting from~\cref{eq:cb_nonlinear_time}, one can construct an entirely analogous theory for the turbulence driven by the collisionless sETG instability, obtaining a result equivalent to \cref{eq:cb_heat_flux}, which reproduces the scaling~\cref{eq:dk_heat_flux_final}, this time for $\alpha = 1$ \citep[see][]{adkins22}. 

Let us discuss the significance of our finding that the outer scale is fixed by the assumption that $\kpar^o L_\parallel \sim 1$. Such a choice goes back to the work by \cite{barnes11}, who conjectured, and numerically verified, that the outer scale of electrostatic, gyrokinetic ITG turbulence in tokamak geometry was set by the connection length $L_\parallel \sim qR$. While in their case, like ours, this was the only scale that could be reasonably viewed as the characteristic system size (the spatial inhomogeneity of the magnetic equilibrium), there was also another, seemingly more physically intuitive, justification available for its role in determining the large-scale cutoff for the ITG turbulence: one could assume that any turbulent structures correlated on parallel scales longer than the connection length would be damped in the stable (``good-curvature") region on the inboard side of the tokamak. Thus, one could believe that the operative reason for the significance of $L_\parallel \sim qR$ was the presence of large-scale dissipation, rather than, as we have now concluded, just the breaking of scale invariance --- in our case, by the finiteness of a periodic box in the parallel direction\footnote{Our system does of course also have parallel dissipation via heat conduction, at the rate $\sim \omega_\parallel$, but $\omega_\parallel$ decreases with increasing parallel scale and, at any rate, does not break scale invariance, so cannot set $\kpar^o$.}. A practical implication of this conclusion for more realistic systems appears to be that any long-scale parallel inhomogeneity should be sufficient to set $\kpar^o$, without the need for it to be tied to an energy sink --- this could matter for the analysis of turbulence in, e.g., edge plasmas \citep{parisi20,parisi22} or in stellarators \citep{robergclark22}, where magnetic fields have parallel structure on scales shorter than the connection length.

\subsection{Two-dimensional spectra}
\label{sec:two_dimensional_spectra}
To provide a more detailed description of the critically balanced cascade (and to provide more evidence that it is indeed a critically balanced cascade), it is interesting to consider the 2D spectra:
\begin{align}
	E^\varphi_{2\text{D}}(\kperp,\kpar) & =  2\pi\kperp \left<|\varphi_{\vec{k}}|^2 \right> \label{eq:cb_2d_spectrum_phi_definition}, \\
	E^T_{2\text{D}}(\kperp,\kpar) & =  2\pi\kperp \left<|{\dTe}_{\vec{k}}/T_{0e}|^2 \right> \label{eq:cb_2d_spectrum_t_definition}.
\end{align}
Unlike in \cref{sec:constant_flux}, we can no longer assume that $E^\varphi_{2\text{D}} \sim E^T_{2\text{D}}$; this was true only for the ``integrated" 1D spectra dominated by the wavenumbers where the critical-balance conjecture~\cref{eq:cb_critical_balance} was assumed satisfied, and the two fields thus had the same scaling~\cref{eq:cb_phi_scaling} for $\omega \sim \omega_\parallel$.    
With this in mind, we will first consider the spectrum of the temperature perturbations, from which the spectrum of the potential perturbations can then be inferred via \cref{eq:cb_phi_scaling}. 

We consider two wavenumber regions, above and below the ``critical-balance line"~\cref{eq:cb_critical_balance_wavenumbers}:
\begin{align}
	E_{2\text{D}}^T(\kperp, \kpar) \sim 
	\left\{
	\begin{array}{ll}
		\displaystyle \kpar^{-a} \kperp^b, &  \displaystyle  \kpar \gtrsim \kperp^{2/3} ,   \\[4mm]
		\displaystyle  \kperp^{-c} \kpar^d, &  \displaystyle  \kpar \lesssim \kperp^{2/3}, 
	\end{array}
	\right. 
	\label{eq:cb_2d_spectrum_initial}
\end{align}
where $a$, $b$, $c$, and $d$ are positive constants to be determined. Here, and in what follows, whenever our expressions appear to be dimensionally incorrect, this is because we have implicitly chosen to normalise our wavenumbers to the outer scale $\kpar/\kpar^o \rightarrow \kpar$, $\kperp/\kperp^o \rightarrow \kperp$ so as to reduce notational clutter. To determine the scaling exponents, we follow the general scheme, which, for MHD turbulence, was laid out by \cite{sch22} (see his appendix C).

Evidently, the scalings in the two regions in \cref{eq:cb_2d_spectrum_initial} must match along the boundary $\kpar \sim \kperp^{2/3}$, giving
\begin{align}
	a + d = \frac{3}{2}(b+c).
	\label{eq:cb_matching_condition}
\end{align}
If $a>1$ and $d> -1$, $\kpar \sim \kperp^{2/3}$ will be the energy-containing parallel wavenumber at a given~$\kperp$. The 1D perpendicular spectrum is, therefore, 
\begin{align}
	E_\perp^T(\kperp) = \int \rmd \kpar \: E_{2\text{D}}^T(\kperp, \kpar) \sim \int_0^{\kperp^{2/3}} \rmd \kpar \: \kperp^{-c}\kpar^d \sim \kperp^{-c + 2(1+d)/3}.
	\label{eq:cb_perp_spectrum}
\end{align}
This must match the scaling \cref{eq:cb_perp_spectrum_scaling} of the 1D perpendicular spectrum derived from the constant-flux conjecture, implying that 
\begin{align}
	c = \frac{2}{3}(1+d) + \frac{7}{3}. 
	\label{eq:cb_matching_condition_flux}
\end{align}

Two further constraints follow from imposing boundary conditions as $\kpar$ or $\kperp \rightarrow 0$ at constant $\kperp$ or $\kpar$, respectively. The scaling of the spectrum as $\kperp \rightarrow 0$ (in the region $\kpar \gg \kperp^{2/3}$) can be determined purely kinematically: the low-$\kperp$ asymptotic behaviour of a homogenous 2D-isotropic field must be $\kperp^3$, implying that 
\begin{align}
	b = 3.
	\label{eq:cb_low_kperp}
\end{align}
This is a fairly standard result\footnote{Though not one that can be taken for granted. For example, \cite{hosking22forced} \citep[see also appendix C of][]{sch22} showed that a $\kperp^1$ scaling could emerge instead through a balance between turbulent diffusion at large scales and the nonlinear `source' that would otherwise give rise to the $\kperp^3$ scaling. Let us estimate the rate of turbulent diffusion in our system. The dominant contribution to the turbulent-diffusion coefficient $D$ will be from $\kperp \sim \kpar^{3/2}$, which, at any given $\kpar$, plays the role of the energy-containing scale. Then $D \sim v_E^2 t_\text{nl} \sim \tilde{\omega}_\parallel/{\tilde{k}_\perp}^2$, where have used the critical-balance condition \cref{eq:cb_critical_balance}, and the tildes denote quantities evaluated at $\tilde{k}_\perp \sim \kpar^{3/2} \gg \kperp$, where $\kperp$ is the wavenumber at which turbulent diffusion is acting. The rate of turbulent diffusion at this wavenumber will thus be $\kperp^2 D \sim \tilde{\omega}_\parallel (\kperp/\tilde{k}_\perp)^2 \ll \omega_\parallel$. Turbulent diffusion is, therefore, negligible, and so we are justified in adopting the $\kperp^3$ scaling. Note that the survival of the $\kperp^3$ scaling is a noteworthy feature of our system, where the dynamics at $\kpar \gg \kperp^{2/3}$ are dominated by parallel dissipation due to thermal conductivity and do not produce significant turbulent diffusion --- unlike waves, which, e.g., in RMHD, do \citep{sch22}.} \citep[see, e.g., appendix A of][]{sch16}. Finally, the scaling as $\kpar \rightarrow 0$ (in the region $\kpar \ll \kperp^{2/3}$) follows from causality. Indeed, in \cref{sec:constant_flux}, we argued that fluctuations become decorrelated for $\omega_\parallel \lesssim t_\text{nl}^{-1}$ because they cannot communicate across parallel distances $\sim \kpar^{-1}$, but such $\kpar$ are also too small for the fluctuations to be erased by thermal conduction. Therefore, the parallel spectrum at $\kpar \ll \kperp^{2/3}$ must be the spectrum of a 1D white noise: 
\begin{align}
	d = 0.
	\label{eq:cb_white_noise}
\end{align}

Combining \cref{eq:cb_low_kperp} and \cref{eq:cb_white_noise} with \cref{eq:cb_matching_condition} and \cref{eq:cb_matching_condition_flux}, we find
\begin{align}
	a = 9, \quad c = 3.
	\label{eq:cb_constants}
\end{align}
This gives us the following scalings for the 2D spectrum of the temperature perturbations\footnote{\cite{sch16} obtained, by a similar method, an analogous result for long-wavelength electrostatic ITG turbulence (which, in this approach, is no different for ETG). Specifically, they found that $a = 5$ and $c = 11/3$ --- this was a consequence of the fact that they considered collisionless turbulence, for which the critical-balance condition is $\wpar \sim \kpar \vthe \sim t_\text{nl}^{-1}$ implying that $ \kpar \sim \kperp^{4/3}$.}:
\begin{align}
	E_{2\text{D}}^T(\kperp, \kpar) \sim 
	\left\{
	\begin{array}{ll}
		\displaystyle \kpar^{-9} \kperp^3, &  \displaystyle  \kpar \gtrsim \kperp^{2/3} ,   \\[4mm]
		\displaystyle  \kperp^{-3} \kpar^0, &  \displaystyle  \kpar \lesssim \kperp^{2/3}. 
	\end{array}
	\right. 
	\label{eq:cb_2d_spectrum_final_t}
\end{align}

Turning now to the 2D spectrum of the potential perturbations, analogously to \cref{eq:cb_2d_spectrum_initial}, the conditions \cref{eq:cb_matching_condition} and \cref{eq:cb_matching_condition_flux} are unmodified --- the spectrum must still be continuous along $\kpar \sim \kperp^{2/3}$, and match the scaling of the 1D perpendicular spectrum that follows from the constant-flux conjecture, which is the same for both the potential and temperature perturbations. Similarly, the scaling of the spectrum as $\kperp \rightarrow 0$ (in the region $\kpar \gtrsim \kperp^{2/3}$) will once again be $\kperp^3$ by the same kinematic argument, implying \cref{eq:cb_low_kperp}. From \cref{eq:cb_matching_condition} and \cref{eq:cb_matching_condition_flux}, we again have $a=9$. However, the causality argument that led to the white-noise scaling \cref{eq:cb_white_noise} at $\kpar \lesssim \kperp^{2/3}$ now no longer holds, because $\varphi$ is not directly decorrelated by the nonlinearity. Instead, it inherits its scaling from $\dTe/T_{0e}$ via the balance \cref{eq:cb_phi_scaling}, viz.,
\begin{align}
	E_{2\text{D}}^\varphi \sim \frac{\omega_\parallel^2}{\omega^2} E_{2\text{D}}^T \sim \frac{\kpar^4 \kperp^{-3}}{\omega^2},
	\label{eq:cb_phi_scaling_spectrum}
\end{align}
where we have used $\omega_\parallel \propto \kpar^2$ and the second expression in \cref{eq:cb_2d_spectrum_final_t}. Now, in the region $\kpar \lesssim \kperp^{2/3}$, we expect thermal conductivity in the temperature equation to be subdominant to the nonlinear rate, and so estimating $\omega \sim t_\text{nl}^{-1} \propto \kperp^{4/3}$ in \cref{eq:cb_phi_scaling_spectrum}, we find that 
\begin{align}
	d=4, \quad c = \frac{17}{3}.
	\label{eq:cb_white_noise_phi}
\end{align}
This gives us the following scalings of the 2D spectrum of the potential fluctuations:
\begin{align}
	E_{2\text{D}}^\varphi(\kperp, \kpar) \sim 
	\left\{
	\begin{array}{ll}
		\displaystyle \kpar^{-9} \kperp^3, &  \displaystyle  \kpar \gtrsim \kperp^{2/3} ,   \\[4mm]
		\displaystyle  \kperp^{-17/3} \kpar^4, &  \displaystyle  \kpar \lesssim \kperp^{2/3}. 
	\end{array}
	\right. 
	\label{eq:cb_2d_spectrum_final_phi}
\end{align}

\begin{figure}
	
	\includegraphics[width=1\textwidth]{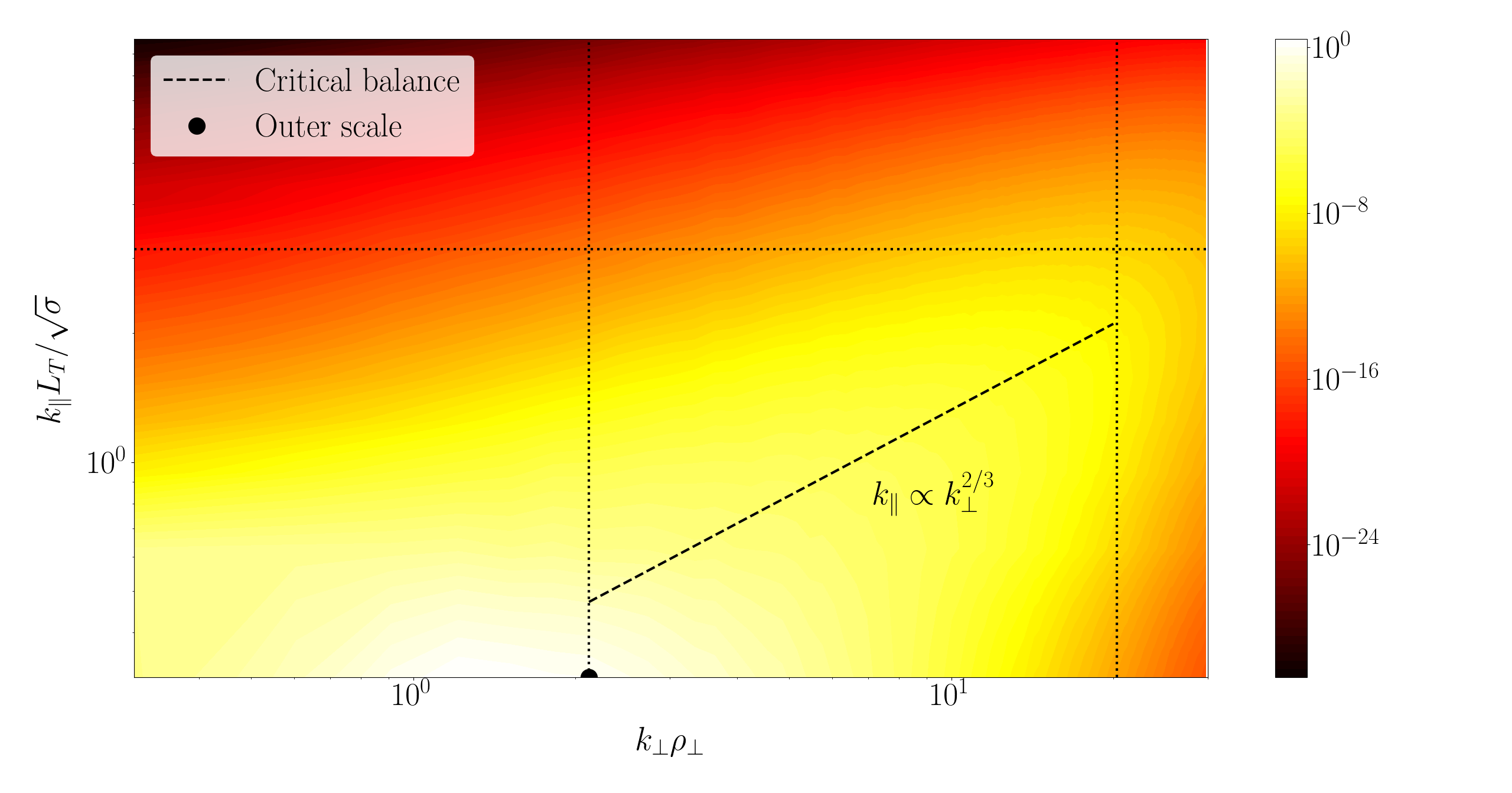}
	
	\centering
	
	\caption[]{A contour plot of the logarithm of the 2D spectrum \cref{eq:cb_2d_spectrum_t_definition} of the temperature perturbations in the $(\kperp,\kpar)$ plane, normalised to its value at the outer scale. The line of critical balance is shown as the dashed black line, while the outer scale is shown by the black dot. The horizontal dotted line shows the upper bound on the parallel-wavenumber cuts plotted in the right panels of \cref{fig:high_res_spectrum_kpar_cuts}. Similarly, the vertical dotted lines show the lower and upper bounds on the perpendicular-wavenumber cuts plotted in the right panels of \cref{fig:high_res_spectrum_kperp_cuts}.}
	\label{fig:high_res_2d_spectrum}
	
\end{figure} 

\begin{figure}[p]
	
	\centering
	\vspace{-0.5cm}
	\begin{tabular}{c}
		
		\hspace{-1cm}	\includegraphics[width=1\textwidth]{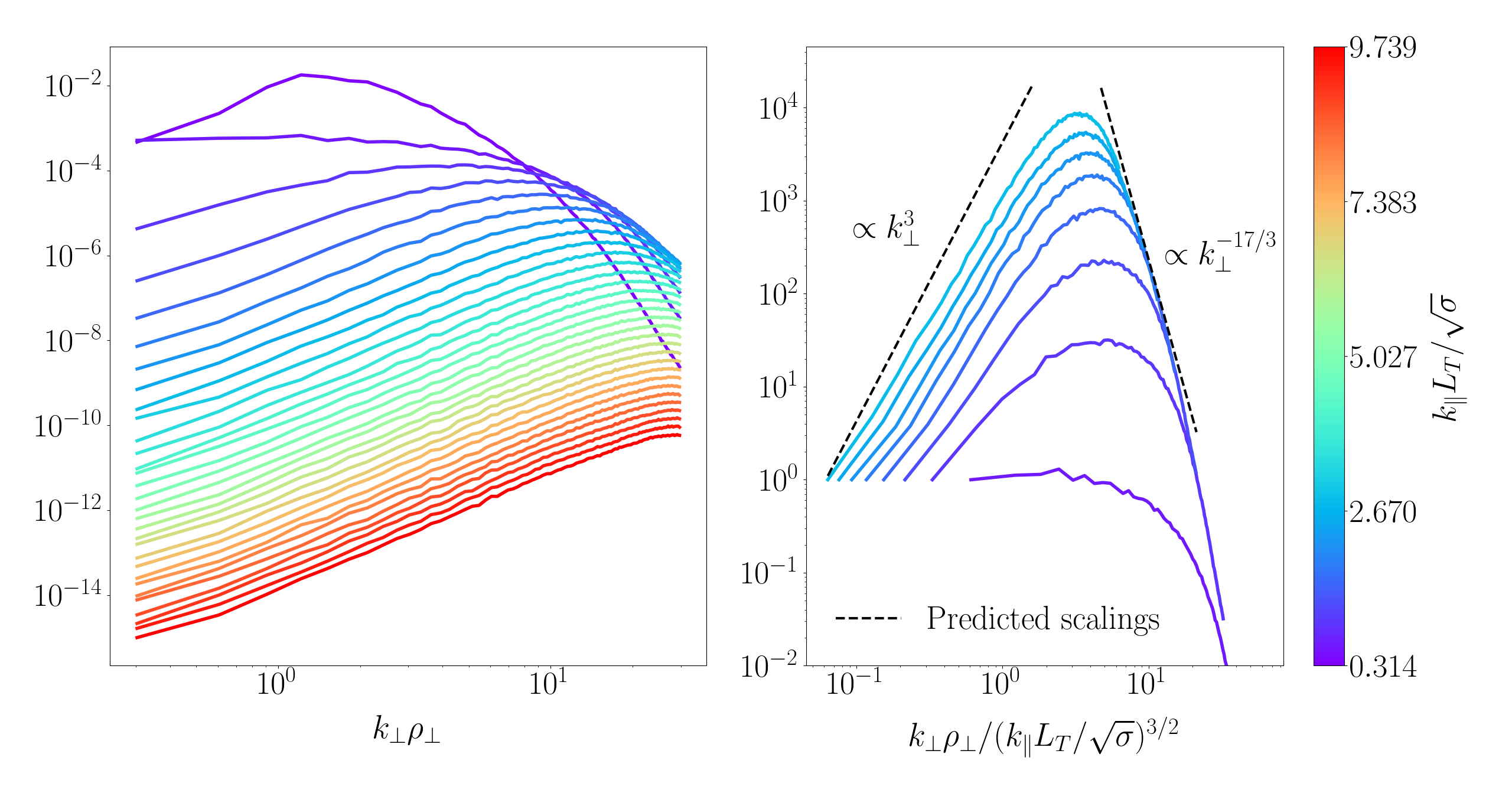}  \\
		(a) $E_{2\text{D}}^\varphi (\kperp,\kpar = \text{const})$\\
		\hspace{-1cm}	\includegraphics[width=1\textwidth]{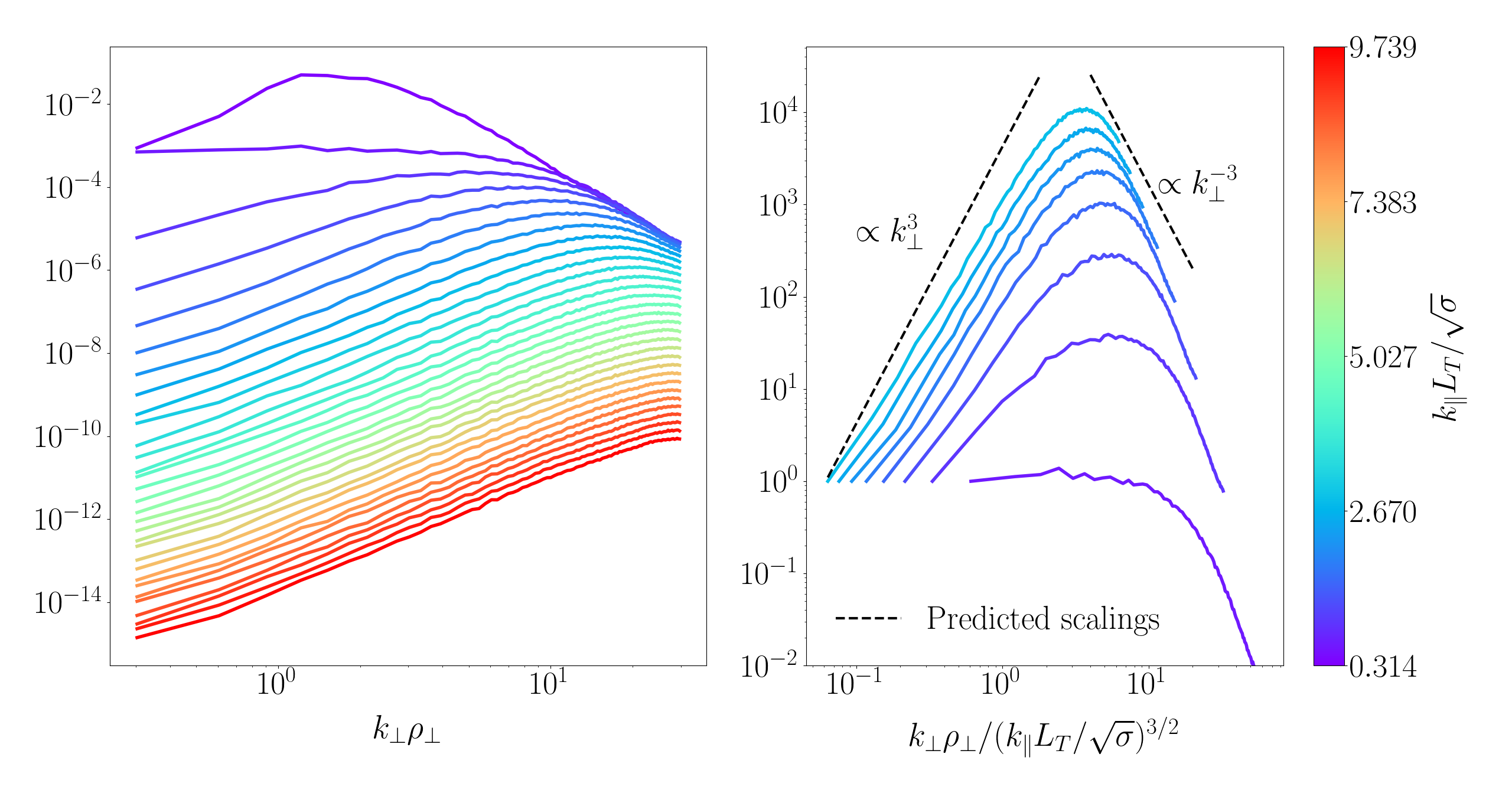}  \\
		(b) $E_{2\text{D}}^T (\kperp,\kpar = \text{const})$
		
	\end{tabular}

	\caption[Cuts of the two-dimensional spectrum of sETG turbulence at constant $\kpar$]{Cuts of the 2D spectra of (a) the electrostatic potential and (b) the temperature perturbations at constant $\kpar$, normalised to $(\rhoperp/L_T)^2$. The colours indicate the value of $\kpar L_T/\sqrt{\sigma}$ for a given cut. The left panels show the entire spectrum plotted as a function of $\kperp \rhoperp$. The right panels show selected cuts for $\kpar L_T$ within the inertial range, with $\kperp$ rescaled according to the critical-balance relation \cref{eq:cb_critical_balance_wavenumbers}. The black dashed lines show the theoretical scalings \cref{eq:cb_2d_spectrum_final_phi} and \cref{eq:cb_2d_spectrum_final_t} in panels (a) and (b), respectively. The spectra show reasonable agreement with theory at both small and large perpendicular scales, despite the effects of hyperviscosity being present at the smallest scales.}
	\label{fig:high_res_spectrum_kpar_cuts}
	
\end{figure}

\begin{figure}[p]
	
	\centering
	\vspace{-0.5cm}
	\begin{tabular}{c}
		
		\hspace{-1cm}	\includegraphics[width=1\textwidth]{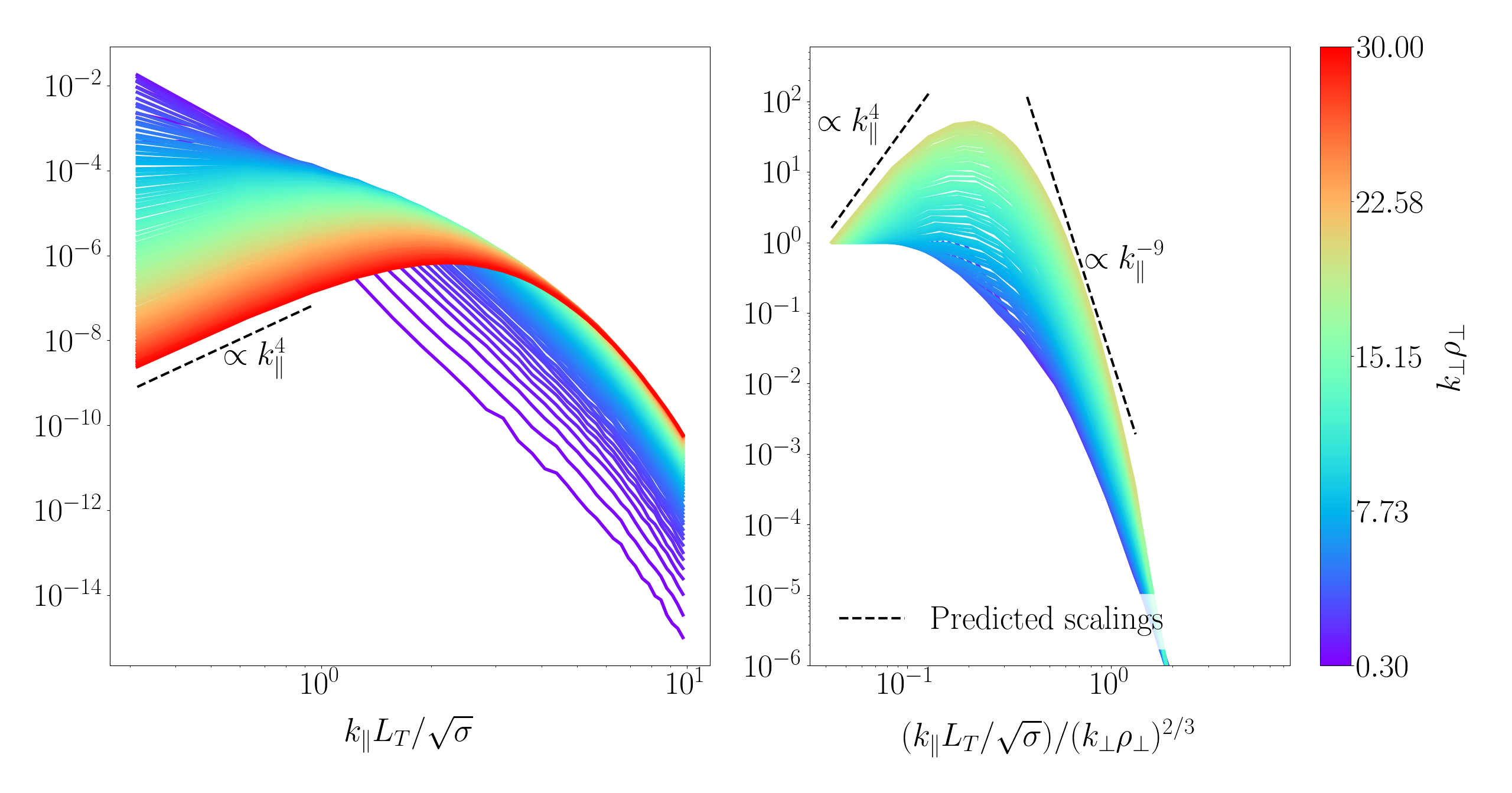}  \\
		(a) $E_{2\text{D}}^\varphi (\kperp = \text{const}, \kpar)$ \\
		\hspace{-1cm}	\includegraphics[width=1\textwidth]{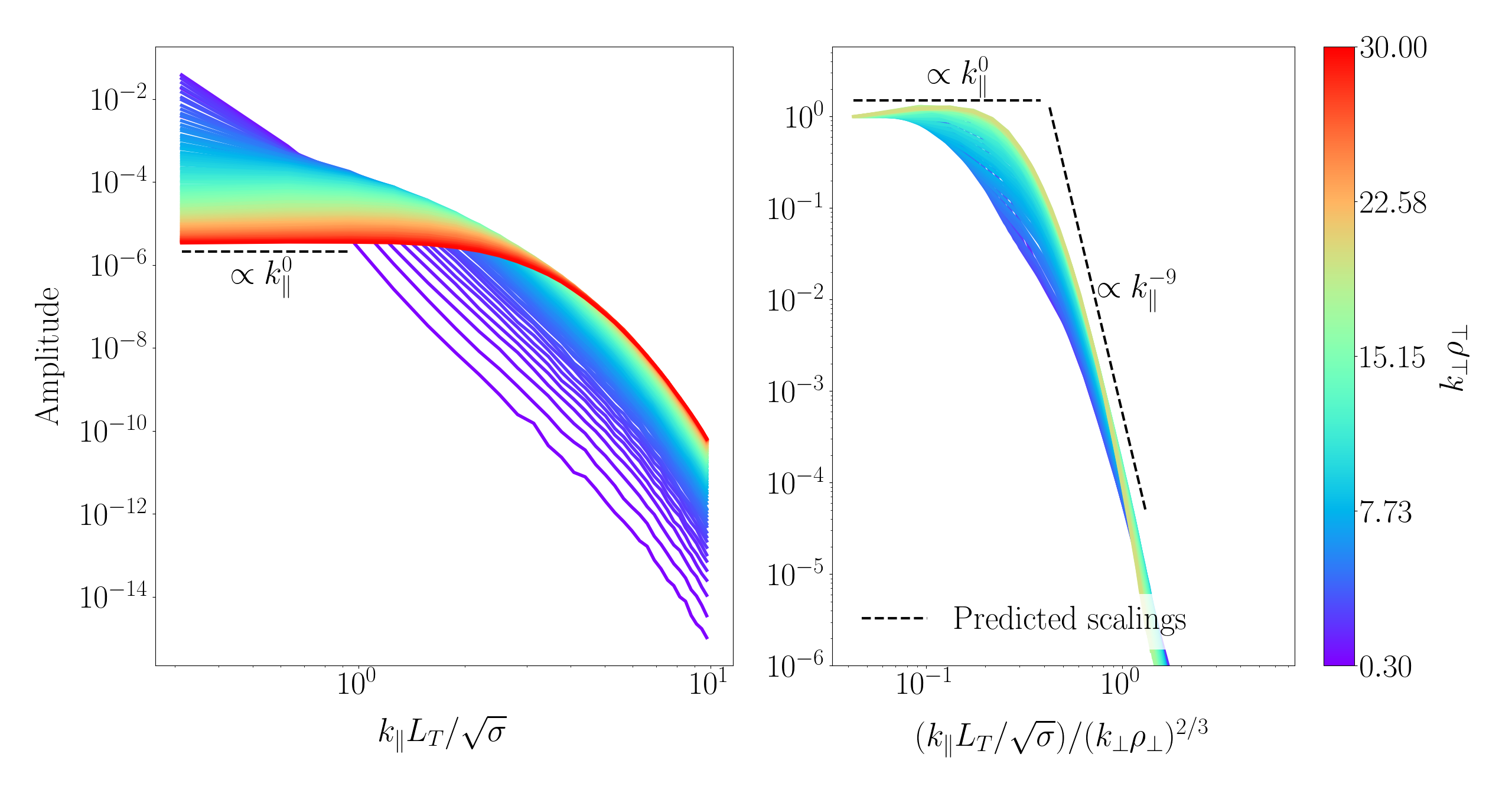}  \\
		(b) $E_{2\text{D}}^T (\kperp = \text{const}, \kpar)$
		
	\end{tabular}

	\caption[Cuts of the two-dimensional spectrum of sETG turbulence at constant $\kperp$]{Cuts of the 2D spectra of (a) the electrostatic potential and (b) the temperature perturbations at constant $\kperp$, normalised to $(\rhoperp/L_T)^2$. The colours indicate the value of $\kperp \rhoperp$ for a given cut. The left panels show the entire spectrum plotted as a function of $\kpar L_T$. The right panels show selected cuts of the spectrum for $\kperp \rhoperp$ within the inertial range, with $\kpar$ rescaled according to the critical-balance relation \cref{eq:cb_critical_balance_wavenumbers}. The black dashed lines show the theoretical scalings \cref{eq:cb_2d_spectrum_final_phi} and \cref{eq:cb_2d_spectrum_final_t} in panels (a) and (b), respectively. There is very good agreement with theory, especially at $\kpar \lesssim \kperp^{2/3}$, where the scalings extend well beyond the inertial range to higher $\kperp \rhoperp$, as can be seen from the left panels --- this is because the causality argument is not sensitive to the precise details of the decorrelation physics.}
	\label{fig:high_res_spectrum_kperp_cuts}
	
\end{figure}

The full 2D spectrum of the temperature perturbations is plotted in \cref{fig:high_res_2d_spectrum}. The organisation of the system about the critical-balance line is manifest here.
Cuts of the 2D spectra \cref{eq:cb_2d_spectrum_phi_definition} and \cref{eq:cb_2d_spectrum_t_definition} at constant $\kpar$ and $\kperp$ are shown in figures \ref{fig:high_res_spectrum_kpar_cuts} and \ref{fig:high_res_spectrum_kperp_cuts}, respectively, for both potential and temperature perturbations, showing good agreement with the theoretical scalings \cref{eq:cb_2d_spectrum_final_t} and \cref{eq:cb_2d_spectrum_final_phi}. The white-noise spectra at $\kpar \lesssim \kperp^{2/3}$, in particular, are another confirmation of the causal nature of the critical balance. 

The extraordinarily steep parallel-wavenumber scaling of the 2D spectra \cref{eq:cb_2d_spectrum_final_t} and \cref{eq:cb_2d_spectrum_final_phi} in the region $\kpar \gtrsim \kperp^{2/3}$ can also be viewed as further evidence for the version of critical balance proposed following \cref{eq:cb_critical_balance}. In terms of timescales, this wavenumber constraint corresponds to $\omega_\parallel \gtrsim t_\text{nl}^{-1}$, and thus to a region of dominant thermal conduction that attempts to erase parallel structure created by the turbulence. The $\kpar$ scaling in this region proves to be so steep that the free-energy sink due to parallel dissipation is ineffective: the free energy cannot be nonlinearly transferred into this region in an efficient way, and instead cascades towards higher perpendicular wavenumbers along the critical-balance line, eventually encountering perpendicular dissipation, introduced, in our model, through hyperviscosity. Parallel dissipation thus acts not as a sink for the cascade, but instead creates the aforementioned ``dissipation ridge", constraining the cascade of energy in wavenumber space to be along the critical-balance line $\kpar \sim \kperp^{2/3}$. 

One could dismiss this feature as being a peculiarity of our collisional model, given that the dissipative nature of collisional sETG instability \cref{eq:lin_setg} is hard-wired into it by construction. However, this picture might not be entirely dissimilar from what is observed in more generic systems of plasma turbulence: e.g., \cite{hatch11prl} observed an overlap of the spatial scales of energy injection and dissipation in electrostatic, ion-scale toroidal gyrokinetic simulations, as did \cite{told15} in the context of Alfv\'enic turbulence. The same behaviour could also be relevant in the context of kinetic ETG-driven turbulence. The growth rate of the collisionless sETG instability is limited by the parallel streaming rate $\wpar \sim \kpar \vthe$ \citep[see, e.g.,][]{adkins22}, which is also the rate of Landau damping; viewed in the context of the current discussion, this suggests, perhaps, that Landau damping could play a dissipative role similar to that of the thermal conduction in determining the way in which the system organises itself in order to support a constant-flux cascade of energy to small scales. Then, the rates of either parallel streaming or thermal conduction appearing in the critical balance $\wpar \sim \omega \sim t_\text{nl}^{-1}$
can also be interpreted as being there because they are the rates of parallel dissipation, rather than of the parallel propagation of information, limiting any further refinement of the parallel scale of the turbulent structures.

\subsection{Perpendicular isotropy}
\label{sec:perpendicular_isotropy}
\begin{figure}
	
	\begin{tabular}{cc}
		\hspace{-1cm} \includegraphics[width=0.6\textwidth]{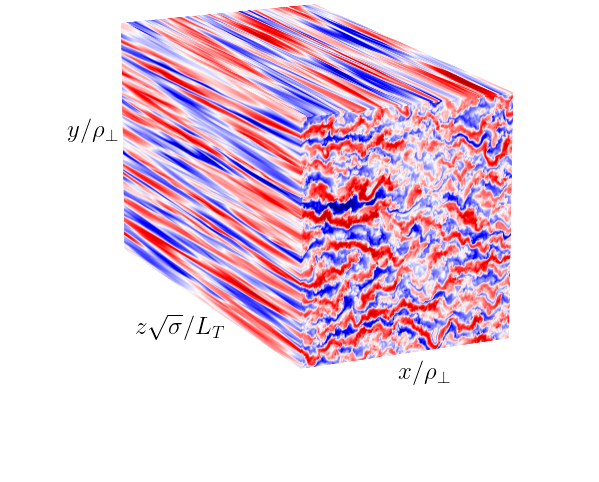} & 
		\hspace{-1cm} \includegraphics[width=0.6\textwidth]{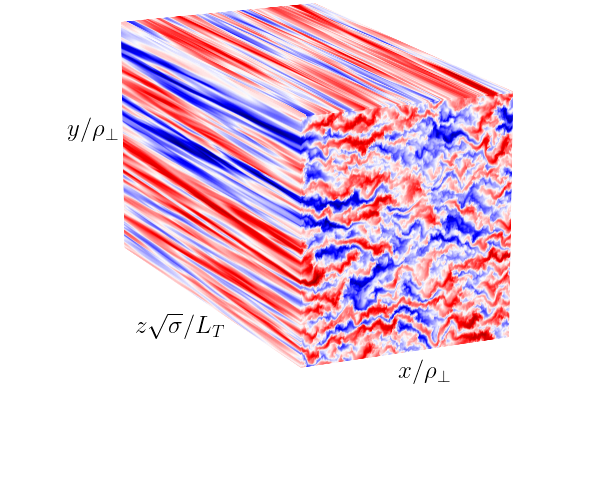} \\[-2.2em]
		(a) $(L_T/\rhoperp)\varphi$ &   (b) $(L_T/\rhoperp)\dTe/T_{0e}$
	\end{tabular}
	
	\centering
	\vspace{0.3cm}
	\caption[]{Real-space snapshots of (a) the electrostatic potential and (b) the temperature perturbations from the higher-resolution simulation at $(\rho_e/\rhoperp)^2\nu_{ei} t/2\arbnorm = 3000$ (see \cref{tab:simulation_parameters}). The coordinate axes are as shown, while the red and blue colours correspond to regions of positive and negative fluctuation amplitudes. The turbulence does not appear to be isotropic on the large scales that are visible in these plots (streamers are manifest), but turns out to be isotropic in the inertial range (see \cref{fig:high_res_isotropy}).}
	\label{fig:high_res_real_space}
	
\end{figure}

Throughout \crefrange{sec:constant_flux}{sec:two_dimensional_spectra}, our theoretical deductions were carried out under the assumption that $k_x \sim k_y \sim \kperp$. This assumption of perpendicular isotropy is not obviously true and must be tested. Indeed, the maximum sETG growth rate \cref{eq:lin_max} is at $k_x=0$, and so the outer-scale energy injection is predominantly into the so-called `streamers': highly anisotropic ($k_x \ll k_y$) structures that can be identified in real space by their alternating pattern of horizontal bands stretched along the radial ($x$) direction \citep{cowley91}. In the context of ITG-driven turbulence, it is often assumed (and usually confirmed numerically) that these streamers are broken apart by zonal flows (see \citealt{barnes11} and references therein), restoring isotropy at the outer scale; isotropy in the inertial range is then assumed as well. In ETG-driven turbulence, however, the role of zonal flows is less obvious \citep[see, e.g.,][]{dorland00,jenko00,colyer17}, and the existence of an isotropic state far from guaranteed --- indeed, the real-space snapshots shown in \cref{fig:high_res_real_space} suggest that the system is in fact dominated by streamer-like structures on the largest scales, and there is little zonal-flow activity. Qualitatively, this is quite similar to what ETG turbulence has been reported to look like in gyrokinetic simulations \citep{joiner06,candy07,roach09,guttenfelder11shear}.

\begin{figure}
	
	\begin{tabular}{cc}
		\multicolumn{2}{c}{\includegraphics[width=1\textwidth]{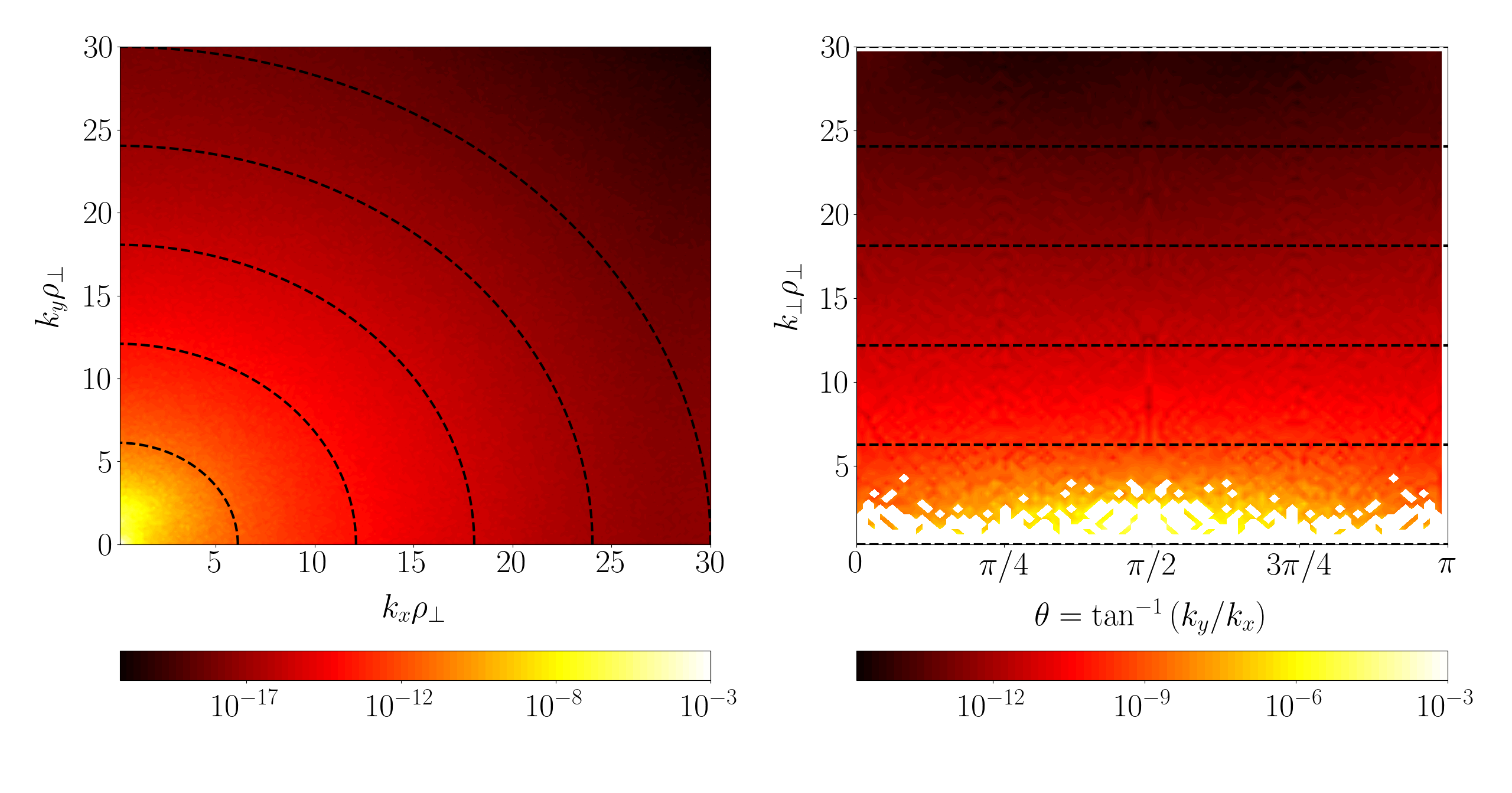}} \vspace*{-0.25cm} \\
		\hspace{2.7cm}	(a) $\log E^T(k_x, k_y)$   &  \hspace{0.5cm}  (b) $\log E^T(\kperp, \theta)$ 
	\end{tabular}
	
	\caption[Perpendicular isotropy of the electrostatic potential perturbations in sETG turbulence]{Contour plots of the two-dimensional spectra of the temperature perturbations, normalised to $(\rhoperp/L_T)^2$: (a) in Cartesian coordinates, with the radial and poloidal wavenumbers plotted on the horizontal and vertical axes, respectively; contours of constant $E^T(k_x, k_y)$ \cref{eq:num_2d_spectra_kx_ky} (black dashed lines) are approximately circular away from the origin, where injection is localised and the presence of streamers is manifested by the spectral power being shifted towards $k_y >k_x$; (b) in polar coordinates, with $\theta = \tan^{-1}(k_y/k_x)$ and $\kperp \rhoperp$ plotted on the horizontal and vertical axes, respectively; contours of constant $E^T(\kperp, \theta)$ \cref{eq:num_2d_spectra_kperp_theta} (black dashed lines) are approximately horizontal far away from $\kperp \rhoperp \lesssim 1$, where injection is localised.}
	\label{fig:high_res_isotropy}
	
\end{figure} 

To assess how isotropic the saturated state is, in particular in the inertial range, we plot the two-dimensional spectra
\begin{align}
	E^T(k_x, k_y) &= \int_{-\infty}^\infty \rmd k_\parallel \left<|{\dTe}_{\vec{k}}/T_{0e}|^2 \right> , \label{eq:num_2d_spectra_kx_ky} \\
	E^T(\kperp, \theta) &= \kperp \int_{-\infty}^\infty \rmd k_\parallel \left<|{\dTe}_{\vec{k}}/T_{0e}|^2 \right> \label{eq:num_2d_spectra_kperp_theta}
\end{align}
in \cref{fig:high_res_isotropy}. Here and in what follows, $\theta = \tan^{-1}(k_y/k_x)$ is the polar angle in the perpendicular wavenumber plane. In both Cartesian and polar representations, we see that the spectrum is approximately isotropic with respect to the perpendicular wavevectors: at scales sufficiently smaller than the outer scale (viz., in the inertial range) contours of constant $E^T$ are either circles, in the case of~\cref{eq:num_2d_spectra_kx_ky}, or horizontal lines, in the case of~\cref{eq:num_2d_spectra_kperp_theta}. The spectra of the potential perturbations, defined analogously to \cref{eq:num_2d_spectra_kx_ky} and~\cref{eq:num_2d_spectra_kperp_theta}, display a similar isotropy. Thus, despite the fact that the largest scales are anisotropic due to the existence of streamers and the lack of vigorous zonal flows to break them apart, isotropy is restored in the inertial range.

The observed lack of zonal-flow activity is linked to the fact that there is nothing on electron scales to give zonal flows privileged status. This makes ETG turbulence quite unlike its ITG cousin on ion scales, where one encounters the modified adiabatic electron response \citep[see, e.g.,][]{hammett93,abel13b}:
\begin{align}
	\frac{\delta n_i}{n_{0i}} = \frac{\dne}{n_{0e}} = \varphi',
	\label{eq:modified_adiabatic_electron_response}
\end{align}
where $\varphi'$ is the non-zonal component of the (non-dimensionalised) electrostatic potential. Indeed, \cref{eq:modified_adiabatic_electron_response} has been found to be crucial for capturing essential zonal-flow physics \citep{rogers00,ivanov20,ivanov22}. Physically, this can be explained by the fact that \cref{eq:modified_adiabatic_electron_response} reserves a special status for zonal flows, in that it allows there to be non-trivial nonlinear interactions between the zonal flows and~$\varphi'$ through the electrostatic $\vec{E} \times \vec{B}$ nonlinearity contained in the convective derivative~\cref{eq:convective_derivative} of the density perturbation. However, as we discussed in~\cref{sec:constant_flux}, the adiabatic ion response~\cref{eq:quasineutrality_final} causes the nonlinearity in the electron continuity equation~\cref{eq:density_moment_num} to vanish identically. Crucially, this means that the system lacks any nonlinearity capable of generating two-dimensional secondary instabilities that are responsible for the generation of zonal flows and destruction of streamer structures \citep{hasegawa78,hasegawa83,terry83,diamond05,ivanov20,zhu20}. A further consequence of this is that the model system \cref{eq:density_moment_num}-\cref{eq:temperature_moment_num} proves to be incapable of generating zonal flows even on longer timescales than those over which the observed isotropisation occurs, unlike what was observed in, e.g., \cite{colyer17} and \cite{tirkas23}. 

\section{Summary and discussion}
\label{sec:summary_and_discussion}
We have considered the transport properties of electrostatic, drift-kinetic plasma turbulence, with a particular focus on the connection between its macroscopic transport properties and microscale (inertial-range) dynamics. In the presence of constant perpendicular equilibrium gradients, it has been observed that the equations of electrostatic drift kinetics possess a symmetry associated with their intrinsic scale invariance, in both the collisionless and collisional limits. Under the assumptions of spatial periodicity, stationarity and locality, this symmetry has been shown to imply a particular scaling \cref{eq:dk_heat_flux_final} of the heat flux $Q_\s$ with the parallel system size $L_\parallel$, viz., $Q_\s \propto L_\parallel$ (or $\propto L_\parallel^2$, in the collisional limit), with the dependence on the equilibrium temperature gradient following from dimensional analysis (in the absence of other equilibrium gradients). This macroscopic transport prediction was then confirmed numerically in an electron-scale, collisional fluid model of electrostatic turbulence driven by the electron-temperature gradient. A critically balanced, constant-flux cascade of energy from some large, outer scale --- at which energy is effectively injected by the (collisional) slab ETG instability --- to small scales was then shown to be the microscale dynamics consistent with these macroscopic transport properties. This is one of only two extant numerical demonstrations of the existence of such a state in gradient-driven turbulence of this kind --- the other being \cite{barnes11}, for gyrokinetic ITG turbulence. 

Two key observations can be made from our results: (i) the effects of dissipation associated with parallel thermal conduction play a key role in determining the saturated state of the turbulence, limiting the cascade of free energy in parallel wavenumbers by clamping it to the ``dissipation ridge", or line of ``critical-balance", in wavenumber space (Landau damping could play an analogous role in the collisionless limit); (ii) the outer scale of the turbulence is determined by the breaking of drift-kinetic scale invariance due to the existence of some long-scale parallel inhomogeneity --- in our case, this was the finite size of our periodic box. In more realistic plasma systems like the tokamak, this could be the connection length $L_\parallel \sim qR$, or some shorter scale associated with inhomogeneities in the equilibrium magnetic field, such as in the tokamak edge or in stellarators. 

These results demonstrate that the details of the (parallel) plasma equilibrium play a central role in determining the microscopic outer scale for the turbulence, and thus the saturated amplitudes to which turbulent fluctuations will grow, which in turn determine the observed macroscopic transport properties. The fact that turbulence in gradient-driven systems appears to behave similarly to those in which energy is injected by explicitly large-scale processes is encouraging from the perspective of theory, as it suggests that existing insights into, and experience of, the latter can be applied to the former, significantly less well-studied case. 

Though the implications of drift-kinetic scale invariance were investigated here in a reduced model of ETG-driven turbulence, they nevertheless have implications for more realistic plasma systems due to strong constraints placed on the system by the resultant symmetry of the governing equations. {Indeed, it must be stressed that while the physical regimes covered by the reduced model are limited in scope --- lacking dynamics associated with, e.g., gradients of the plasma density or equilibrium magnetic field, kinetic effects, etc. --- the scaling \cref{eq:dk_heat_flux_final} of the heat flux suffers from no such limitations since it follows directly from the scale invariance of the electrostatic drift-kinetic system of equations.} The existence of {this scaling}, however, is predicated on the adoption of the drift-kinetic limit. Restoring FLR effects by reverting to the (electrostatic) gyrokinetic equation will evidently break the scale invariance, as the scales $\kperp \sim \rho_\s^{-1}$ will now appear explicitly in the equations through the Bessel functions \citep[see, e.g.,][]{abel13}. Apart from some interesting exceptions \citep{parisi20,parisi22}, the general effect of these Bessel functions is to provide a cutoff for instabilities at large perpendicular wavenumbers, restricting the region of instability on the ultraviolet side and providing a sink of energy (dissipation) beyond the wavenumbers where the sETG growth rate peaks. The constant-flux arguments of \cref{sec:inertial_range_dynamics} assumed that there was sufficient separation between the outer scale and these dissipation regions in order to allow an inertial range to develop at the intermediate scales.  Should such a separation exist, the system will effectively be drift-kinetic in the inertial range and, crucially, at the outer scale, where our results will continue to apply, despite the system being fully gyrokinetic. In other words, even if drift-kinetic scale invariance is broken at small scales, the assumption behind \cref{eq:dk_heat_flux_final} is that the transport is set by the outer scale, which is in the drift-kinetic limit, and the relevant breaking of scale-invariance is done by~$L_\parallel$. This is indeed what we observed in \cref{sec:inertial_range_dynamics}: despite the breaking of scale invariance at small perpendicular scales due to the introduction of hyperviscosity, our simulation results still confirmed the scaling \cref{eq:dk_heat_flux_final} as the well-defined outer scale was still set by~$L_\parallel$. We acknowledge, however, that the scale separation required for such a state is far from guaranteed: non-zero magnetic shear, for example, can create long-wavelength modes with binormal wavenumbers $k_y \rho_i \sim 1$ but narrow radial structures near mode-rational surfaces on the scale $k_x \rho_e \sim 1$ \citep{hardman22,hardman23beta,parisi22}. Whether our results are robust to the effects of significant magnetic shear and other forms of equilibrium shaping that can amplify the importance of FLR --- and thus undermine the possible separation between FLR effects and a putative outer scale --- is a subject for future work.

A key assumption behind the scaling \cref{eq:dk_heat_flux_final} is that the heat flux is able to reach a (statistical) steady state. The existence of such a state, however, is less assured than one might think: indeed, it has been known for some time that nonlinear saturation can fail to occur in simulations of electron-scale, gradient-driven turbulence due to the persistence of large-scale streamer structures in the absence of flow shear or a non-adiabatic ion response \citep{joiner06,candy07,roach09,guttenfelder11shear}. In our simple fluid model, we too find that introducing magnetic drifts associated with an inhomogeneous equilibrium magnetic field is sufficient to reproduce this behaviour. In such simulations, the curvature-mediated ETG instability \citep{horton88,adkins22}, absent in a straight magnetic field, gives rise to nonlinearly robust, large-scale streamer structures that cause unbounded growth of the heat flux with time. Further details of these simulations can be found in \cref{app:finite_magnetic_field_gradients}. This behaviour is consistent with the view that the adiabatic ion response \cref{eq:quasineutrality_final} is insufficient to saturate ETG-scale turbulence in the presence of finite magnetic-field gradients \citep{hammett93}. It must be stressed that this lack of saturation does not break the drift-kinetic scale invariance \cref{eq:dk_invariance}, which is valid for any constant perpendicular equilibrium gradients, including those of the equilibrium magnetic field --- it merely demonstrates that the steady state required to deduce \cref{eq:dk_heat_flux_final} from \cref{eq:dk_invariance} may not always be achievable in the regime of interest. Indeed, if we had been able to find a case of turbulence driven by the curvature-mediated ETG instability that saturated, we would have expected the scaling \cref{eq:dk_heat_flux_final} for the corresponding heat flux (although not necessarily the same detailed inertial-range structure as for the sETG turbulence that we studied above), but in any event, no such saturated cases have so far presented themselves.

Another limiting assumption of our work was the electrostatic nature of the turbulence. The existence of finite electromagnetic perturbations also formally breaks the (electrostatic) drift-kinetic symmetry observed in \cref{sec:drift_kinetic_scale_invariance} (this is manifest on inspection of the equations of electromagnetic gyrokinetics). However, this does not necessarily imply that the heat-flux scaling \cref{eq:dk_heat_flux_final} can never be realised in systems with finite beta. Indeed, we argued above that this scaling would still hold in the presence of FLR effects if the outer scale of the turbulence remained within the drift-kinetic limit, despite scale invariance being formally broken at the smallest spatial scales. A similar argument is applicable here. If the outer scale lies at scales sufficiently \textit{smaller} than those on which electromagnetic effects are important \citep[the `flux-freezing scale', determined by the electron inertia in the collisionless limit, or resistivity in the collisional one; see][]{adkins22}, then the scaling \cref{eq:dk_heat_flux_final} will continue to hold as, once again, the assumption behind it is that the transport is set by the outer scale located in the electrostatic, drift-kinetic limit, and the relevant breaking of scale invariance is done by $L_\parallel$, rather than the flux-freezing scale. For example, \cite{chapman22} performed nonlinear, electromagnetic simulations of JET-ILW pedestals for $\kperp \rho_i \gtrsim 1$, and observed the same scaling of the heat flux with $L_T$ as \cref{eq:dk_dimensional_analysis} at gradients sufficiently far above the linear threshold. However, if the outer scale lies on scales larger than the flux-freezing scale, i.e., if the turbulence is truly electromagnetic, then the constraints imposed by the scale invariance of electrostatic drift kinetics is lifted. Given that such regimes will likely be realised within tokamak-relevant reactor scenarios \citep[see, e.g.,][]{shimomura01,sips05,patel21} due to higher experimental values of the plasma beta (the ratio of the thermal to magnetic pressures), a central focus of ongoing research is the extent to which any of the general physical conclusions of this paper carry over into truly electromagnetic systems of tokamak turbulence.

\section*{Acknowledgements}
We are indebted to 
G. Acton,
M. Barnes,
W. Clarke,
S. Cowley,
and
W. Dorland
for helpful discussions and suggestions at various stages of this project.

\section*{Funding}
This work was supported by the Engineering and Physical Sciences Research Council (EPSRC) [EP/R034737/1]. TA was previously supported by a UK EPSRC studentship. His work was carried out within the framework of the EUROfusion Consortium and has received funding from the Euratom research and training programmes 2014–2018 and 2019–2020 under Grant Agreement No. 633053, and from the UKRI Energy Programme (EP/T012250/1). The views and opinions expressed herein do not necessarily reflect those of the European Commission. The work of AAS was also supported in part by a grant from STFC (ST/W000903/1) and by the Simons Foundation via a Simons Investigator award.

\section*{Declaration of interests}
The authors report no conflicts of interest.


\begin{appendix}
	
\section{Derivation of scale invariance}
\label{app:derivation_of_drift_kinetic_scale_invariance}
In this appendix, we demonstrate explicitly that electrostatic drift kinetics remains invariant under the transformation \cref{eq:dk_invariance}, which leads to the heat-flux scaling \cref{eq:dk_heat_flux_final}.

We take as our starting point the electrostatic drift-kinetic system, in which the perturbed distribution function for species $\s$ consists of a Boltzmann and gyrokinetic parts
\begin{align}
	\df_\s(\vec{r}, \vec{v}, t) = - \frac{q_\s \phi}{T_{0\s}} f_{0\s}(\vec{r}, \vec{v}) + h_\s (\vec{r},\vec{v}, t)	,
	\label{eq:delta_f}
\end{align}
and $h_s$ evolves according to 
\begin{align}
	& \frac{\partial }{\partial t} \left(h_\s - \frac{q_\s \phi }{T_{0\s}} f_{0\s} \right) + \left( \vpar \vec{b}_0 + \vec{v}_{d\s} \right) \cdot \gradd h_\s  + \frac{c}{B_0} \vec{b}_0 \cdot \left[ \gradd \phi \times \gradd \left( h_\s + f_{0\s} \right) \right] = \sum_{\s'} C_{\s\s'}^{(l)}[h_\s].\label{eq:invariance_drift_kinetics}   
\end{align}
Here, and in what follows, $f_{0\s}$ is the Maxwellian equilibrium distribution of species $s$ with density $n_{0\s}$ and temperature $T_{0\s}$, and $\phi$ is the perturbed electrostatic potential. The magnetic-drift velocity arising from the inhomogeneities in the equilibrium magnetic field is given by
\begin{equation}
	\vec{v}_{d\s} = \frac{\vec{b}_0}{\Omega_\s} \times \left( \vpar^2 \vec{b}_0\cdot \grad\vec{b}_0 + \frac{1}{2}\vperp^2 \grad\log B_0 \right),
	\label{eq:magnetic_drifts}
\end{equation}
where $\vec{b}_0 = \vec{B}_0/B_0$ is the direction of the equilibrium magnetic field, $B_0 = |\vec{B}_0|$ is its magnitude, and $\Omega_\s = q_\s B_0/m_\s c$, $q_\s$ and $m_\s$ are the Larmor frequency, charge and mass of species $\s$, respectively. The collision term on the right-hand side of \cref{eq:invariance_drift_kinetics} is the linearised Landau collision operator
\begin{align}
	C_{\s \s'}^{(l)}\left[h_\s \right] = \frac{\gamma_{\s \s'}}{m_\s} \gradd_v \cdot \int \rmd^3 \vec{v}' \: & f_{0\s}(v) f_{0\s '}(v') (\gradd_w \gradd_w w ) \label{eq:landau_operator} \\
	& \cdot \left[\frac{1}{m_\s} \gradd_v  \frac{h_\s(\vec{v})}{f_{0\s}(v)}  -\frac{1}{m_{\s '}}\gradd_{v'}  \frac{h_{\s '}(\vec{v}')}{f_{0\s '}(v')} \right], \nonumber
\end{align}
where $w= |\vec{w}|$, $\vec{w} = \vec{v} - \vec{v}'$, $ \gamma_{\s \s'} = 2\pi q_\s^2 q_{\s '}^2 \log \Lambda $, and all velocity derivatives are evaluated at constant position $\vec{r}$. Finally, \cref{eq:invariance_drift_kinetics} is closed by quasineutrality:
\begin{align}
	0  = \sum_{\s} q_\s \delta n_\s & = \sum_\s q_\s \left[ -\frac{q_\s \phi}{T_{0\s}}  n_{0\s}    + \int \rmd^3 \vec{v}   h_\s \right].
	\label{eq:quasineutrality} 
\end{align}

In what follows, it will be useful to decompose $h_\s$ into parts that are even and odd in the parallel velocity $\vpar$, viz.,
\begin{align}
	h_\s^\text{even} (\vec{r}, \vpar, \vperp, t) &= \frac{1}{2 }\left[h_\s(\vec{r}, \vpar, \vperp, t) + h_\s(\vec{r}, -\vpar, \vperp, t)\right],
	\label{eq:invariance_h_even} \\
	h_\s^\text{odd} (\vec{r}, \vpar, \vperp, t) & = \frac{1}{2 }\left[h_\s(\vec{r}, \vpar, \vperp, t) - h_\s(\vec{r}, -\vpar, \vperp, t)\right]. 
	\label{eq:invariance_h_odd}
\end{align}
It follows straightforwardly from \cref{eq:invariance_drift_kinetics} and \cref{eq:landau_operator} that $h_\s^\text{even}$ and $h_\s^\text{odd}$ satisfy, respectively,
\begin{align}
	& \frac{\partial }{\partial t} \left(h_\s^\text{even} - \frac{q_\s \phi }{T_{0\s}} f_{0\s} \right) +\vpar \vec{b}_0 \cdot \gradd h_\s^\text{odd} + \vec{v}_{d\s} \cdot \gradd h_\s^\text{even} \label{eq:invariance_h_even_equation} \\
	& \quad \quad + \frac{c}{B_0} \vec{b}_0 \cdot \left[ \gradd \phi  \times \gradd h_\s^\text{even}\right] + \frac{c}{B_0} \vec{b}_0 \cdot \left[\gradd \phi  \times \gradd f_{0\s}\right] = \sum_{\s'} C_{\s\s'}^{(\ell)} \left[h_\s^\text{even} \right], \nonumber   \\
	&\frac{\partial h_\s^\text{odd}}{\partial t} +\vpar \vec{b}_0 \cdot \gradd h_\s^\text{even} + \vec{v}_{d\s} \cdot \gradd h_\s^\text{odd} + \frac{c}{B_0} \vec{b}_0 \cdot \left[\gradd \phi \times \gradd h_\s^\text{odd}\right] = \sum_{\s'} C_{\s\s'}^{(\ell)} \left[h_\s^\text{odd} \right]. \label{eq:invariance_h_odd_equation}  
\end{align}
The quasineutrality condition \cref{eq:quasineutrality} becomes
\begin{align}
	0  = \sum_\s q_\s \left[ -\frac{q_\s \phi}{T_{0\s}}  n_{0\s}    + \int \rmd^3 \vec{v} \:   h_\s^\text{even} \right].
	\label{eq:invariance_quasineutrality}
\end{align}
Note that in \cref{eq:invariance_h_even_equation} and \cref{eq:invariance_h_odd_equation}, we have assumed that the (radial) gradient of the equilibrium distribution function $\gradd f_{0\s}$ is an even function of $\vpar$ --- this is only the case in systems without any equilibrium flows.  

We now wish to consider transformations of the system of equations \cref{eq:invariance_h_even_equation}-\cref{eq:invariance_quasineutrality} that can be made whilst preserving the size of perpendicular equilibrium gradients. It is obvious from considering, e.g., the magnetic-drift velocity \cref{eq:magnetic_drifts} that any rescaling of the velocity variables $\vpar$ and $\vperp$ --- at fixed equilibrium magnetic-field strength --- would require a compensatory rescaling of $\gradd \log B_0$ and $|\vec{b}_0 \cdot \gradd \vec{b}_0|$ in order to preserve the magnitude and direction of $\vec{v}_{d\s}$. Therefore, we will henceforth restrict ourselves to transformations involving only the spatial and time coordinates. In a similar vein to \cite{connor97}, we consider the following one-parameter transformation:
\begin{align}
	\tilde{h}_\s^\text{even} & = \lambda^{a_\mathrm{e}} \: h_\s^\text{even} (x/\lambda^{a_\perp}, y/\lambda^{a_\perp}, z/\lambda^{a_\parallel},  t/\lambda^{a_t}), \label{eq:invariance_transformation_h_even} \\
	\tilde{h}_\s^\text{odd} & = \lambda^{a_\mathrm{o}} \: h_\s^\text{odd} (x/\lambda^{a_\perp}, y/\lambda^{a_\perp}, z/\lambda^{a_\parallel},  t/\lambda^{a_t}),\label{eq:invariance_transformation_h_odd}  \\
	\tilde{\phi} & = \lambda^{a_\mathrm{e}} \: \phi (x/\lambda^{a_\perp}, y/\lambda^{a_\perp}, z/\lambda^{a_\parallel},  t/\lambda^{a_t}), \label{eq:invariance_transformation_phi}
\end{align}
where $x$, $y$ and $z$ are the radial, binormal and parallel (to the magnetic field) coordinates, respectively, the tildes indicate the transformed distribution functions and fields, and $a_i$ are real constants parametrising the transformation. Quasineutrality \cref{eq:invariance_quasineutrality} implies that the amplitudes of the `even' fields must be rescaled in the same way, as in \cref{eq:invariance_transformation_h_even} and \cref{eq:invariance_transformation_phi}, while the rescaling of the amplitude of $h_\s^\text{odd}$ remains unconstrained. The spatial and time coordinates can then be rescaled independently, with the caveat that the radial and binormal coordinates should be rescaled in the same way in order not to rule out perpendicular isotropy. The rescaling \cref{eq:invariance_transformation_h_even}-\cref{eq:invariance_transformation_phi} is the most general one-parameter transformation of electrostatic drift kinetics that can be made while allowing (although not requiring) the spatial isotropy of structures in the perpendicular plane.

The constants $a_i$ can be fixed by demanding that the transformation leave \cref{eq:invariance_h_even_equation} and~\cref{eq:invariance_h_odd_equation} invariant. In the collisionless limit, the collision operator can be neglected and it is easy to show that $a_\mathrm{e} = a_\mathrm{o} = a_\perp = a_\parallel = a_t$ is the only choice that fulfils this condition. The collisional limit is somewhat more subtle. As we have done throughout this paper, we order $\omega \sim (\kpar \vthe)^2/\nu_{\s\s'} \ll \nu_{\s\s'}$ and $\omega h_\s^{\text{even}} \sim \kpar \vths h_\s^\text{odd}$. In the resultant collisional expansion, the collision operator is forced to vanish at leading order (see \cref{app:zeroth_order}), and can only survive at higher order due to the presence of finite-Larmor-radius effects (see \cref{app:second_order}), {\tiny }which are neglected within the drift-kinetic approximation. At first order, one obtains, from \cref{eq:invariance_h_odd_equation}, a balance between the parallel streaming of $h_\s^{\text{even}}$ and the collision operator acting on $h_\s^\text{odd}$ (see \cref{app:first_order}). At second order, one evolves $h_\s^\text{even}$ via \cref{eq:invariance_h_even_equation} with the collision operator neglected (see \cref{app:second_order}). One can then show that $a_\mathrm{e} = 2 a_\mathrm{o} = a_\perp = 2 a_\parallel = a_t$ is the only choice of parameters that leaves the drift-kinetic equations invariant. Any constraints on $a_i$ inferred from \cref{eq:invariance_transformation_h_even}-\cref{eq:invariance_quasineutrality} in this way are only valid to second order within the collisional expansion, and not to any higher orders. However, given that the solvability conditions \cref{eq:zeroth_solvability} and \cref{eq:first_solvability} guarantee that a closed system can be obtained solely from these two orders, this is not a problematic limitation.

Thus, it follows from the above discussion that electrostatic drift kinetics is invariant under the transformation 
\begin{align}
	\tilde{h}_\s^\text{even} & = \lambda^{2} \: h_\s^\text{even} (x/\lambda^{2}, y/\lambda^{2}, z/\lambda^{2/\alpha},  t/\lambda^{2}), \label{eq:invariance_transformation_h_even_electrostatic} \\
	\tilde{h}_\s^\text{odd} & = \lambda^{2/\alpha} \: h_\s^\text{odd} (x/\lambda^{2}, y/\lambda^{2}, z/\lambda^{2/\alpha},  t/\lambda^{2}),\label{eq:invariance_transformation_h_odd_electrostatic}  \\
	\tilde{\phi} & = \lambda^{2} \: \phi (x/\lambda^{2}, y/\lambda^{2}, z/\lambda^{2/\alpha},  t/\lambda^{2}), \label{eq:invariance_transformation_phi_electrostatic} 
\end{align}
where we have chosen $a_\mathrm{e} = 2$ without loss of generality, and $\alpha = 1,2$ in the collisionless and collisional limits, respectively. The transformation of the odd and even parts of the distribution function $h_\s$ is inherited by its moments that are odd and even in $\vpar$; e.g., the temperature perturbation, being a velocity moment that is even in $\vpar$, viz., 
\begin{align}
	\frac{\delta T_\s}{T_{0\s}} = \frac{2}{3n_{0s}} \int \rmd^3 \vec{v} \left(\frac{m_\s v^2}{2T_{0\s}} - \frac{3}{2} \right) h_\s^{\text{even}},
	\label{eq:invariance_temperature_perturbation}
\end{align}
will transform according to \cref{eq:invariance_transformation_h_even_electrostatic}. This, when combined with the transformation \cref{eq:invariance_transformation_phi_electrostatic} of the electrostatic potential $\phi$ gives exactly \cref{eq:dk_invariance}, which is the starting point for the deductions presented in the main text. 

\section{Derivation of collisional fluid model}
\label{app:derivation_of_collisional_fluid_model}
This appendix details a self-contained derivation of the electron-fluid equations \cref{eq:density_moment}-\cref{eq:temperature_moment}. An alternative route to these via a subsidiary expansion of a more general system of (electromagnetic) equations can be found in \cite{adkins23thesis} \citep[see also appendix G of][]{adkins22}. In what follows, \cref{app:collisional_electron_scale_ordering} describes and physically motivates our electron-scale, collisional ordering, which is then implemented to derive equations describing our ion and electron dynamics in appendices~\ref{app:ion_kinetics_and_field_equations} and \ref{app:electron_fluid_equations}, respectively. Although the magnetic geometry adopted throughout the majority of this paper is that of a conventional slab (see \cref{sec:collisional_fluid_model}), we shall here consider the more general case in which the equilibrium (mean) magnetic field $\vec{B}_0$ is assumed to have the scale length and radius of curvature
\begin{align}
	L_B^{-1} = - \frac{1}{B_0} \frac{\rmd B_0}{\rmd x}, \quad R^{-1} = \left| \vec{b}_0 \cdot \gradd \vec{b}_0 \right|,
	\label{eq:magnetic_gradients}
\end{align}
assumed constant across the domain. Doing so will allow us to capture the effects of the magnetic drifts on our plasma while retaining most of the simplicity associated with conventional slab gyrokinetics \citep{howes06,newton10,ivanov20,ivanov22,adkins22}. Note that for a low-beta plasma, $R = L_B$.

\subsection{Collisional, electron-scale ordering}
\label{app:collisional_electron_scale_ordering}
In our model, we would like to be able to capture, at a minimum, the physics associated with drift waves, perpendicular advection by both magnetic drifts and $\vec{E}\times \vec{B}$ flows, and parallel heat conduction. Therefore, we postulate an asymptotic ordering in which the frequencies $\omega$ of the perturbations in the plasma are comparable to the characteristic frequencies associated with these phenomena, viz.,
\begin{align}
	\nu_{ee} \sim \nu_{ei} \gg \omega  \sim \omega_{*\s} \sim \omega_{d\s} k_\perp v_E \sim \kappa k_\parallel^2,  
	\label{eq:ordering_timescales_initial}
\end{align} 
where 
\begin{align}
	\omega_{*\s} = \frac{k_y \rho_\s \vths}{2L_{T_\s}}, \quad \omega_{d\s} = \frac{k_y \rho_\s \vths}{2L_B}
	\label{eq:definition_timescales}
\end{align}
are the drift and magnetic-drift frequencies, respectively, $\vec{v}_E = c \vec{E} \times \vec{B}/B^2$ is the $\vec{E} \times \vec{B}$ drift velocity ($c$ is the speed of light), $\kappa \sim \vthe^2/\nu_{ei}$ is the electron thermal diffusivity, and  
\begin{align}
	\nu_{ei} = \frac{4\sqrt{2\pi}}{3} \frac{e^4 n_{0e} \log  \Lambda}{m_e^{1/2} T_{0e}^{3/2}} ,\quad  \nu_{ee}= \frac{\nu_{ei}}{Z}
	\label{eq:definition_collision_frequencies}
\end{align}
are the electron-ion and electron-electron collision frequencies, respectively, $\log\Lambda$ being the Coulomb logarithm \citep{braginskii65,helander05}.

The ordering of the parallel conduction rate with respect to the drift frequency gives us a constraint relating parallel and perpendicular wavenumbers:
\begin{align}
	\kappa \kpar^2 \sim \omega_{*\s} \sim \omega_{d\s} \sim \kperp \rho_e \frac{\vthe}{L} \quad \Rightarrow \quad (\kpar L)^2 \sim \frac{L}{\lambdae} \kperp \rho_e ,
	\label{eq:ordering_conduction_vs_drift_waves}
\end{align}
where $\lambdae = \vthe/\nu_{ei}$ is the electron-ion mean free path and $L$ is some (perpendicular) equilibrium length scale, $L\sim L_{T_\s} \sim L_B \sim R$. The ordering of the parallel conduction rate with respect to the $\vec{E} \times \vec{B}$ drifts determines the size of perpendicular flows within our system:
\begin{align}
	\kappa \kpar^2 \sim \kperp v_E \quad \Rightarrow \quad \frac{v_E}{\vthe} \sim \frac{\kpar}{\kperp} \kpar \lambdae \sim \frac{\rho_e}{L} \equiv \epsilon,
	\label{eq:ordering_exb_flows}
\end{align}
where $\epsilon = \rho_e/L$ is the gyrokinetic small parameter \citep[see, e.g.,][]{abel13}, mandating small-amplitude, anisotropic perturbations. The frequency of these perturbations is small compared to the Larmor frequencies of both the electrons and ions:
\begin{align}
	\frac{\omega}{\Omega_e} \sim \frac{\kperp v_E}{\Omega_e} \sim \kperp \rho_e \epsilon, \quad \frac{\omega}{\Omega_i} = \frac{m_i}{Z m_e} \frac{\omega}{\Omega_e}\sim \kperp \rho_e \epsilon \frac{ m_i}{m_e}.
	\label{eq:ordering_frequencies_initial}
\end{align}
The ordering \cref{eq:ordering_exb_flows} of $v_E$ relative to the electron thermal velocity allows us to order the amplitude of the perturbed scalar potential $\phi$:
\begin{align}
	\frac{v_E}{\vthe} \sim \frac{c}{B_0} \frac{k_\perp \phi}{\vthe} \sim k_\perp\rho_e \frac{e\phi}{T_{0e}} \quad \Rightarrow \quad \frac{e\phi}{T_{0e}} \sim \frac{ \epsilon}{\kperp \rho_e}.
	\label{eq:ordering_phi}
\end{align}
The density perturbations $\dn_\s$ are ordered anticipating a Boltzmann density response and the temperature perturbations $\delta T_s$ are assumed comparable to them:
\begin{align}
	\frac{\dTe}{T_{0e}} \sim \frac{\delta T_i}{T_{0i}} \sim \frac{\delta n_{i}}{n_{0i}} =\frac{\dne}{n_{0e}}  \sim \frac{e\phi}{T_{0e}} \sim \frac{ \epsilon}{\kperp \rho_e}.
	\label{eq:ordering_field_amplitudes_electrostatic}
\end{align}
Finally, for the ordering of perpendicular magnetic-field perturbations, we demand that the effects of Lorentz tension (equivalently, of parallel compressions) must always be large enough to have an effect on the electron density perturbation, viz. [cf. \cref{eq:density_moment}],
\begin{align}
	\frac{\rmd}{\rmd t} \frac{\dne}{n_{0e}} \sim \gpar u_{\parallel e} \sim \frac{c}{4\pi e n_{0e}} \gpar \left[ \vec{b}_0 \cdot \left( \gradd_\perp \times \dBperp \right) \right] \quad \Rightarrow \quad  \frac{\dBperp}{B_0} \sim \beta_e\frac{\kpar \lambdae}{\kperp \rho_e} \frac{e\phi}{T_{0e}},
	\label{eq:ordering_dBperp}
\end{align}
where $\beta_e = 8\pi n_{0e} T_{0e}/B_0^2$ the electron plasma beta.
The (compressive) parallel magnetic-field perturbations are ordered anticipating pressure balance:
\begin{align}
	\frac{\dBpar}{B_0}  = \frac{4\pi}{B_0^2} \delta \left( \frac{B^2}{8\pi} \right) \sim \frac{4\pi}{B_0^2} \delta (n_\s T_\s) \sim \beta_e \frac{\dTe}{T_{0e}} \sim \frac{\epsilon \beta_e}{\kperp \rho_e}.
	\label{eq:ordering_dBpar}
\end{align}

The orderings \cref{eq:ordering_conduction_vs_drift_waves}-\cref{eq:ordering_dBpar} still allow for a choice of ordering for perpendicular wavenumbers $\kperp$ with respect to the electron and ion Larmor radii. Given that we would like to obtain a set of electrostatic equations that exhibit the scale invariance discussed in \cref{sec:drift_kinetic_scale_invariance}, we consider wavenumbers 
\begin{align}
	\beta_e\frac{\lambdae}{L} \ll \kperp \rho_e  \ll \frac{\lambdae}{L},
	\label{eq:ordering_sigma}
\end{align}
for which the physical motivation is discussed in \cref{sec:scale_invariance}. 
In terms of time scales, \cref{eq:ordering_sigma} is equivalent to demanding that
\begin{align}
	(\kperp \rho_e)^2 \nu_{ee} \ll \omega  \sim \omega_{*\s} \sim k_\perp v_E \sim \kappa k_\parallel^2 \ll (\kperp d_e )^2 \nu_{ei},
	\label{eq:ordering_timescales_sigma}
\end{align}
where $d_e = \rho_e/\sqrt{\beta_e}$ is the electron inertial scale. We shall formalise \cref{eq:ordering_sigma} by demanding that $\kperp \rho_e \sim \arbnorm \lambdae/L$, where $\arbnorm$ is a placeholder constant satisfying $\beta_e \ll \sigma \ll 1$. In other words, $\kperp \rhoperp \sim 1$, where $\rhoperp = \rho_e L/\lambdae \arbnorm$, an ``intermediate" spatial scale [cf. \cref{eq:col_wavenumber_range_rhoperp}].

To summarise, \cref{eq:ordering_conduction_vs_drift_waves}-\cref{eq:ordering_sigma} imply the following ordering of frequencies:
\begin{align}
	\frac{\omega}{\Omega_e} \sim \arbnorm \frac{\lambdae}{L} \epsilon, \quad \frac{\omega}{\Omega_i} \sim \frac{m_i}{m_e} \arbnorm \frac{\lambdae}{L} \epsilon,
	\label{eq:ordering_frequencies}
\end{align}
length scales:
\begin{align}
	 \kperp \rho_i \sim \arbnorm \frac{\lambdae}{L} \sqrt{\frac{m_i}{m_e}}, \quad \kperp \rho_e  \sim \arbnorm \frac{\lambdae}{L}, \quad \kpar L \sim \sqrt{\arbnorm}, \quad \frac{\kpar}{\kperp} \sim \frac{L}{\sqrt{\arbnorm}\lambdae} \epsilon,
	\label{eq:ordering_lengthscales}
\end{align} 
and amplitudes:
\begin{align}
	\frac{e\phi}{T_{0e}} \sim \frac{\dne}{n_{0e}} \sim \frac{\delta n_{i}}{n_{0i}} \sim \frac{\dTe}{T_{0e}} \sim \frac{\delta T_i}{T_{0i}} \sim \frac{L}{\arbnorm\lambdae} \epsilon, \quad \frac{\dBperp}{B_0} \sim \frac{\beta_e}{\sqrt{\arbnorm}} \frac{e \phi}{T_{0e}}, \quad \frac{\dBpar}{B_0}\sim \beta_e \frac{e \phi}{T_{0e}}.
	\label{eq:ordering_amplitudes}
\end{align}
All relevant quantities are thus naturally ordered with respect to some combination of $m_e/m_i$, $\arbnorm$, $\lambdae/L$, and the gyrokinetic small parameter $\epsilon = \rho_e/L$. The above ordering of frequencies, length scales and amplitudes with respect to $\epsilon$ is the standard gyrokinetic ordering \citep[see, e.g.,][]{abel13}. We choose to treat the ordering in $\lambdae/L$ --- the fact that this should be formally small following straightforwardly from, e.g., $\nu_{ei} \gg \omega_{*\s}$ --- as subsidiary to both the orderings in $\epsilon$ and in the mass ratio [see the first expression in \cref{eq:ordering_lengthscales}], meaning that the formal hierarchy of our expansions is
\begin{align}
	\epsilon \ll \sqrt{\frac{m_e}{m_i}} \ll \arbnorm \frac{\lambdae}{L} \ll 1,
	\label{eq:ordering_hierarchy}
\end{align}
with all other dimensionless parameters treated as finite.

\subsection{Ion kinetics}
\label{app:ion_kinetics_and_field_equations}
Given that the ordering of perpendicular wavenumbers \cref{eq:ordering_lengthscales} implies that $\kperp \rho_i \gg 1$ under the expansion in the mass ratio, the ion distribution function $h_i$ will satisfy the gyrokinetic equation \citep[see, e.g.,][]{abel13}, rather than the drift-kinetic one \cref{eq:invariance_drift_kinetics}. It is straightforward to show [by, e.g., expanding the Bessel functions therein for $\kperp \rho_i \gg 1$] that, to leading order in the mass-ratio expansion, the gyrokinetic equation is solved by
\begin{align}
	h_i = 0.
	\label{eq:ion_adiabatic_solution}
\end{align} 
The contributions to quasineutrality \cref{eq:quasineutrality} arising from the next-order solution will be of the size 
\begin{align}
	\frac{\left< h_i \right>_{\vec{r}}}{f_{0i}} \sim \left<\left<\varphi \right>_{\vec{R}_i} \right>_{\vec{r}} \sim \frac{\varphi}{\kperp \rho_i},
	\label{eq:ion_inhomogenous_size}
\end{align}
which can be safely neglected. Thus, the ion dynamics do not enter anywhere into our equations, which is the approximation of `adiabatic ions'. Given that no further reference will be made to the ion temperature gradient $L_{T_i}$, we henceforth denote the electron temperature gradient $L_{T_e} = L_T$.

\subsection{Electron fluid equations}
\label{app:electron_fluid_equations}
We now proceed with our derivation of the electron fluid equations. It will turn out that the ordering \cref{eq:ordering_lengthscales}  of perpendicular length scales means that no finite-Larmor-radius (FLR) effects need be retained within our equations --- these can only enter at second order within our expansion, but they are negligible even at this order (see \cref{app:second_order}). Furthermore, the ordering of the perpendicular and parallel magnetic field perturbations~\cref{eq:ordering_amplitudes} implies that both $\dBperp$ and $\dBpar$ can be neglected at all orders in our expansion. We thus adopt the drift-kinetic equation \cref{eq:invariance_drift_kinetics} for $s=e$ as the starting point, expanding our distribution function $h_e$ in $\arbnorm \lambdae/L \ll 1$ as
\begin{align}
	h_e = \sum_{n=0}^{\infty} h_e^{(n)}, \quad h_e^{(n)}  \sim  \left(\arbnorm \frac{\lambdae}{L} \right)^{n} \frac{e \phi}{T_{0e}} f_{0e}. 
	\label{eq:electron_distribution_function_expansion}
\end{align}

\subsubsection{Zeroth order: perturbed Maxwellian}
\label{app:zeroth_order}
Given the ordering of timescales \cref{eq:ordering_timescales_initial}, the collision operator on the right-hand side of~\cref{eq:invariance_drift_kinetics} is dominant to leading order:
\begin{align}
	C_{ee}^{(l)}\left[ h_e^{(0)} \right] + \mathcal{L}_{ei}\left[ h_e^{(0)} \right] = 0,
	\label{eq:zeroth_collision_operator}
\end{align}
where $C_{ee}^{(l)}$ is given by \cref{eq:landau_operator} for $\s = \s' = e$, and 
\begin{align}
	\mathcal{L}_{ei} \left[ h_e \right] = \frac{\gamma_{ei} n_{0e}}{m_e^2} \gradd_v \left[ f_{0e} \cdot \left(\gradd_v \gradd_v v \right) \cdot \gradd_v  \frac{h_e}{f_{0e}} \right]
	\label{eq:lorentz_collision_operator}
\end{align}
is the pitch-angle scattering (Lorentz) collision operator, valid to leading order in the mass ratio. We multiply \cref{eq:zeroth_collision_operator} by $h_e^{(0)}/f_{0e}$ and integrate over the entire phase space, yielding 
\begin{align}
	\int \frac{\rmd^3 \vec{r}}{V} \int \rmd^3 \vec{v}	 \: \frac{h_e^{(0)}}{f_{0e}} C_{ee}^{(l)}\left[ h_e^{(0)} \right] + \int \frac{\rmd^3 \vec{r}}{V} \int \rmd^3 \vec{v}	 \: \frac{h_e^{(0)}}{f_{0e}}\mathcal{L}_{ei}\left[ h_e^{(0)} \right] = 0.
	\label{eq:zeroth_h_theorem}
\end{align}
Both terms in \cref{eq:zeroth_h_theorem} are negative definite and must vanish individually, meaning that the solution is constrained to be a perturbed Maxwellian with no mean flow \citep{helander05}, viz., 
\begin{align}
	h_e^{(0)} = \left[\frac{\dne}{n_{0e}} - \varphi + \frac{\dTe}{T_{0e}} \left(\frac{v^2}{\vths^2} - \frac{3}{2} \right) \right] f_{0e},
	\label{eq:zeroth_solution}
\end{align}
where $\varphi = e\phi/T_{0e}$, and we have imposed the solvability conditions
\begin{align}
	\int \rmd^3 \vec{v} \: h_e^{(n)} = \int \rmd^3 \vec{v} \: v^2 h_e^{(n)} = 0, \quad n \geqslant 1, 
	\label{eq:zeroth_solvability}
\end{align}
in order to determine uniquely the density $\dne$ and temperature $\dTe$ moments in \cref{eq:zeroth_solution}. Note that, in general, the Lorentz collision operator constrains the electron distribution function to be isotropic in the frame moving with the parallel ion velocity. However, the parallel ion velocity is zero to all orders within our expansion in $\arbnorm \lambdae/L$ [given the adiabatic ion solution \cref{eq:ion_adiabatic_solution}], meaning that the electron distribution function will have no parallel velocity moment to leading order.

We are now in a position to simplify the quasineutrality constraint \cref{eq:quasineutrality}. Using the solutions~\cref{eq:ion_adiabatic_solution} and \cref{eq:zeroth_solution}, it straightforwardly becomes \cref{eq:quasineutrality_final}.

\subsubsection{First order: parallel flows}
\label{app:first_order}
The parallel flows are determined self-consistently from the leading-order perturbations at the next order in our expansion, viz., $h_e^{(1)}$ is determined by the solution of the Spitzer-H\"arm problem \citep{spitzer53,braginskii65,helander05}:
\begin{align}
	\vpar \frac{\partial}{\partial z} \left[\left(\frac{\dne}{n_{0e}} -\varphi + \frac{\dTe}{T_{0e}} \right) + \left( \frac{v^2}{\vthe^2} - \frac{5}{2} \right) \frac{\dTe}{T_{0e}} \right]f_{0e} =  C_{ee}^{(l)}\left[ h_e^{(1)} \right] + \mathcal{L}_{ei}\left[ h_e^{(1)} \right].
	\label{eq:first_spizter_problem}
\end{align}

This can be inverted for $h_e^{(1)}$ by means of a standard variational method. We define the functional:
\begin{align}
	\Sigma[h_e] = & - \left<h_e, C_{ee}^{(l)} \left[ h_e\right] \right> - \left<h_e , \mathcal{L}_{ei}\left[ h_e\right] \right> \nonumber \\
	&+ 2 \left<h_e, \vpar \frac{\partial}{\partial z} \left[\left(\frac{\dne}{n_{0e}} -\varphi + \frac{\dTe}{T_{0e}} \right) + \left( \frac{v^2}{\vthe^2} - \frac{5}{2} \right) \frac{\dTe}{T_{0e}} \right] f_{0e} \right>,
	\label{eq:first_functional}
\end{align}
where $\left< \dots, \dots \right>$ denotes an inner product in velocity space weighted by the inverse of the electron (Maxwellian) equilibrium $f_{0e}$. Then, considering small variations $h_e = h_\text{min} + \delta h$ and using the self-adjointness of the linearised collision operator, it is straightforward to show that the functional $\Sigma[h_e]$ has a minimum at $h_\text{min} = h_e^{(1)}$, for any variation $\delta h$ \citep[see, e.g., ][]{helander05}. Given that the spherical harmonics are eigenfunctions of the linearised collision operator, we choose to expand our distribution function in terms of spherical coordinates in velocity space $(x, \alpha, \beta)$, with $x = v^2 /\vthe^2$, as 
\begin{align}
	h_e^{(1)} = \sum_{p=0}^\infty a_p L_p^{(3/2)}(x) \vpar f_{0e}(v) = \sum_{p=0}^\infty a_p L_p^{(3/2)}(x) v \cos \alpha f_{0e}(v),
	\label{eq:first_distribution_expansion}
\end{align}
where $L_p^{(3/2)}(x)$ are the generalised Laguerre polynomials and $a_p$ coefficients to be determined. Using this in \cref{eq:first_functional}, one obtains 
\begin{align}
	\Sigma \left[h_e^{(1)} \right] = n_{0e} \vthe^2 & \left[\sum_{p=0}^\infty \sum_{q=0}^\infty \frac{a_p a_q}{2}  \left( \nu_{ee} K_{pq}^{ee} + \nu_{ei} K_{pq}^{ei} \right) \right.\label{eq:first_functional_integrals}\\
	&\quad\quad\quad\quad\quad\quad \left. + a_0  \frac{\partial}{\partial z} \left(\frac{\dne}{n_{0e}} -\varphi + \frac{\dTe}{T_{0e}} \right) - \frac{5}{2} a_1 \frac{\partial}{\partial z} \frac{\dTe}{T_{0e}}\right], \nonumber
\end{align}
where 
\begin{align}
	K_{pq}^{ee} & = - \frac{2}{n_e \nu_{ee}} \left< x^{1/2} L_p^{(3/2)}(x) f_{0e}(v) \cos \alpha, C_{ee}^{(l)} \left[x^{1/2} L_q^{(3/2)}(x) f_{0e}(v) \cos \alpha \right] \right>, \label{eq:first_kpq_ee} \\
	K_{pq}^{ei} & = -\frac{2}{n_e \nu_{ei}} \left< x^{1/2} L_p^{(3/2)}(x) f_{0e}(v) \cos \alpha, \mathcal{L}_{ei} \left[x^{1/2} L_q^{(3/2)}(x) f_{0e}(v) \cos \alpha \right] \right> \label{eq:first_kpq_ei}
\end{align}
are the coefficients calculated in, e.g., \cite{hardman22} (and references therein). Truncating \cref{eq:first_distribution_expansion} at $p = 3$, and demanding that the functional \cref{eq:first_functional_integrals} be stationary with respect to variations in the coefficients $a_p$, we find that 
\begin{align}
	h_e^{(1)} = \left[ a_0 + a_1 L_1^{(3/2)}(x) + a_2 L_2^{(3/2)}(x) \right] \vpar f_{0e},
	\label{eq:first_solution}
\end{align} 
where the coefficients are given by 
\begin{align}
	\nu_{ei} a_0  & = - \frac{\frac{217}{64} + \frac{151}{8\sqrt{2}Z} + \frac{9}{2Z^2}}{1 + \frac{61}{8 \sqrt{2}Z} +  \frac{9}{2Z^2}}\frac{\partial}{\partial z}\left(\frac{\dne}{n_{0e}} -\varphi + \frac{\dTe}{T_{0e}} \right) - \frac{\frac{5}{2}\left( \frac{33}{16} + \frac{45}{8 \sqrt{2}Z} \right)}{1 + \frac{61}{8 \sqrt{2}Z} +  \frac{9}{2Z^2}} \frac{\partial}{\partial z} \frac{\dTe}{T_{0e}}, 
	\label{eq:a_0}\\
	\nu_{ei} a_1 & = \frac{\frac{33}{16} + \frac{45}{8 \sqrt{2}Z} }{1 + \frac{61}{8 \sqrt{2}Z} +  \frac{9}{2Z^2}}\frac{\partial}{\partial z} \left(\frac{\dne}{n_{0e}} -\varphi + \frac{\dTe}{T_{0e}} \right) + \frac{ \frac{5}{2} \left( \frac{13}{4} + \frac{45}{8\sqrt{2}Z} \right) }{1 + \frac{61}{8 \sqrt{2}Z} +  \frac{9}{2Z^2}} \frac{\partial}{\partial z} \frac{\dTe}{T_{0e}}, \label{eq:a_1} \\
	\nu_{ei} a_2 & = - \frac{ \frac{3}{8} - \frac{3}{2\sqrt{2}Z}}{1 + \frac{61}{8 \sqrt{2}Z} +  \frac{9}{2Z^2}}\frac{\partial}{\partial z} \left(\frac{\dne}{n_{0e}} -\varphi + \frac{\dTe}{T_{0e}} \right)- \frac{\frac{5}{2 }\left( \frac{3}{2} + \frac{3}{2\sqrt{2}Z} \right)}{1 + \frac{61}{8 \sqrt{2}Z} +  \frac{9}{2Z^2}} \frac{\partial}{\partial z} \frac{\dTe}{T_{0e}}, \label{eq:a_2}
\end{align}
which can easily be shown to satisfy the \cite{onsager31} relations. 

The solution \cref{eq:first_solution} allows us to determine, subject to the solvability condition
\begin{align}
	\int \rmd^3 \vec{v} \: \vpar h_{e}^{(n)} = 0, \quad n \geqslant 2,
	\label{eq:first_solvability}
\end{align}
the parallel electron flow:
\begin{align}
	u_{\parallel e} = \frac{1}{n_{0e}} \int \rmd^3 \vec{v} \: \vpar h_e^{(1)}.
	\label{eq:first_amperes_law}
\end{align}
Using \cref{eq:first_solution} for $h_e^{(1)}$ in \cref{eq:first_amperes_law} and defining the (ion-charge-dependent) coefficients [cf., for $Z=1$, (C16) and (C17) in \cite{hardman22}]
\begin{equation}
	c_1 = \frac{\frac{217}{64} + \frac{151}{8\sqrt{2}Z} + \frac{9}{2Z^2}}{1 + \frac{61}{8 \sqrt{2}Z} +  \frac{9}{2Z^2}}, \spc c_2  = \frac{\frac{5}{2}\left( \frac{33}{16} + \frac{45}{8 \sqrt{2}Z} \right)}{1 + \frac{61}{8 \sqrt{2}Z} +  \frac{9}{2Z^2}}, \spc c_3 = \frac{\frac{25}{4} \left( \frac{13}{4} + \frac{45}{8 \sqrt{2} Z} \right)}{1 + \frac{61}{8 \sqrt{2}Z} + \frac{9}{2Z^2}} - \frac{c_2^2}{c_1},
	\label{eq:charge_coefficients}
\end{equation}
we obtain \cref{eq:velocity_moment}. 

\subsubsection{Second order: density and temperature evolution}
\label{app:second_order}
At second order, the electron drift-kinetic equation 
\begin{align}
	&\frac{\rmd}{\rmd t} \left( h_e^{(0)} +  \varphi f_{0e} \right) + \vpar \frac{\partial h_e^{(1)}}{\partial z}  + \vec{v}_{de} \cdot \gradd_\perp h_e^{(0)} + \frac{\rho_e \vthe}{2 L_T} \frac{\partial \varphi}{\partial y} \left( \frac{v^2}{\vthe^2} - \frac{3}{2} \right) f_{0e} \nonumber \\
	& = C^{(l)}_{e e }\left[ h_e^{(2)} \right] +  \mathcal{L}_{e i } \left[h_e^{(2)} \right], \label{eq:second_gk_equation}
\end{align}
describes the evolution of the density and temperature perturbations in \cref{eq:zeroth_solution}.
When taking the density and temperature moments of \cref{eq:second_gk_equation}, the contributions from the collision operator on the right-hand side vanish, as the electron-electron and Lorentz collision operators conserve particle number and energy to this order in our expansion. An observant reader may have noticed, however, that in starting our expansion from the drift-kinetic equation \cref{eq:invariance_drift_kinetics}, we ruled out the possibility of retaining higher-order collisional terms due to FLR motions of the electrons. Indeed, if we had instead considered the gyrokinetic equation \citep[see, e.g.,][]{abel13} and expanded the oscillatory exponential factors $ e^{\pm i \vec{k} \cdot \vec{\rho}_e}$ arising from the presence of the gyroaverages of the collision operator on its right-hand side, we would have obtained terms of the form $\sim \nu_{ee} \rho_e^2 \gradd_\perp^2 h_e^{(0)}$ at order $(\kperp \rho_e)^2$. These represent electron thermal diffusion \citep[cf.][]{newton10,ivanov20,hardman22}. Recalling the ordering of timescales \cref{eq:ordering_timescales_sigma}, however, it is clear that these terms are negligible in comparison to those on the left-hand side of \cref{eq:second_gk_equation} --- this justifies \textit{post factum} the choice to perform our expansion starting from the drift-kinetic equation \cref{eq:invariance_drift_kinetics}, rather than from the gyrokinetic one. 

Thus, taking the density moment of \cref{eq:second_gk_equation}, employing the identity [see \cref{eq:magnetic_drifts} and~\cref{eq:magnetic_gradients}]
\begin{align}
	\vec{v}_{de} \cdot \gradd_\perp h_e^{(0)} = \frac{\rho_e \vthe}{2} \left( \frac{2}{R} \frac{\vpar^2}{\vths^2} + \frac{1}{L_B} \frac{\vperp^2}{\vths^2} \right) \frac{\partial  h_e^{(0)}}{\partial y},
\end{align}
and making use of the fact that $R = L_B$ for a low-beta plasma, we find:
\begin{align}
	\frac{\rmd}{\rmd t}  \frac{\dne}{n_{0e}}+ \frac{\partial u_{\parallel e}}{\partial z}  + \frac{\rho_e \vthe}{L_B} \frac{\partial}{\partial y} \left( \frac{\dne}{n_{0e}} - \varphi + \frac{\dTe}{T_{0e}} \right) =0.
	\label{eq:density_moment_appendix}
\end{align}
For the temperature moment, we first note that, from \cref{eq:first_solution},
\begin{align}
	\frac{1}{n_{0e}} \int \rmd^3 \vec{v} \: \vpar \left( \frac{v^2}{\vthe^2} - \frac{3}{2}\right) h_e^{(1)} = \left( 1 + \frac{c_2}{c_1} \right) u_{\parallel e} + \frac{\delta q_e}{n_{0e} T_{0e}},
	\label{eq:second_temperature_vpar_moment}
\end{align}
where $\delta q_e$ is defined in \cref{eq:collisional_heatflux}. Therefore, taking the temperature moment of \cref{eq:second_gk_equation} and dividing throughout by $3/2$ yields
\begin{align}
	& \frac{\rmd}{\rmd t} \frac{\dTe}{T_{0e}}  + \frac{2}{3} \frac{\partial}{\partial z} \frac{\delta q_e}{n_{0e} T_{0e}} + \frac{2}{3} \left( 1+ \frac{c_2}{c_1} \right) \frac{\partial u_{\parallel e}}{\partial z} + \frac{2}{3} \frac{\rho_e \vthe}{L_B} \frac{\partial}{\partial y} \left( \frac{\dne}{n_{0e}} - \varphi + \frac{7}{2} \frac{\dTe}{T_{0e}} \right) \nonumber \\
	& = -\frac{\rho_e \vthe}{2 L_T} \frac{\partial \varphi}{\partial y}.
	\label{eq:temperature_moment_appendix}
\end{align} 
Neglecting the magnetic drifts, one straightforwardly obtains \cref{eq:density_moment} and \cref{eq:temperature_moment} from \cref{eq:density_moment_appendix} and \cref{eq:temperature_moment_appendix}, respectively.

\section{Case with finite magnetic-field gradients}
\label{app:finite_magnetic_field_gradients}
This appendix details the behaviour of our model system in the presence of magnetic drifts associated with an inhomogeneous equilibrium magnetic field. Assuming its scale length $L_B$ [see \cref{eq:magnetic_gradients}] to be constant across the domain, our evolution equations for the electrostatic potential and temperature perturbations are now [see \cref{eq:density_moment_appendix} and \cref{eq:temperature_moment_appendix}]:
\begin{align}
	&\frac{\partial}{\partial t} \taubar^{-1} \varphi - \frac{c_1 \vthe^2}{2 \nu_{ei}}  \frac{\partial^2 }{\partial z^2}\left[\left(1 + \frac{1}{\taubar} \right)\varphi   - \left( 1 + \frac{c_2}{c_1} \right) \frac{\dTe}{T_{0e}} \right] \label{eq:density_moment_finite} \\
	&\quad \quad \quad \:\: + \frac{\rho_e \vthe}{L_B} \frac{\partial}{\partial y} \left[ \left( 1 + \frac{1}{\taubar} \right) \varphi - \frac{\dTe}{T_{0e}} \right]= 0, \nonumber \\
	&\frac{\rmd}{\rmd t} \frac{\dTe}{T_{0e}} + \frac{2}{3 } \frac{c_1 \vthe^2}{2 \nu_{ei}} \frac{\partial^2}{\partial z^2} \left\{\left( 1 + \frac{1}{\taubar} \right)\left( 1 + \frac{c_2}{c_1} \right) \varphi - \left[\frac{c_3}{c_1} + \left( 1 + \frac{c_2}{c_1} \right)^2 \right]  \frac{\dTe}{T_{0e}} \right\} \label{eq:temperature_moment_finite} \\
	& \quad \quad \quad - \frac{2}{3} \frac{\rho_e \vthe}{L_B} \frac{\partial}{\partial y} \left[ \left( 1 + \frac{1}{\taubar} \right) \varphi - \frac{7}{2}\frac{\dTe}{T_{0e}}\right]= - \frac{\rho_e \vthe}{2 L_T}\frac{\partial \varphi}{\partial y}. \nonumber 
\end{align}
The reduction of these equations to \cref{eq:density_moment_num}-\cref{eq:temperature_moment_num} occurs for very steep electron-temperature gradients, in the limit $L_B/L_T \rightarrow \infty$.

\begin{figure}
	
	\includegraphics[width=1\textwidth]{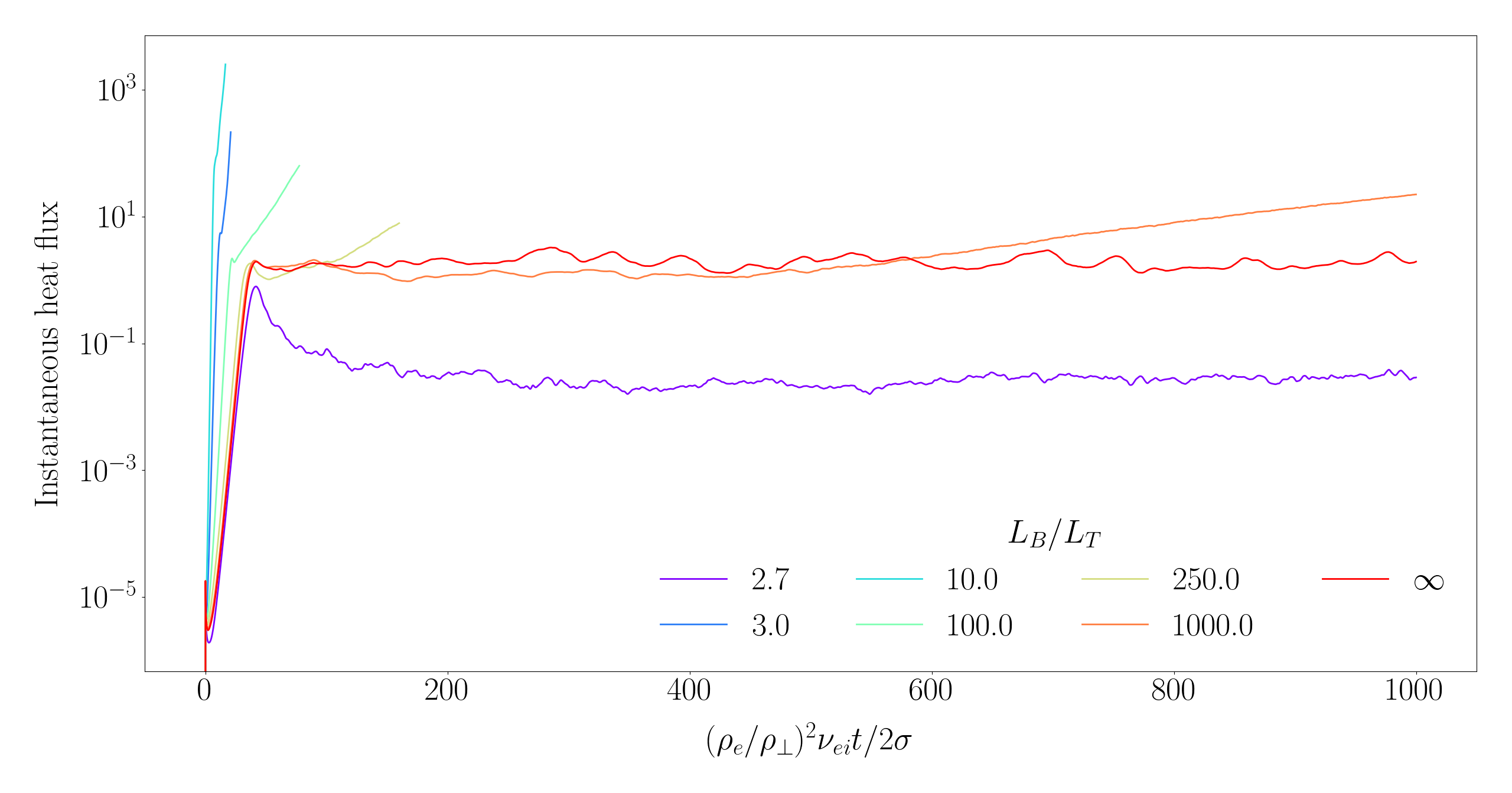}
	
	\centering
	
	\caption[]{Time traces of the instantaneous heat flux from simulations with finite $L_B/L_T$, with the limit of $L_B/L_T \rightarrow \infty$ shown for comparison. All parameters are the same as the baseline simulation (see \cref{tab:simulation_parameters}), and the heat flux is normalised to $(\rhoperp/\rho_e) Q_{\text{gB}e}$. The heat flux grows without bound in all simulations with (finite) $L_B/L_T$ above the linear critical gradient \cref{eq:critical_gradient} ($\approx 2.72$ for $\taubar= 1$), with the rate of divergence decreasing as $L_B/L_T$ is increased.}
	\label{fig:heat_flux_with_kappaB}
	
\end{figure}

The presence of the magnetic-drift terms in \cref{eq:density_moment_finite}-\cref{eq:temperature_moment_finite} introduces another instability into the system, the curvature-mediated ETG (cETG) instability \citep{horton88,adkins22}, which can, and generally does, modify its turbulent-transport properties. In particular, the turbulence theory of \cref{sec:inertial_range_dynamics} assumed that the sETG instability was the dominant source of energy injection; this is only the case at sufficiently large $L_B/L_T$, meaning that we would expect departures from the behaviour observed in \cref{sec:inertial_range_dynamics} to be most significant for $L_B/L_T$ of order unity. A series of simulations were conducted in which $L_B/L_T$ was varied, with all other parameters being kept the same as in the baseline simulation (see \cref{tab:simulation_parameters}); the heat flux from these simulations is plotted in \cref{fig:heat_flux_with_kappaB}. It is readily apparent that the introduction of finite magnetic-field gradients leads to a failure of saturation for all simulations where $L_B/L_T$ is above the linear critical gradient for the cETG instability: 
\begin{align}
	\frac{L_B}{L_T} > \frac{1}{2} \left( \taubar + \frac{40}{9} \frac{1}{\taubar^2} \right),
	\label{eq:critical_gradient}
\end{align}
a threshold that can be derived straightforwardly from \cref{eq:density_moment_finite} and \cref{eq:temperature_moment_finite} in the two-dimensional limit. 
This lack of saturation appears to persist irrespective of changes in box size, aspect ratio, and resolution in any (or all) of the coordinate directions. 

\begin{figure}
	
	\begin{tabular}{cc}
		\hspace{-1.5cm} \includegraphics[width=0.6\textwidth]{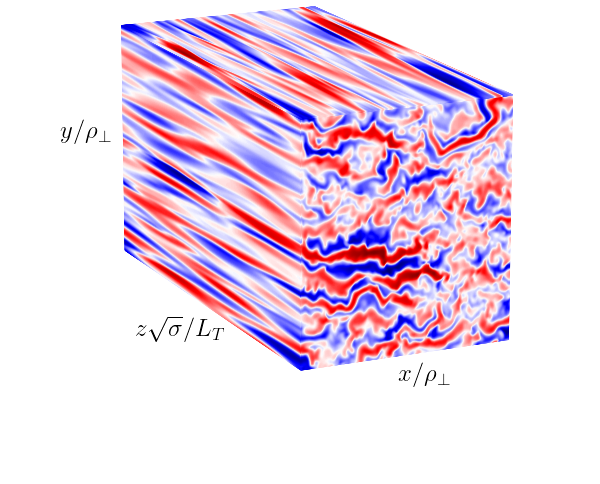} & 
		\hspace{-1.5cm} \includegraphics[width=0.6\textwidth]{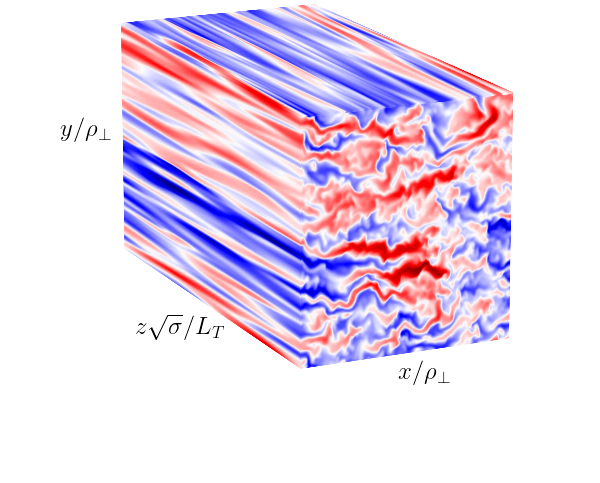} \\[-2.2em]
		(a) $(L_T/\rhoperp)\varphi$ &   (b) $(L_T/\rhoperp)\dTe/T_{0e}$
	\end{tabular}
	
	\centering
	\vspace{0.3cm}
	\caption[]{Real-space snapshots of the (a) electrostatic potential and (b) temperature perturbations from the simulation with $L_B/L_T = 1000$ from \cref{fig:heat_flux_with_kappaB}, taken at $(\rho_e/\rhoperp)^2\nu_{ei} t/2\arbnorm = 200$. The coordinate axes are as shown, while the red and blue colours correspond to regions of positive and negative fluctuation amplitude. At these early times, the turbulence appears similar to that of saturated sETG turbulence for $L_B/L_T \rightarrow \infty$ (cf. \cref{fig:high_res_real_space}), despite the eventual lack of saturation (see \cref{fig:real_space_with_kappaB}).}
	\label{fig:real_space_with_kappaB_early}
	
	\bigskip
	\bigskip
	
	\begin{tabular}{cc}
		\hspace{-1.5cm} \includegraphics[width=0.6\textwidth]{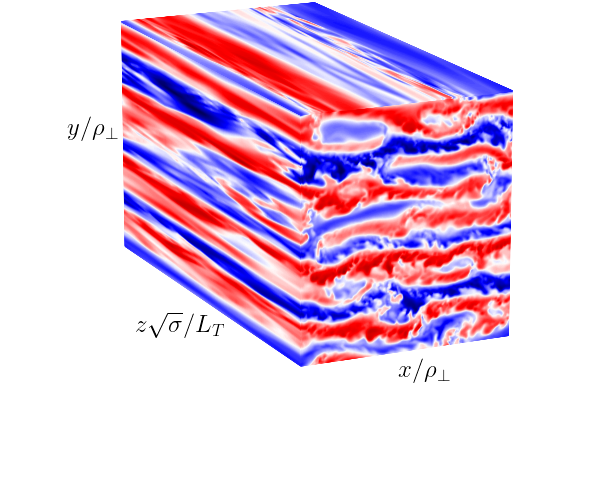} & 
		\hspace{-1.5cm} \includegraphics[width=0.6\textwidth]{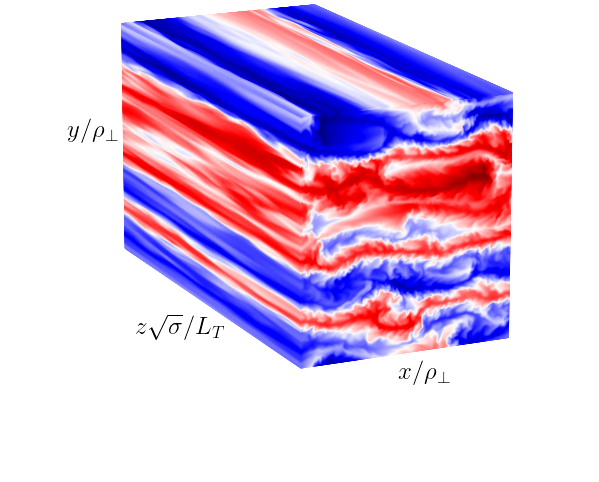} \\[-2.2em]
		(a) $(L_T/\rhoperp)\varphi$ &   (b) $(L_T/\rhoperp)\dTe/T_{0e}$
	\end{tabular}
	
	\centering
	\vspace{0.3cm}
	\caption[]{The same as \cref{fig:real_space_with_kappaB_early}, except taken at $(\rho_e/\rhoperp)^2\nu_{ei} t/2\arbnorm = 1000$.  The unbounded growth of the heat flux is associated with the formation of large-scale, approximately two-dimensional streamer structures that appear to be immune to all types of nonlinear shearing.}
	\label{fig:real_space_with_kappaB}
	
\end{figure}

As discussed in \cref{sec:perpendicular_isotropy}, the adiabatic ion response \cref{eq:quasineutrality_final} causes the nonlinearity in the electron-scale continuity equation \cref{eq:density_moment_num} to vanish identically, a property that is shared by \cref{eq:density_moment_finite}, meaning that the system lacks any nonlinearity capable of generating two-dimensional secondary instabilities that are responsible for the generation of zonal flows and destruction of streamer structures (see references in \cref{sec:perpendicular_isotropy}). Indeed, the lack of saturation in the case of our ETG simulations appears to be due to the inability of the system to break apart the streamers created by the cETG instability; the existence of such streamers causes the heat flux to diverge as they `short circuit' the heat transport across the radial domain. Even if the simulation initially appears to saturate after the linear phase, it eventually forms these large-scale streamers, which appear to be immune to all types of nonlinear shearing, as seen clearly in the real-space snapshots of cETG turbulence shown in figures \ref{fig:real_space_with_kappaB_early} and \ref{fig:real_space_with_kappaB}.
\\

This perhaps confirms the view of \cite{hammett93} that the adiabatic ion response~\cref{eq:quasineutrality_final} is insufficient to saturate ETG-scale turbulence in the presence of finite magnetic-field gradients, and one may have to resort to more inclusive closures for the ions. One such closure including scales comparable to the ion-Larmor radius is \citep[see, e.g.,][]{adkins22}
\begin{align}
	\frac{\dne}{n_{0e}} = - \taubar^{-1} \varphi + \frac{1}{n_{0i}} \int \rmd^3 \vec{v} \: \left< g_i \right>_{\vec{r}},
	\label{eq:quasineutrality_taubar}
\end{align}
where $\taubar^{-1}$ is now an operator defined as follows:
\begin{align}
	- \taubar^{-1} \varphi = - \frac{Z}{\tau} (1 - \hat{\Gamma}_0) \varphi \approx \left\{
	\begin{array}{cc}
		\displaystyle\frac{Z}{2\tau} \rho_i^2 \gradd_\perp^2 \varphi, & \displaystyle k_\perp \rho_i \ll 1,\\[4mm]
		\displaystyle - \frac{Z}{\tau} \varphi, & \displaystyle k_\perp \rho_i \gg 1,
	\end{array}
	\right.
	\label{eq:taubar_definition}
\end{align}
and the operator $\hat{\Gamma}_0$ can be expressed in Fourier space in terms of the modified Bessel function of the first kind: $\Gamma_0 = I_0(\alpha_i)e^{-\alpha_i}$, where $\alpha_i = (k_\perp \rho_i)^2/2$. The presence of the non-adiabatic ion distribution function $g_i$ in \cref{eq:quasineutrality_taubar}, however, means that one would have also to include a self-consistent treatment of ions in order to make use of this closure ($g_i =0 $ is not a solution to the ion gyrokinetic equation in the presence of finite magnetic drifts). Should this, or other similar closures, allow for saturation, this would imply that one must always appeal to (elements of) ion-scale physics for saturation of electrostatic cETG-driven turbulence. The extent to which such considerations are practically relevant, however, depends on whether or not the system being considered contains any electromagnetic physics, and thus on the value of the (electron) plasma beta $\beta_e$. Indeed, for $\beta_e \gtrsim m_e/m_i$, the ``flux-freezing scale" $d_e = \rho_e/\sqrt{\beta_e}$ is encountered before (i.e., is smaller than) the ion Larmor radius $\rho_i$ when moving towards larger perpendicular scales. Provided that the wavenumber interval between $d_e$ and $\rho_i$ is sufficiently wide to allow for the presence of electron-scale, electromagnetic instabilities, the mechanisms of saturation in such a system could be vastly different than in the electrostatic regime. This is a subject of ongoing research.

\end{appendix}

\bibliography{bibliography.bib}{}
\bibliographystyle{jpp}

\end{document}